\documentclass[fleqn,10pt,usenames,dvipsnames]{wlscirep}

\usepackage[utf8]{inputenc} %unicode support
\usepackage[T1]{fontenc}
\usepackage{amsmath,amssymb,amsfonts}
\usepackage{graphicx} % for dealing with figures.

\usepackage{amsthm}
\usepackage{xspace}
\usepackage{cite}
\usepackage{xr-hyper} % referenze a documenti terzi
\usepackage{hyperref}
\usepackage{nameref}
\usepackage{url}
\usepackage{booktabs,tabularx,ltablex,longtable}
\usepackage{eucal} % to render the mathcal font "correctly"
\usepackage{bbold} % for mathbb numbers
\usepackage{orcidlink} % to put ORCID links

\usepackage{multirow}
\usepackage[oldenum]{paralist}
\usepackage{enumitem}
\usepackage[left]{lineno}

\usepackage[utf8]{inputenc}
\usepackage{graphicx,epsfig}% Include figure files
\usepackage{times}
\usepackage{listings}

\usepackage{dcolumn,bm,epic,eepic,float}
\usepackage{amssymb,amsmath,multirow,rotate}
\usepackage{xcolor}
\usepackage{colortbl}
\usepackage{paralist}
\usepackage{booktabs,tabularx,ltablex,longtable}
\usepackage{soul}
\usepackage{fancybox}
\usepackage{textcomp}
\usepackage[normalem]{ulem}
\usepackage{calc}
\usepackage{blindtext}
\usepackage{bbold} % for mathbb numbers
\usepackage{subfig}
\usepackage{pythonhighlight}

\DeclareMathAlphabet{\mathcal}{OMS}{cmsy}{m}{n}

\begin{document}

%\documentclass[fleqn,10pt,usenames,dvipsnames]{wlscirep}

%\usepackage[utf8]{inputenc} %unicode support
%\usepackage[T1]{fontenc}
%\usepackage{amsmath,amssymb,amsfonts}
%\usepackage{graphicx} % for dealing with figures.

%\usepackage{amsthm}
%\usepackage{xspace}
%\usepackage{cite}
%\usepackage{xr-hyper} % referenze a documenti terzi
%\usepackage{hyperref}
%\usepackage{nameref}
%%\usepackage{algorithmic} for describing algorithms
%%\usepackage{subfig} for getting the subfigures e.g., "Figure 1a and 1b" etc.
%\usepackage{url}
%%\usepackage[usenames,dvipsnames]{xcolor}
%%\usepackage{natbib}
%\usepackage{booktabs,tabularx,ltablex,longtable}
%\usepackage{eucal} % to render the mathcal font "correctly"
%\usepackage{bbold} % for mathbb numbers
%\usepackage{orcidlink} % to put ORCID links

%% \usepackage{xr}
%\usepackage{multirow}
%\usepackage[oldenum]{paralist}
%\usepackage{enumitem}
%\usepackage[left]{lineno}

%%%% CUSTOM MACROS %%%%%

\newcommand{\avg}[1]{\left\langle #1 \right\rangle}
\newcommand{\norm}[1]{\bigl\lVert #1 \bigr\rVert}
\newcommand{\modu}[1]{\bigl\lvert #1 \bigr\rvert}
\newcommand{\abs}[1]{\lvert #1 \rvert}

\newcommand{\ie}{i.e.\xspace}
\newcommand{\eg}{e.g.\xspace}
\newcommand{\et}{et al.\xspace}

\newcommand{\NM}[1]{{\sf NM#1}\xspace}

\newcommand{\mariana}[1]{\textcolor{orange}{#1}}
\newcommand{\alessio}[1]{\textcolor{Plum}{#1}}
\newcommand{\ronaldo}[1]{\textcolor{blue}{#1}}

%% command to reset mathcal to default font
%\DeclareMathAlphabet{\mathcal}{OMS}{cmsy}{m}{n}

% graphic paths
\graphicspath{{figures/}}

% %%%% COMMANDS TO ALLOW CROSS-REFERENCING OVERLEAF
%\makeatletter
%\newcommand*{\addFileDependency}[1]{% argument=file name and extension
%	\typeout{(#1)}
%	\@addtofilelist{#1}
%	\IfFileExists{#1}{}{\typeout{No file #1.}}
%}
%\makeatother

%\newcommand*{\myexternaldocument}[2]{%
%	\externaldocument[#2]{#1}%
%	\addFileDependency{#1.tex}%
%	\addFileDependency{#1.aux}%
%}

% % references to the supplementary material
%\myexternaldocument{supplementary}{S-}

\title{The parenthood effect in urban mobility}

%%% Authors
\author[1,2,*]{Mariana Macedo\,\orcidlink{0000-0002-7071-379X}}
\author[3,4]{Ronaldo Menezes\,\orcidlink{0000-0002-6479-6429}}
\author[5,6,7]{Alessio Cardillo\,\orcidlink{0000-0003-4811-9978}}

\affil[1]{Constructor University, Bremen, Germany}
\affil[2]{Data Science Department, Northeastern University London, London, United Kingdom}
\affil[3]{BioComplex Laboratory, Department of Computer Science, University of Exeter, Exeter, United Kingdom}
\affil[4]{Department of Computer Science, Federal University of Cear\'a, Brazil}
\affil[5]{Department of Condensed Matter Physics, University of Barcelona, E-08028 Barcelona, Spain}
\affil[6]{University of Barcelona Institute of Complex Systems (UBICS), University of Barcelona, E-08028 Barcelona, Spain}
\affil[7]{GOTHAM Lab -- Institute for Biocomputation and Physics of Complex Systems (BIFI), University of Zaragoza, E-50018 Zaragoza, Spain\vspace{.3cm}}
\affil[*]{Corresponding author: mariana.macedo@nulondon.ac.uk}

\begin{abstract}
    We investigate how parenthood and marriage (two major life events) reshape urban mobility patterns, an aspect overlooked in traditional `average citizen' mobility models. Leveraging US census data, we analyse whether these life transitions create distinct urban experiences. Parenthood introduces new priorities including caregiving responsibilities, work-life balance adjustments, and access to family-friendly environments. Similarly, marriage introduces new dynamics including shared household decision-making, potential dual-income benefits, combined residential preferences, and shifts in social networks and lifestyle patterns. Our analysis demonstrates that cities vary significantly in how mobility can be accommodated by different household arrangements: some better accommodate either single individuals (Houston, Virginia Beach) or married people (Atlanta, Baltimore), whereas others favour parents (Cincinnati, Chicago). This classification becomes increasingly relevant for individuals and families as remote work expands relocation possibilities. We find that parents and married individuals face different mobility costs and amenity access patterns compared to their counterparts, with variations consistent across multiple null model tests. This research advances urban planning discourse by advocating for tailored design strategies addressing diverse demographic needs rather than one-size-fits-all approaches.
\end{abstract}

%\begin{document}

\flushbottom
\maketitle

\thispagestyle{empty}

% begin line numbering
%\begin{linenumbers} % begin of line numbering

\section*{Introduction}
\label{sec:intro}

Human mobility is a fundamental aspect of societal functioning, representing the movement of people to satisfy a variety of needs, whether it is commuting to work, seeking medical care, or engaging in leisure activities. Understanding and modelling this mobility is crucial for several reasons, including its impact on urban infrastructure, economic productivity, and environmental sustainability~\cite{Barbosa2018}. Mobility modelling allows researchers and policymakers to predict travel patterns, optimise transportation systems, and plan for future urban growth~\cite{alessandretti2020scales}. As urban populations continue to grow, particularly in developing regions, the importance of accurate mobility studies accounting for the diversity observed in today's world has never been greater.

Although mobility can be captured in various contexts, the vast majority of mobility studies pertains to urban environments. This focus is justified by the demographic reality that over 55\% of the global population currently resides in cities (as of 2018), a figure expected to rise to 68\% by 2050 according to the United Nations~\cite{UN2018}. In the United States, approximately 82\% of the population lived in urban areas as of the 2010 census, with this proportion expected to increase further~\cite{UN2018}. Urban areas are not only densely populated but also generate copious amounts of data~\cite{gallotti2015multilayer,lenormand2016towards,kujala2018collection,volpati2018spatial,reia2022spatial}, making them fertile ground for data-driven studies. Additionally, urbanites face unique mobility challenges and opportunities (\eg, navigating congested road networks, adapting to multi-modal transportation systems, access to a variety of resources) that differ significantly from those in rural areas, further reinforcing the urgent need to concentrate on cities when modelling human movement.

Mobility modelling is closely connected to problems addressed by multiple United Nations Sustainable Development Goals (SDGs)~\cite{SDG2015}, such as reducing carbon emissions (SDG 7 and 13), improving access to public transportation (SDG 11), and promoting inclusive and sustainable urbanisation (SDG 7 and 11). These goals underscore the need to also consider the diversity of urban residents in mobility to create cities that are not only efficient but also equitable and environmentally responsible.

Cities are inherently diverse and cosmopolitan places, composed of individuals from varied backgrounds, with different needs, preferences, and resources. Consequently, modelling urban mobility by averaging the behaviour of a generic citizen encourages oversimplification and can lead to policies that do not account for the diverse realities of city dwellers. Equity in urban planning demands that we move beyond one-size-fits-all models and, instead, develop approaches accounting for the differences in how various demographic groups experience and navigate cities.

Life-changing events are known to fundamentally alter an individual's priorities, routines, and interactions with their environment~\cite{raymore2001leaving,larouche2020effect}. Such events include major milestones like completing education, entering the workforce, retirement, and perhaps most significantly, the decision to become a parent. Parenthood introduces a new set of responsibilities and constraints profoundly reshaping daily life, including mobility patterns~\cite{mccarthy2021trajectories}. Parents may prioritise proximity to schools, childcare facilities, and safe, family-friendly neighbourhoods when deciding where to settle. The concept of a `friendly' city becomes particularly relevant in this context, as parents increasingly seek environments that cater to their needs, especially in an era where remote work has made relocating more feasible. Furthermore, it has become evident that parents' own well-being improves when they spend more time with their children~\cite{nomaguchi2020parenthood}.

Another significant life change is the decision to have a partner, whether in the form of marriage or other long-term living arrangements. The transition to being married also brings changes in lifestyle and mobility needs, such as the need for housing that accommodates two adults, increased economic status due to possible dual incomes, and shared household responsibilities. While these changes might not be as transformative as those induced by parenthood, they still represent a shift that could influence mobility behaviour. In this paper, we will refer to individuals as \emph{married} if they declared living together and indicated their marital status as married in the survey. We consider both homosexual and heterosexual couples, although we do not account for their gender in our analysis. We rely on the data collected by the American Community Survey (ACS), and we avoid counting flatmates as married individuals. 

The recognition that different groups, whether defined by parenthood, marriage, gender, or socioeconomic status, experience cities differently has gained traction in recent years. Early mobility models often treated travellers as homogeneous, indistinguishable entities, failing to grasp the diversity inherent in human populations~\cite{song2010limits,simini2012universal,barbosa2015effect}. However, this general approach has been increasingly challenged as researchers have begun to explore how these demographic differences manifest in mobility patterns, leading to more nuanced and effective urban planning strategies~\cite{lenormand2015influence,lotero2016rich,gauvin2020gender,macedo2020gender,macedo2022differences,battiston-epj_ds-2023}.

In addition to these demographic factors, the concept of multimodality and multiple scales in mobility (where individuals use multiple forms of transportation within a single trip) further complicates the picture~\cite{brockmann-nature-2006,battyscience2008, alessandretti2020scales,molkenthin-prl-2020,franca-complexity-2016, lotero2016rich,alessandretti-env_plan_b-2022,kurant-prl-2006, aleta-scirep-2017}. Different demographic groups exploit these transportation modes in varying ways, depending on factors like income, gender, and family status. For instance, parents may rely more heavily on cars for the convenience of transporting children, whereas younger, single individuals might prefer public transit or cycling.

Parenthood, in particular, brings significant costs and considerations. The financial burden of raising children, the time demands of caregiving, and the need for access to specific amenities such as schools and healthcare facilities can all influence individuals' mobility patterns. Research indicates that parents are more likely to prioritise living near essential services even if it means longer commutes to work~\cite{stgeorge2012time} or less competitive schools for the kids~\cite{bell2007space}. According to the United States Department of Agriculture, the cost of raising a child to the age of 18 in the United States averages around US\$ 233,610, underscoring the significant impact of parenthood on household decision-making~\cite{Lino:Expenditures:2017}.

Despite the importance of parenthood and marriage in shaping individuals' lives and mobility patterns, there has been little systematic study of how these life events alter their mobility, access to amenities, and overall urban experiences. This paper aims to fill that gap by examining how parenthood and marriage influence mobility patterns across several metropolitan areas in the United States.
Using American Community Survey data~\cite{data_usa_acs}, we analyse how these life-changing events reshape urban experiences and how cities can be characterized depending on their level of mobility diversity (access to amenities) and cost (commuting travel time) for parents and married people. We use variables related to individuals (e.g., marital status and parental status), their home and work locations, commuting travel times, and weighting factors that map the sample to a representative population. This dataset combined with OpenStreetMap information about the geolocation of amenities allows us to study the amenity accessibility from home and workplace locations and the mobility costs incurred during home-to-work commutes. Individual classification into married, non-married, parent, and non-parent groups follows Census Bureau definitions as detailed in the Methods section.

Our findings indicate that cities like Cincinnati and Chicago are more accommodating to parents, whereas cities like Houston and Virginia Beach are better suited for single individuals. These results are robust when considering the distribution of groups, travels, and travel distances (tested by five null cases). We also found that non-parents and single people tend to be less widely distributed within metropolitan regions, indicating a tendency to live in particular regions. Leisure and work, for instance, are types of amenities that tend to be more spatially concentrated than education and health, which might influence the decision on where non-parents and single people live more than parents and married people. These insights contribute to the growing body of research advocating for more nuanced, equitable, and effective urban planning.

\section*{Results}
\label{sec:results}

We begin by characterising the spatial distribution of amenities and population (\ie, travellers) within the 17 United States metropolitan areas (also referred to as CBSAs) included in our study (see Methods and Table~\ref{tab:data} for details). To this aim, using spatial information theory, we computed a quantity named \emph{diversity}, $H$, ranging from 0 to 1 (see Methods). A value of $H = 0$ indicates complete spatial concentration, where a given feature is present in only a single zone. Conversely, $H = 1$ indicates perfect spatial homogeneity, with the feature evenly distributed across all zones. Intermediate values of $H$ reflect non-homogeneous spatial distributions, with lower (higher) values signifying greater concentration (dispersion).

\begin{figure}[h!]
        \centering
        \includegraphics[width=0.95\textwidth]{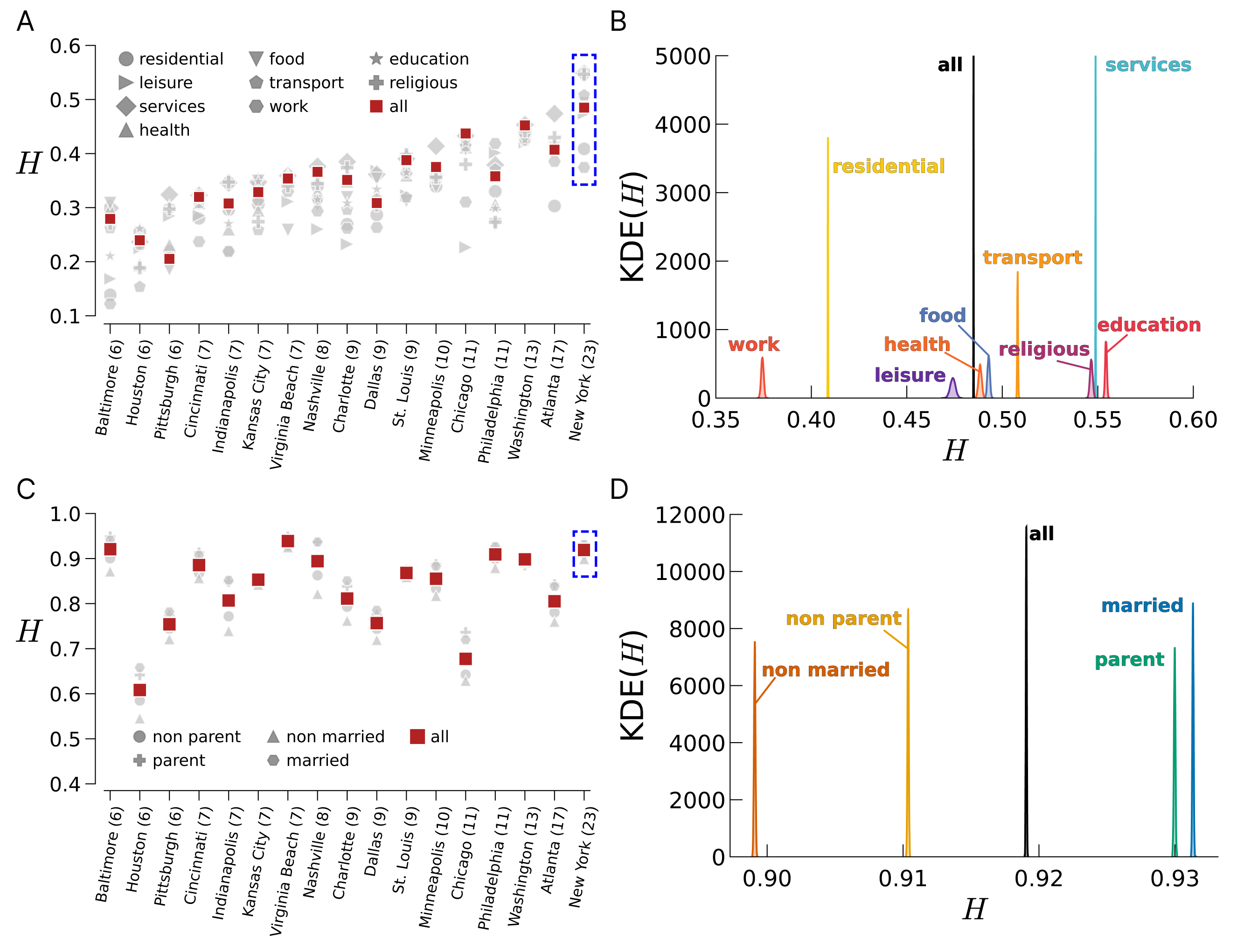}
        \caption{Spatial characterisation of amenities and population distributions over the metropolitan areas. We plot the values of the diversity, $H$, quantifying the static spatial distribution of amenity types (panel A) and residential locations of traveller groups (panel C) across zones within each metropolitan area. Higher values of $H$ indicate more homogeneous spatial distributions. The dashed boxes around New York in panels A and C indicate that this metropolitan area is shown in detail in panels B and D, respectively. Panels B and D display the Kernel Density Estimator (\emph{KDE}) using the default Gaussian kernel from the seaborn library in Python, illustrating the distribution of $H$ values across different amenity types (panel B) and traveller groups (panel D) for the New York metropolitan area. The displayed values of $H$ are obtained by bootstrapping $80\%$ of the data over 5,000 realisations.}
        % \caption{Spatial characterisation of amenities and population distributions over the metropolitan areas. We plot the values of the diversity, $H$, of the spatial distribution of both the types of amenities (panel A) and the travellers (panel C). Panels B and D display the Kernel Density Estimator (\emph{KDE}) using the default Gaussian kernel used by the seaborn library in Python. The displayed values of $H$ is obtained by bootstrapping $80\%$ of the data for the New York metropolitan area.}
        \label{fig:spatial_characterisation}
\end{figure}

Looking at the distribution of amenities (Figure~\ref{fig:spatial_characterisation}A), we notice that the values of diversity, $H$, span approximately between 0.1 and 0.6. Such a range of values indicates that metropolitan areas like Baltimore and Pittsburgh have (on average) more inhomogeneously distributed amenities than, for instance, Atlanta and New York. Such inhomogeneity depends marginally on the type of amenity considered (\eg, leisure), whereas it seems to depend more on the number of zones into which the metropolitan area is divided. In fact, metropolitan areas with more zones exhibit more homogeneous distributions, as confirmed by the analysis of the bootstrapped values of $H$. For instance, for the New York area (Figure~\ref{fig:spatial_characterisation}B), we have found that \emph{work} and \emph{residential} amenities are more inhomogeneously distributed than amenities classified as \emph{education} or \emph{services} (see Methods and Supplementary Materials for more details). 

Concerning the population's distribution (Figure~\ref{fig:spatial_characterisation}C), the overall values of $H$ fall in a range between 0.5 and 1.0, indicating a more homogeneous distribution, and the differences between distinct household arrangements within the same metropolitan area are narrower than for amenities. Taking the New York area as an example (Figure~\ref{fig:spatial_characterisation}D), the difference between the average values of the most inhomogeneously distributed group (\emph{non-married}) and the most homogeneously distributed one (\emph{married}) is approximately equal to 0.03, thus confirming the weaker, but statistically significant (using Kolmogorov-Smirnov test described in Section S9), dependence on household arrangement. These findings raise the question of whether differences in the distribution of amenities and sociodemographic groups within metropolitan areas may be amplified by mobility patterns.

%
%  FIGURE 1: SPATIAL CHARACTERISATION
%

\begin{figure}[h!]
        \centering
        \includegraphics[width=0.925\textwidth]{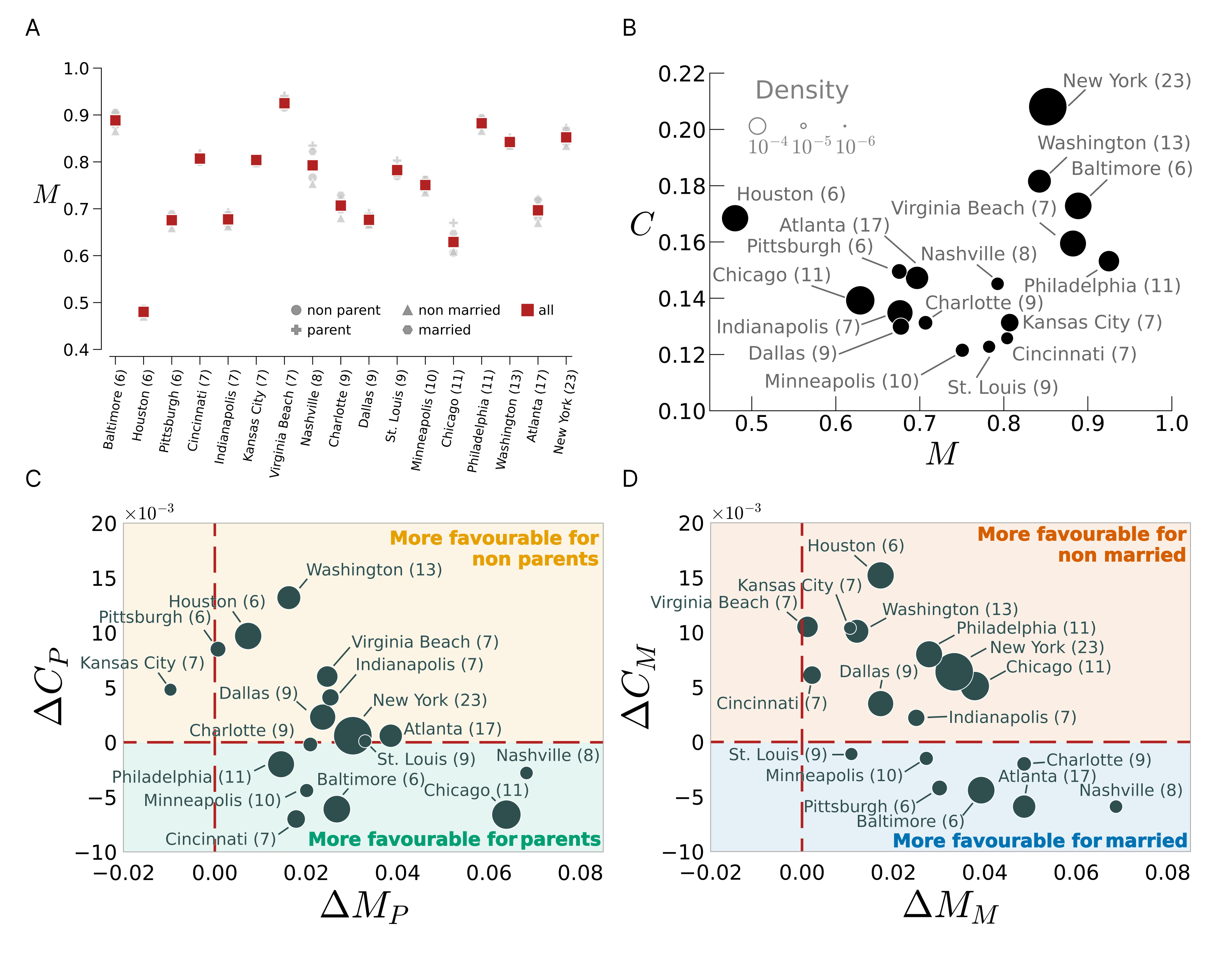}
        \caption{Characterisation of the mobility diversity and cost for parent and married travellers. Mobility diversity, $M$, measures how evenly travellers access amenities across destination zones (typically workplaces) based on their commuting patterns, representing the dynamic counterpart to the static spatial diversity $H$ shown in Figure~\ref{fig:spatial_characterisation}. For each urban area, we display: the mobility diversity, $M$, for travellers of different household arrangements (panel A); the average value of travel cost, $C$, and mobility diversity, $M$ (panel B); the differences in cost, $\Delta C$, and mobility diversity, $\Delta M$, between parents and non-parents (panel C); and between married and non-married travellers (panel D). The size of the dots in panels B, C, and D denotes the population density of each metropolitan area.}
        % \caption{Characterisation of the mobility diversity and cost for parent and married travellers. For each urban area, we display: the mobility diversity, $M$, for travellers of different types (panel A); the average value of travel cost, $C$, and mobility diversity, $M$ (panel B); the differences in cost, $\Delta C$, and mobility diversity, $\Delta M$, for parents/non-parents (panel C); and for married/non-married travellers (panel D). The size of the dots in panels B, C, and D denotes the population's density for each urban area.}
        \label{fig:mobility_characterisation}
\end{figure}

Now that we understand the static distribution of amenities and their visitation patterns by groups (Figure \ref{fig:spatial_characterisation}), we can switch our attention to mobility. In Figure~\ref{fig:mobility_characterisation}A, we display the values of \emph{mobility diversity} $M$ (a mobility counterpart of $H$, see Methods) for travellers of different household arrangements. At first glance, we notice that the values of $M$ span a wider range than those of $H$ displayed in Figure~\ref{fig:spatial_characterisation}C. Moreover, we do not observe the same dependency on the number of zones as for the case of $H$, and different types of travellers do not display significantly different values of $M$. However, these values are in line with those available in the literature on human mobility~\cite{Barbosa2018} and urban systems~\cite{battyscience2008}. The relationship between diversity, $M$, and average cost, $C$, displayed in Figure~\ref{fig:mobility_characterisation}B suggests that (except for Houston) metropolitan areas with higher values of $M$ (\eg, Baltimore) are those with higher values of $C$. Such a relationship does not seem to depend strongly neither on the population's density nor on the population itself (which, have a Spearman correlation with each other equal to 0.664), as areas like Chicago display average costs inferior to those of areas like Nashville, which are significantly less populated (see Tab.~\ref{tab:data}). However, in agreement with previous results~\cite{levinson1997density}, we do observe correlation between the average cost, $C$, and the population's density (Spearman correlation equal to 0.743). 
If we compute the differences of entropy $\Delta M_{P}$ and cost $\Delta C_{P}$ between non-parents and parents travellers (Figure~\ref{fig:mobility_characterisation}C; statistical tests using Kolmogorov-Smirnov described in Section S17--S18), we observe variation across metropolitan areas, with some showing higher costs for parents and others showing the opposite. The statistical significance of these differences varies across null model tests (Table~\ref{tab:null_models_parenthood}), with some cities showing robust differences across all five null models ($p < 0.05$), and others showing sensitivity to specific assumptions about travel patterns. Overall, in 10 of 17 urban areas, the empirical differences are positive ($\Delta C_{P}>0$), suggesting a general tendency toward higher mobility costs for non-parents, though the robustness of this pattern varies by metropolitan area.

% If we compute the differences of entropy $\Delta M_{P}$ and cost $\Delta C_{P}$ between non-parents and parents travellers (Figure~\ref{fig:mobility_characterisation}C; statistically significant using Kolmogorov-Smirnov test described in Section S7--S8), we observe that areas like Nashville, Baltimore, Minneapolis, and Chicago appear more favourable to parents travellers. However, areas like Washington, Houston, Virginia Beach, and Indianapolis display the opposite tendency, appearing more favourable to non-parents. Finally, areas like St. Louis and Charlotte are more neutral towards parenthood status ($\abs{\Delta C_{P}} < 0.002$), and there is only one area with $\Delta M_P < 0$: Kansas City, indicating that non-parents from this region tend to explore the metropolitan area more homogeneously than parents.
% The probable reason for these phenomena is that parents tend to live preferentially in residential areas (with more availability of bigger houses, children's entertainment, and/or closer to schools)~\cite{wolf2017places}. 

%
%  FIGURE 2: MOBILITY
%

For married travellers (Figure~\ref{fig:mobility_characterisation}D; statistical tests using Kolmogorov-Smirnov described in Section S17--S18), we similarly observe variation across metropolitan areas. In 10 urban areas the empirical differences are positive, indicating that those areas show patterns consistent with higher costs for non-married travellers. However, as shown in Table~\ref{tab:null_models_marriage}, the statistical significance of these patterns varies for \NM{3} and \NM{5}, with some metropolitan areas showing consistent effects across all tests whereas others are sensitive to model assumptions about travel time and distance distributions.

% If we now look at the presence of inequalities for the case of married travellers (Figure~\ref{fig:mobility_characterisation}D; statistically significant using Kolmogorov-Smirnov test described in Section S7--S8), the scenario changes slightly. For both cases, parenthood and marriage status, we observe that in 10 urban areas the differences are positive, indicating that those areas tend to be more favourable for non-parents and non-married travellers, respectively. We notice the presence of areas neutral towards the marital status of travellers, like Cincinnati and Dallas, whereas areas like Washington, Atlanta, and Baltimore seem to favour the mobility of married people. Taking the two perspectives together, it might be possible that Indianapolis, Minneapolis, and Charlotte are good places for parents and married people ($\Delta C_{P} < 0$ and $\Delta C_{M} < 0$), showing also, in Figure~\ref{fig:mobility_characterisation}B, a low mobility cost ($C$) and average mobility diversity ($M$).

All the differences observed in Figure~\ref{fig:mobility_characterisation} raise the following question: `Are these differences meaningful, or are they just the product of noise or chance?' To address this question, we designed five null hypotheses/models: %
\begin{inparaenum}[\sf NM1:]
\item We keep the travels (\ie, their origin, destination, and time), but we shuffle the household arrangement feature (\ie, married/non-married or parent/non-parent). 
\item We keep the travels' origin, destination, and travellers' type, but the travel time is extracted from a probability distribution obtained by fitting the empirical values.
\item We keep the travels' origin, time, and travellers' type, but the travel distance is extracted from a probability distribution obtained by fitting the empirical travel distances. If there is only one zone at such a distance from the origin, we select it as the destination zone. On the contrary, if multiple zones are located at the same distance, we select one of them uniformly at random.
\item We keep the travels' origin, destination, and travellers' type, but we shuffle the travel time feature.
\item We keep the travels' origin, time, and travellers' type, but we pick a random travel destination.
\end{inparaenum}
For each null case, we average the results over an ensemble of 5,000 realisations by bootstrapping 80\% of the data (see Methods). \NM{1} examines whether the cost differences observed, $\Delta C$, are due exclusively to the type of travellers (\eg, parent) and their distribution across the metropolitan area. \NM{2} and \NM{3} test whether the distributions of travel time and distance explain most of the differences. \NM{4} aims to understand whether the differences are due to the `speed' at which travels occur. Finally, \NM{5} checks whether the differences are due to the `spatial tessellation' by which zones are defined. More details are available in the SM.

An important strength of our null model approach is that it provides some robustness against potential confounding from sociodemographic covariates that may correlate with household arrangements (see Section~\ref{sec:confounding}) The fact that our findings remain statistically significant across multiple null models—each disrupting different potential confounding mechanisms—strengthens our confidence that the observed patterns genuinely reflect differences in how household arrangements experience urban mobility.

In Tables~\ref{tab:null_models_parenthood} and \ref{tab:null_models_marriage}, we report the analysis comparing the average differences in mobility cost observed in our dataset, with those obtained from the null models mentioned above (check Section~\ref{sec:nm_complete_stats} of the SM for the detailed statistics). We computed the $p$-values corresponding to the empirical average differences on the probability distribution of values generated by each null model. Such a procedure allow us to quantify the likelihood that the empirical differences are explained by the null model.%

%
%  TABLE 1: COMPARISON WITH NM FOR PARENTS
%
%
\begin{table}[ht!]
    \centering
    \caption{Comparison between the empirical average mobility cost differences between parents and non-parents travellers, $\Delta C_P$, and the correspondent values computed using each null model \NM{x} (with $x \in \{1,2,3,4,5\}$). For each urban area, beside the cost difference we provide also the $p$-value. Additional statistics for the confidence intervals and $t$-tests are presented in Section S2.1 of the Supplementary Material.}
    \label{tab:null_models_parenthood}
    \resizebox{0.85\linewidth}{!}{
    \begin{tabular}{rrrcrcrcrcrc}
    \toprule
    \multicolumn{1}{c}{\multirow{2}{*}{Urban area}} & \multicolumn{1}{c}{DATA} & \multicolumn{2}{c}{\NM{1}} & \multicolumn{2}{c}{\NM{2}} & \multicolumn{2}{c}{\NM{3}} & \multicolumn{2}{c}{\NM{4}} & \multicolumn{2}{c}{\NM{5}} \\
      & \multicolumn{1}{c}{$\Delta C_P$} & \multicolumn{1}{c}{$\Delta C_P$} & $p$-val & \multicolumn{1}{c}{$\Delta C_P$} & $p$-val & \multicolumn{1}{c}{$\Delta C_P$} & $p$-val & \multicolumn{1}{c}{$\Delta C_P$} & $p$-val &  \multicolumn{1}{c}{$\Delta C_P$} & $p$-val \\
    \midrule
    Baltimore (6) & -6.10e-03 & -1.27e-06 & 0.0000 & -1.39e-06 & 0.0000 & 1.59e-03 & 0.0702 & 9.02e-06 & 0.0000 & -4.79e-03 & 0.0000 \\
    Houston (6) & 9.70e-03 & 4.32e-06 & 0.0000 & -4.34e-05 & 0.0000 & 7.39e-04 & 0.1540 & 1.55e-05 & 0.0000 & 9.99e-03 & 0.0002  \\
    Pittsburgh (6) & 8.50e-03 & -1.55e-05 & 0.0000 & 7.87e-06 & 0.0000 & 8.69e-03 & 0.4510 & -9.08e-06 & 0.0000 & 1.65e-02 & 0.3160 \\
    Cincinnati (7) & -7.00e-03 & -1.23e-05 & 0.0000 & -1.98e-05 & 0.0000 & 2.20e-02 & 0.0000 & 1.13e-06 & 0.0000 & -2.50e-03 & 0.0000  \\
    Indianapolis (7) & 4.10e-03 & -1.59e-05 & 0.0000 & 8.46e-06 & 0.0000 & -4.02e-03 & 0.0652 & 1.90e-05 & 0.0000 & 4.43e-03 & 0.0000  \\
    Kansas City (7) & 4.80e-03 & -1.24e-05 & 0.0000 & 5.32e-06 & 0.0000 & 1.21e-02 & 0.1570 & 1.57e-05 & 0.0000 & 4.44e-03 & 0.0000  \\
    Virginia Beach (7) & 6.00e-03 & -1.16e-05 & 0.0000 & -6.88e-06 & 0.0000 & 9.61e-03 & 0.1830 & -1.47e-05 & 0.0000 & 6.73e-03 & 0.0000  \\
    Nashville (8) & -2.80e-03 & -4.48e-06 & 0.0000 & 4.01e-06 & 0.0000 & 2.15e-02 & 0.0000 & -3.75e-06 & 0.0000 & 2.36e-04 & 0.0000  \\
    Charlotte (9) & -2.00e-04 & 1.05e-05 & 0.0124 & 2.28e-05 & 0.0150 & -7.34e-03 & 0.2140 & 1.29e-05 & 0.0096 & -8.00e-05 & 0.0104 \\
    Dallas (9) & 2.30e-03 & -3.40e-06 & 0.0000 & 2.93e-06 & 0.0000 & 2.73e-03 &  0.4900 & 4.43e-06 & 0.0000 & 2.75e-03 & 0.0000  \\
    St. Louis (9) & 1.00e-04 & 8.69e-06 & 0.0976 & 2.34e-06 & 0.1100 & 1.29e-02 & 0.0000 & 1.29e-05 & 0.1070 & 3.45e-03 & 0.0000  \\
    Minneapolis (10) & -4.40e-03 & 6.73e-06 & 0.0000 & -1.33e-05 & 0.0000 & 6.49e-03 & 0.0026 & 2.70e-05 & 0.0000 & -2.05e-03 & 0.0000  \\
    Chicago (11) & -6.60e-03 & -5.26e-06 & 0.0000 & 2.66e-06 & 0.0000 & 1.60e-02 & 0.0000 & -1.33e-05 & 0.0000 & -6.33e-03 & 0.0000  \\
    Philadelphia (11) & -2.00e-03 & -3.09e-06 & 0.4430 & -2.17e-05 & 0.4420 & 9.89e-03 & 0.0000 & -3.74e-03 & 0.4790 & -3.43e-03 & 0.0000  \\
    Washington (13) & 1.32e-02 & -8.51e-06 & 0.0000 & 1.25e-05 & 0.0000 & 5.23e-03 & 0.0524 & -5.86e-06 & 0.0000 & 1.19e-02 & 0.0000  \\
    Atlanta (17) & 6.00e-04 & -1.78e-06 & 0.0000 & -6.37e-06 & 0.0000 & 8.96e-03 & 0.1300 & -1.15e-05 & 0.0000 & 3.61e-03 & 0.0000  \\
    New York (23) & 6.00e-04 & -2.46e-05 & 0.0000 & -9.91e-06 & 0.0000 & 6.14e-03 & 0.1470 & 6.97e-06 & 0.0000 & 6.00e-04 & 0.4530 \\
    \bottomrule
    \end{tabular} 
} %   
\end{table}
Taking the case of parenthood as an example (Table~\ref{tab:null_models_parenthood}), $p$-values below 0.05 indicate that it is highly unlikely for the observed differences to occur under the specific null case being tested. The sign of the observed value indicates the direction of the difference, with positive values corresponding to higher mobility cost for parents than non-parents.

% Taking the case of parenthood as an example (Table~\ref{tab:null_models_parenthood}), $p$-values below 0.05 indicate that it is highly unlikely for the observed differences to occur under the null case. Conversely, when the $p$-value is equal to or greater than 0.10, we cannot reject the hypothesis that the observed differences could result from simple travel or random processes. Moreover, the sign of the observed value indicates the direction of the difference, with positive values corresponding to higher mobility cost for parents than non-parents. By comparing the signs of empirical and null models' differences, we can identify whether the hierarchies of mobility costs change or not.

Tables~\ref{tab:null_models_parenthood} and~\ref{tab:null_models_marriage} reveal heterogeneity in how metropolitan areas perform across different null model specifications. For null models \NM{1}, \NM{2}, \NM{4}, and \NM{5}, the majority of metropolitan areas show $p$-values well below $0.05$, indicating that the observed differences cannot be explained by reshuffling household arrangements (\NM{1}), by the marginal distributions of travel time (\NM{2}), by travel speeds (\NM{4}), or by spatial tessellation (\NM{5}). However, \NM{3} (which preserves travel distance distributions and mimics known mobility behaviour~\cite{Barbosa2018}) shows stronger similarities with observed data across several metropolitan areas, as expected given that this model captures fundamental distance-dependent mobility patterns. The metropolitan areas maintaining $p < 0.05$ across all five null models demonstrate the most robust household arrangement effects, suggesting that these differences cannot be explained by any of the tested confounding mechanisms alone.

Overall, our analysis reveals patterns consistent with previous findings related to the travel time and travel distance distributions across urban areas~\cite{Barbosa2018,macedo2020gender}, demonstrating that mobility costs are not merely artefacts of random processes. This evidence underscores that the differences in cost captured in Figure~\ref{fig:mobility_characterisation} are not stochastic noise but rather, a hallmark of parenthood and marital effects on mobility.

%
%  TABLE 2: COMPARISON WITH NM FOR MARRIAGE
%
%
\begin{table}[h!]
    \centering
    \caption{Comparison between the empirical average mobility cost differences between married and non-married travellers, $\Delta C_M$, and the correspondent values computed using each null model \NM{x} (with $x \in \{1,2,3,4,5\}$). For each urban area, beside the cost difference we provide also the $p$-value. Additional statistics for the confidence intervals and $t$-tests are presented in Section S2.1 of the Supplementary Material.}
    \label{tab:null_models_marriage}
    \resizebox{0.85\linewidth}{!}{
    \begin{tabular}{rrrcrcrcrcrc}
    \toprule
    \multicolumn{1}{c}{\multirow{2}{*}{Urban area}} & \multicolumn{1}{c}{DATA} & \multicolumn{2}{c}{\NM{1}} & \multicolumn{2}{c}{\NM{2}} & \multicolumn{2}{c}{\NM{3}} & \multicolumn{2}{c}{\NM{4}} & \multicolumn{2}{c}{\NM{5}} \\
      & \multicolumn{1}{c}{$\Delta C_M$} & \multicolumn{1}{c}{$\Delta C_M$} & $p$-val & \multicolumn{1}{c}{$\Delta C_M$} & $p$-val & \multicolumn{1}{c}{$\Delta C_M$} & $p$-val & \multicolumn{1}{c}{$\Delta C_M$} & $p$-val &  \multicolumn{1}{c}{$\Delta C_M$} & $p$-val \\
    \midrule
    Baltimore (6) & -4.40e-03 & 1.69e-05 & 0.0000 & -1.00e-05 & 0.0000 & -4.56e-03 & 0.4640 & -3.17e-05 & 0.0000 & -4.16e-03 & 0.0000 \\
    Houston (6) & 1.52e-02 & 2.64e-05 & 0.0000 & -1.03e-05 & 0.0000 & 1.60e-03 & 0.0194 & 2.34e-05 & 0.0000 & 1.55e-02 & 0.0000 \\
    Pittsburgh (6) & -4.20e-03 & 7.24e-06 & 0.0000 & 2.34e-05 & 0.0000 & 1.94e-03 & 0.1390 & 1.81e-05 & 0.0000 & -3.39e-03 & 0.0000 \\
    Cincinnati (7) & 6.10e-03 & -1.71e-05 & 0.0000 & -5.54e-06 & 0.0000 & 6.75e-03 & 0.4540 & 1.55e-05 & 0.0000 & 6.33e-03 & 0.0000 \\
    Indianapolis (7) & 2.20e-03 & 2.45e-05 & 0.0000 & 7.25e-06 & 0.0000 & -1.86e-02 & 0.0000 & -2.16e-06 & 0.0000 & 6.79e-03 & 0.0000 \\
    Kansas City (7) & 1.04e-02 & -6.96e-06 & 0.0000 & -2.98e-05 & 0.0000 & 9.01e-04 & 0.0322 & -9.98e-06 & 0.0000 & 9.26e-03 & 0.0000 \\
    Virginia Beach (7) & 1.05e-02 & -1.21e-05 & 0.0000 & 4.88e-06 & 0.0000 & 7.72e-03 & 0.2220 & -2.97e-05 & 0.0000 & 1.25e-02 & 0.0000 \\
    Nashville (8) & -5.90e-03 & -5.48e-06 & 0.0000 & 2.21e-05 & 0.0000 & -2.02e-03 & 0.2230 & 8.75e-06 & 0.0000 & -7.51e-04 & 0.0000 \\
    Charlotte (9) & -2.00e-03 & -4.23e-06 & 0.0000 & 6.01e-06 & 0.0000 & -3.55e-03 & 0.3720 & 9.90e-06 & 0.0000 & 8.38e-04 & 0.0000 \\
    Dallas (9) & 3.50e-03 & 2.11e-05 & 0.0000 & -1.29e-05 & 0.0000 & 7.58e-03 & 0.1620 & 1.00e-05 & 0.0000 & 7.25e-03 & 0.0000 \\
    St. Louis (9) & -1.10e-03 & -7.57e-06 & 0.0000 & -2.30e-05 & 0.0000 & 2.02e-02 & 0.0000 & -5.37e-06 & 0.0000 & 6.00e-04 & 0.0000 \\
    Minneapolis (10) & -1.50e-03 & -1.63e-05 & 0.0000 & 9.37e-06 & 0.0000 & -7.75e-03 & 0.1110 & 2.39e-05 & 0.0000 & 5.38e-04 & 0.0000 \\
    Chicago (11) & 5.10e-03 & -1.43e-05 & 0.0000 & -5.45e-06 & 0.0000 & 1.21e-02 & 0.4050 & 1.05e-05 & 0.0000 & 4.01e-03 & 0.0000 \\
    Philadelphia (11) & 8.00e-03 & -5.91e-06 & 0.0000 & -1.50e-05 & 0.0000 & 2.96e-03 & 0.0626 & 1.62e-05 & 0.0000 & 5.43e-03 & 0.0000 \\
    Washington (13) & 1.01e-02 & 7.73e-06 & 0.0000 & 1.36e-05 & 0.0000 & 1.11e-02 & 0.5000 & -5.87e-06 & 0.0000 & 1.02e-02 & 0.1750 \\
    Atlanta (17) & -5.90e-03 & -7.56e-06 & 0.0000 & -1.53e-06 & 0.0000 & -5.04e-04 & 0.0708 & 7.76e-06 & 0.0000 & -6.78e-04 & 0.0000 \\
    New York (23) & 6.40e-03 & -1.81e-06 & 0.0000 & -5.91e-07 & 0.0000 & 8.09e-03 & 0.2880 & -5.42e-06 & 0.0000 & 6.39e-03 & 0.4690 \\
    \bottomrule
    \end{tabular} 
    }
\end{table}

%
%  FIGURE 3: NULLMODEL
%
% \begin{figure}[ht!]
%         \centering
%         \includegraphics[width=\textwidth]{figures/Figure303012025.png}
%         % \includegraphics[width=\textwidth]{figures/Figure3_entropy.png}
%         \caption{Quantifying the parenthood/marital effect on mobility in several null cases. For each urban area, we compute the ratio, $R$, between the differences in cost in the empirical and null case. The hue of the bars (dots), and the sign in the urban area's label, denotes the sign of the ratio $R$. We consider three null cases: \NM{1} (panels A and B), \NM{2} (panel C), and \NM{3} (panel D). The top row accounts for the case of parent travellers, whereas the bottom row for the case of married travellers. To ensure a better visualization, we use a logarithmic scale.}
%         \label{fig:null_models}
% \end{figure}
%

\section*{Discussion}
\label{sec:discussion}

People do not experience urban systems in a similar manner. Some works have shown that differences in mobility reveal inequalities faced by specific sociodemographic groups. For instance, works have found that socioeconomic status~\cite{barbosa2020uncovering}, gender~\cite{macedo2022differences}, race~\cite{anderson2021racial}, and age~\cite{lenormand2015influence} unveil differences in mobility that should be further studied for addressing health and safety issues~\cite{luca2023crime,Moreira2020,tiznado2024uncovering} and avoiding the reinforcement of inequalities in the labour market~\cite{giupponi2024labour}. That is why data-driven policies and interventions are crucial for improving urban systems and ensuring good accessibility for everyone, as cities become more challenging, expensive, and complex to manage/intervene~\cite{volpati2018spatial,hrelja2020create,Moreira2020}. Some works have shown successful policies and interventions addressing barriers in urban systems, benefiting the overall transportation system~\cite{chacon2020role}, women~\cite{d2023data,perez2019invisible,roy2024understanding}, and individuals with disabilities~\cite{kapsalis2024disabled}.

Our study examines the mobility and urban experiences of individuals in different household arrangements, specifically focusing on whether being a parent and/or married plays a pivotal role in shaping urban mobility patterns. Our findings align with the existing literature~\cite{mccarthy2021trajectories}, which highlights how life circumstances such as parenthood can significantly impact mobility choices---for example, by increasing reliance on cars for commuting or, conversely, forcing some families to sell a car due to financial constraints. However, our work goes beyond mere confirmation, offering a nuanced understanding of these changes in the context of access to amenities, which plays a crucial role in enabling individuals to fulfil their daily responsibilities~\cite{hanson2010gender,hail2021concept,pereira2017distributive}.

In this paper, we show that both access to amenities and associated travel costs are affected by parenthood and marriage, underscoring the heightened importance of decisions regarding residential and workplace locations in certain metropolitan areas. For instance, across most examined US cities, the empirical data show patterns where non-parents and unmarried individuals spend less time travelling ($\Delta C > 0$ in Figure~\ref{fig:mobility_characterisation}), whereas parents and married individuals tend to live in closer proximity to a range of essential amenities. Notably, these observed patterns show no clear correlation with either population size or city scale. Moreover, our comparisons with null models confirm that the results are robust to potential fluctuations in travel distributions, household characteristics, and spatial delineations, reinforcing the validity of our conclusions.

% One limitation of our work is that we do not account for the age of the children which it was shown to impact differently the travel patterns of parents~\cite{mccarthy2021trajectories}. ~\cite{mccarthy2021trajectories} shows that the primary caregiver has the highest impact, regardless of gender. Public transportation time is also challenging to fulfil responsibilities. System to support families: investing child friendly infrastructure, car slow down next to schools, parks in close distance, bike and walk safely.

Our findings indicate that in most cities, parents and married individuals work closer to a broader range of amenities. However, in some metropolitan areas, this advantage is coupled with longer average commuting times, suggesting a need for tailored interventions to alleviate the mobility burden on these demographic groups. In the literature, we also observe that changes in household arrangements and status increase gender and race differences in commuting travel time more significantly~\cite{hu2021gender}.

These insights offer valuable guidance for policymakers and urban planners aiming to refine public transportation, reduce inequalities in access to services, foster more sustainable urban environments, and strengthen community cohesion. Moreover, cities that successfully accommodate diverse household arrangements may serve as models for other regions seeking to optimise their urban ecosystems. 

Urban environmental changes offer significant potential to improve neighbourhoods and family life, even if their impacts can vary~\cite{audrey2015healthy,schipperijn2024role}. For instance, enhancing or creating parks may not always increase usage, but such interventions are associated with reduced injury rates and decreased screen time for children. Similarly, improvements in walkability, road speed reduction, and road signage near schools have successfully encouraged more families to walk or cycle, fostering active and healthier lifestyles. Broader urban interventions, such as the inclusion of cable cars in Bogotá, have demonstrated positive impacts by reducing travel time, increasing satisfaction with public transportation, promoting physical activity, improving access to amenities, and decreasing perceptions of insecurity~\cite{baldovino2023effects,rubio2023impacts}. Despite these successes, the concept of a `child-friendly' city remains a topic of debate among practitioners~\cite{powell2024child}, and mobility consistently emerges as a crucial factor in shaping environments that support children's and parents' well-being.

The findings of this study underscore the critical importance of acknowledging demographic diversity in urban mobility frameworks. While broad, generalised models can offer initial guidance, they often fail to capture the complexity arising---among others---from demographic, socio-economic, and cultural variables. As cities continue to expand and evolve, traditional approaches treating populations as homogeneous units are increasingly inadequate. Crucially, the growing availability of comprehensive data sources now enables more refined analyses that can incorporate a range of factors (\eg, household status, income levels, and cultural norms) into our understanding of mobility patterns. By examining how life-altering events like parenthood and marriage shape these dynamics, we highlight the necessity of designing cities that are both efficient and inclusive. Recognising that parents, cohabitants, and single individuals navigate urban spaces differently not only empowers policymakers to develop more targeted interventions, but also contributes to building more sustainable, equitable, and resilient urban systems.

Our work also contributes to the broader discourse on urban planning and sustainability. 
By revealing the diverse mobility needs and preferences of various demographic groups, our study provides insights that can guide policymakers in creating urban spaces that cater to the realities of contemporary urban living. 
As cities strive to become more adaptable and responsive to their residents, the findings presented here offer a valuable framework for integrating demographic considerations into the design and management of urban areas, promoting a more just and sustainable future for all.

While current data sources provide initial insights, they often lack the necessary granularity to capture the evolving mobility experiences of different demographic groups. Additionally, our datasets reflect a reality in which the majority of travels are performed by men, married people, and non-parent individuals. This predominance has been recognised as a challenge in data collection, as highlighted by~\cite{d2023data,perez2019invisible}, pointing out that datasets are often biased in favour of majorities. This underscores the importance of works like ours in advancing the understanding of how minority-specific groups behave and may face disadvantages. Furthermore, as we strive to integrate more detailed information, it becomes essential to prioritise ethical considerations and privacy protections. Developing frameworks combining advanced anonymisation techniques, secure data-sharing protocols, and strict governance can help maintain public trust while granting researchers and policymakers access to critical evidence. With such safeguards in place, it becomes possible to craft policies and interventions that truly reflect the complex, multifaceted realities of urban life, resulting in more inclusive, efficient, and resilient cities.

Despite the insights gained, this study is not exempt from limitations. One key constraint lies in the nature of the data employed. Although the census and related datasets offer valuable information on household arrangements and work-related travel, they do not fully capture the complexity and dynamics of individuals’ mobility decisions. Factors such as non-work trips, nuanced cultural influences, and the role of social networks remain underexplored. Moreover, the predominance of data reflecting majority groups, as discussed earlier, limits the understanding of how minority groups navigate urban systems. We also acknowledge that we have only considered four groups of household arrangements, without examining the particularities of individuals in other categories, such as polyamorous relationships, married couples living separately, divorced individuals, or widow(er)s. Furthermore, our analysis draws from bounded questionnaires which may be subject to defined classifications and aggregations~\cite{kenny2024evaluating} and response biases~\cite{rietveld1999relationship}, and from a limited number of metropolitan areas in the United States, leaving open whether these findings generalise to other global contexts or urban systems. While we have considered household-level distinctions and demographic diversity, other aspects, including income~\cite{fingerman2015ll}, gender~\cite{macedo2022differences}, age~\cite{anthony2025examining}, and access to various transportation options~\cite{chacon2020role}, may also influence mobility in ways not fully accounted for here.

Our data sources further restrict the ability to examine long-term trends or the evolving nature of urban mobility with the level of detail that emerging longitudinal and multimodal datasets might provide. This temporal limitation makes it challenging to understand how life events or policy changes shape travel patterns over time; as individuals transition between different household arrangements or life stages, the patterns and demands of urban mobility may shift, emphasising the need for dynamic, long-term analyses capturing the evolving character of urban life. The lack of high-resolution, privacy-preserving information on trip purpose or travel modes may have constrained the scope of our analysis.

Our analysis is also sensitive to the quality and availability of amenity data extracted from OpenStreetMap. First, some tags are misspelled (e.g., chu, com, gar) or contain multiple, ambiguous entries (e.g., “apartments;hotel;office”), making accurate classification of amenities into categories challenging. Second, certain spatial factors (such as the presence of multiple floors within a building) are not considered, as this information is often missing. Lastly, temporal information about when an amenity was established, is currently operating, or has closed is largely unavailable. Despite these limitations, OpenStreetMap remains a rich and widely used data source in urban science~\cite{botta-pone-2021,boeing2025modeling}, and we acknowledge these constraints in our analysis.

Additionally, our reliance on reported marital status may introduce ambiguity in distinguishing between different types of partnerships. For unmarried cohabiting partners the available categories may not accurately reflect their living arrangements. Consequently, some unmarried cohabiting partners likely select `married' as the closest approximation to their situation, meaning our `married' category de facto includes both legally married couples and an unknown proportion of unmarried cohabiting partnerships. Furthermore, major life transitions such as marriage and parenthood involve planning and anticipation phases, during which mobility adjustments may begin to occur before these events are formally reported in census data. The observation of statistically significant mobility differences at the point of formal reporting thus suggests that the realisation of these life events introduces changes exceeding what anticipatory adjustments alone would have been put in place.

We also acknowledge that our analysis considers only two dimensions influencing the mobility of various household arrangements and does not incorporate other significant factors, including housing affordability, the efficiency of public transportation, income inequality, and the availability of childcare support. Our analysis and interpretations are limited by the dataset and its spatial, temporal, and social coverage; other factors and mobility metrics could provide additional insights into the characteristics of the cities (\eg, accessibility, inclusivity and efficiency). Moreover, our selection of metropolitan areas with higher numbers of zones post-merging (6--23 zones), whilst necessary for robust entropy-based measures, may bias the sample towards larger or more spatially fragmented areas and exclude compact cities that achieve household-friendliness through spatial integration rather than zonal differentiation.

Nevertheless, we believe that the core findings remain both robust and instructive. Acknowledging these constraints highlights avenues for future research, which could draw on richer data sources and more advanced analytical methods. Such efforts would deepen our understanding of urban mobility and better equip policymakers and planners to build more adaptive, equitable, and inclusive urban environments.

\section*{Methods}
\label{sec:methods}

\subsection*{Data}
\label{ssec:data}

Our data are a combination of mobility data obtained from the US census and data on amenities obtained via Open Street Maps. In the following, we present a description of how the data have been extracted, processed, and compiled together.

\subsubsection*{Mobility data and cost}
\label{sssec:mobility_data}

Our data were collected using the United States Census Bureau's API~\cite{us_census_api} for all the urban regions with information available on mobility and demographics under the 2019 edition of the so-called American Community Survey~\cite{data_usa_acs} (ACS). For the purpose of our study, we have considered only trips related to work. The variables collected by the ACS are: the state, city, metropolitan region, individual (traveller) characteristics (\eg gender, age, marital status), mobility characteristics (\eg work location, commuting travel time), and expansion factors\cite{dictionary_acs}. Expansion factors (also called weighting factors or survey weights in some fields) vary in terminology across disciplines and are used to ensure that the sample is representative of the whole population. 
For each individual, the ACS reports information such as: gender, household status, socioeconomic status, and age. For all categories, individuals must be at least 15 years old. Individuals are classified as \emph{married} if they report being currently married (MSP = 1 for married, spouse present, or \emph{MSP} $= 2$ for married, spouse absent). Individuals are classified as \emph{non-married} if they report being widowed, divorced, separated, or never married. It is important to note that the `married' category likely includes not only legally married individuals but also a portion of unmarried cohabiting partners who selected 'married' as the closest approximation to their living arrangement, given that the alternative categories may not accurately reflect their situations. ACS documentation indicates that persons reporting common-law marriages are coded as 'married', and allocation procedures acknowledge that unmarried partners may have marital status values of 'married'. We do not account for the gender of people and consider indiscriminately both homosexual and heterosexual individuals. For the parent category, individuals are classified based on the number of children (\emph{NOC} $> 0$).
% For each individual, the ACS survey reports information such as: gender, household status, socioeconomic status, and age. In particular, we classify individuals as \emph{married} if they live together and report \texttt{married} in the survey (\ie, $MSP = 1 or 2$). Individuals younger than 15 years are excluded, and those classified as non-married include divorced, never married, widowed, or separated respondents. We rely on the data collected by the ACS, and avoid counting flatmates as married individuals. Moreover, we do not account for the gender of people and consider indiscriminately both homosexual and heterosexual individuals. Following a similar reasoning, we refer to people as \emph{parents} if they replied \texttt{parent} in the survey with any number of children higher than zero (i.e., $NOC > 0$).

For each urban area, the space is divided into zones. However, the spatial tessellations of home and work zones provided by the U.S. Census do not perfectly align, which constitutes a limitation of the data. In particular, multiple home zones often map onto a single work zone, requiring us to collapse some home zones. As a result, the effective number of zones per urban area is reduced from 30--60 to approximately 6--23. This limitation in spatial resolution is due to privacy considerations, which are a key aspect of data-sharing regulations. Nevertheless, this preliminary step avoids inconsistencies in the comparison of differences in residential and mobility concentration, albeit it diminishes considerably the number of available zones. We decided to use the metropolitan areas classified as such by the US census~\cite{data_usa_acs} as our reference, as they are well-recognised functional area in which people are more likely to work. Each urban area in our analysis can be identified using its `MET2013\_label' descriptor equivalent to census `CBSA'~\cite{cbsa-census}. Importantly, each `CBSA' corresponds to a single, unique `MET2013\_label'. For instance, the urban area `Pittsburgh' represents one CBSA and is composed of six zones, as described in Table~\ref{tab:data}.

The cost of one travel, $c$, is given by the ratio between the time needed to reach the destination, $t$, and the maximum travel time, $\mathcal{T} = \max(t)$. By rescaling the travel time by its maximum, $\mathcal{T}$, we ensure a fair comparison between urban areas of different sizes. That said, other rescaling factors could be used as well (\eg{,} the so-called Marchetti constant~\cite{Marchetti1994}). The average cost of travels made by travellers of type $X$, $C^X$, is given by

%
%  Eq. Travel's cost
%
\begin{equation}
\label{eq:travel_cost}
C^X = \dfrac{1}{{N_T}^X} \sum_{i} c_i = \dfrac{1}{{N_T}^X} \sum_{i} \dfrac{t_i}{\mathcal{T}} = \dfrac{1}{\mathcal{T}\,{{N_T}^X}} \sum_{i} t_i \,,
\end{equation}
where ${N_T}^X$ is the total number of travels made by travellers of type $X$. The average cost of all travels made, $C$, is simply $C = \tfrac{1}{\mathcal{T}\,{N_T}} \sum_{i} t_i \; \forall X$, with ${N_T}$ being the total number of travels made by all classes.

The statistics of our collected dataset are displayed in Table~\ref{tab:data}. We selected the 17 metropolitan areas with the highest number of zones available after the merging process, resulting in urban areas having a number of zones, $N_Z$, ranging from 6 to 23, which provides sufficient spatial resolution for robust entropy-based measures. Although our data on spatial tessellation is limited, this paper leverages a rich dataset to study sociodemographic groups—an area where such data is typically scarce and constrained~\cite{perez2019invisible,d2023data}.

% US data only work related trips.
% From NY, we have the purpose of the trips.
% For each zone, we know the amount of amenities and which type of amenities. We divided the amenities in to 10 categories: 9 categories plus overall.

\subsubsection*{Amenities}
\label{sssec:amenities_data}

For each of our zones, we extracted from Open Street Maps~\cite{OpenStreetMap2017} all amenities located inside those zones and classified as: \texttt{amenity}, \texttt{highway}, \texttt{building}, or \texttt{healthcare}. To obtain homogeneous and meaningful groups of amenities, we manually grouped them into these categories: \texttt{work}, \texttt{residential}, \texttt{leisure}, \texttt{health}, \texttt{food}, \texttt{transport}, \texttt{religious}, \texttt{education}, and \texttt{services}. Among the amenities available, we excluded those that we could not easily map into any of the aforementioned categories. A sample of the categories that we removed includes: \texttt{karaoke box}, \texttt{compressed air}, \texttt{binoculars}, \texttt{concussion}, and \texttt{show house}. The detailed list of all the amenities included and excluded is available in the Supplementary Materials.

%
%  TAB 3: DATASETS' TABLE
%
\begin{table}[ht!]
\caption{Main characteristics of the urban areas considered. For each area, we report its reference name, the names of the urban areas associated with it, the US states to which they belong (even if only partially), the number of zones $N_{Z}$, the number of travels for the entire population accounting for the expansion factor $N_{T}$, the number of people accounting for the expansion factor $N_P$, the fraction of travels performed by women, married people, and parents ($t_{\text{W}}$, $t_{\text{Married}}$,$t_{\text{Parent}}$) and the fraction of women, married, and parent ($f_{\text{W}}$, $f_{\text{Married}}$,$f_{\text{Parent}}$) travellers. Areas are sorted by increasing number of zones and population.}
\label{tab:data}
\centering
\resizebox{\linewidth}{!}{
\begin{tabular}{rrll|rr|rrr}
    \toprule
     & \multicolumn{1}{c}{Reference} & \multicolumn{1}{c}{Urban areas} & \multicolumn{1}{c|}{State(s)} & \multicolumn{1}{c}{$N_{Z}$} & \multicolumn{1}{c|}{$N_{T}$ $\bigl( N_P \bigr)$} & \multicolumn{1}{c}{$t_{\text{W}}$ $\bigl( f_{\text{W}}\bigr)$} & \multicolumn{1}{c}{$t_{\text{Married}}$ $\bigl( f_{\text{Married}} \bigr)$} & \multicolumn{1}{c}{$t_{\text{Parent}}$ $\bigl(f_{\text{Parent}}\bigr)$}\\
    \midrule
    1 & Pittsburgh & Pittsburgh & PA & 6 & 1,078,656 (2,275,783) & 0.48 (0.51) & 0.51 (0.41) & 0.29 (0.35)\\
    2 & Baltimore & Baltimore $\bullet$ Columbia $\bullet$ Towson & MD & 6 & 1,211,966 (2,749,210) & 0.50 (0.52) & 0.49 (0.38) & 0.33 (0.40) \\
    3 & Houston & Houston $\bullet$ The Woodlands $\bullet$ Sugar Land & TX & 6 & 3,258,783 (6,980,075) & 0.45 (0.50) & 0.53 (0.39) & 0.39 (0.49) \\
    4 & Virginia Beach & Virginia Beach $\bullet$ Norfolk $\bullet$ Newport News & VA \& NC & 7 & 828,636 (1,672,613) & 0.48 (0.51) & 0.50 (0.38) & 0.33 (0.41) \\
    5 & Indianapolis & Indianapolis $\bullet$ Carmel $\bullet$ Anderson & IN & 7 & 988,100 (2,076,620) & 0.49 (0.51) & 0.51 (0.37) & 0.36 (0.44) \\
    6 & Cincinnati & Cincinnati & OH, KY, \& IN & 7 & 1,014,468 (2,150,810) & 0.49 (0.51) & 0.52 (0.39) & 0.36 (0.41) \\
    7 & Kansas City & Kansas City & MO \& KS & 7 & 1,080,449 (2,208,182) & 0.48 (0.51) & 0.53 (0.40) & 0.37 (0.45)\\
    8 & Nashville & Nashville $\bullet$ Davidson county $\bullet$ Murfreesboro $\bullet$ Franklin & TN & 8 & 1,059,418 (2,106,031) & 0.48 (0.51) & 0.54 (0.41) & 0.35 (0.43) \\
    9 & Charlotte & Charlotte $\bullet$ Concord $\bullet$ Gastonia & NC \& SC & 9 & 1,277,111 (2,649,709) & 0.49 (0.52) & 0.52 (0.39) & 0.36 (0.43)\\
    10 & St. Louis & St. Louis & MO \& IL & 9 & 1,352,422 (2,834,644) & 0.49 (0.51) & 0.53 (0.40) & 0.35 (0.40)\\    
    11 & Dallas & Dallas $\bullet$ Fort Worth $\bullet$ Arlington & TX & 9 & 3,701,991 (7,503,488) & 0.46 (0.51) & 0.53 (0.39) & 0.38 (0.48) \\
    12 & Minneapolis & Minneapolis $\bullet$ St. Paul $\bullet$ Bloomington & MN \& WI & 10 & 1,924,212 (3,700,304) & 0.48 (0.50) & 0.53 (0.40) & 0.37 (0.45)\\
    13 & Philadelphia & Philadelphia $\bullet$ Camden $\bullet$ Wilmington & PA, NJ, DE, \& MD & 11 & 2,815,605 (6,146,130) & 0.49 (0.52) & 0.50 (0.37) & 0.32 (0.40)\\
    14  & Chicago & Chicago $\bullet$ Naperville $\bullet$ Elgin & IL, IN, \& WI & 11 & 4,634,178 (9,408,177) & 0.48 (0.51) & 0.50 (0.38) & 0.35 (0.43)\\
    15 & Washington & Washington $\bullet$ Arlington $\bullet$ Alexandria & DC, VA, MD, \& WV & 13 & 3,157,880 (6,170,921) & 0.48 (0.51) & 0.51 (0.39) & 0.36 (0.45)\\
    16  & Atlanta & Atlanta $\bullet$ Sandy Springs $\bullet$ Roswell & GA & 17 & 2,848,348 (5,996,700) & 0.49 (0.52) & 0.51 (0.38) & 0.35 (0.44)\\
    17 & New York & New York $\bullet$ Newark $\bullet$ Jersey City & NY, NJ, \& PA & 23 & 9,619,820 (19,839,535) & 0.48 (0.51) & 0.51 (0.38) & 0.33 (0.41)\\
    \bottomrule
    \end{tabular}
} % end of resizebox
\end{table}

\subsection*{Quantifying diversity}
\label{ssec:diversity}

We employ Shannon entropy-based diversity measures to quantify the spatial evenness of distributions, where high diversity indicates homogeneous spatial distribution across zones, distinct from individual-level mobility diversity measuring trip variety~\cite{wang2018spatial}. Given a metropolitan area divided into $N_Z$ zones, one can compute the \emph{diversity}, $H^{X}$, of the spatial coverage of a given feature $X$ over such an area $A^{X}_{i}$~\cite{batty2010space,cocchi2014spatial,battyjourgeosys2014}. The latter is---up to a multiplicative factor---the Shannon entropy of the coverage, yielding:
\begin{equation}
\label{eq:entropy_space}
H^{X} = - \frac{1}{\log_2 N_{Z}} \sum^{N_Z}_{i=1} p^{X}_{i} \; \log_2 \frac{A^{X}_{i}}{p^{X}_{i}} \,.
\end{equation}
Being $p^{X}_{i}$ the probability of observing feature $X$ (\eg hospitals) in zone $i$ which, in turn, is given by:
\begin{equation}
\label{eq:prob_spatial_feature}
p^{X}_{i} = \frac{n^{X}_{i}}{N^{X}} \,,
\end{equation}
where $n^{X}_{i}$ is the number of entities (\ie, amenities or travellers) of type $X$ in zone $i$, and $N^{X} = \sum_{i}^{N_z} n^{X}_{i}$ is the total number of such entities in the whole metropolitan area. Diversity $H^{X} \in [0,1]$, where $H^{X} = 0$ corresponds to the case in which feature $X$ is concentrated in a single zone, whereas $H^{X} = 1$ denotes the case in which feature $X$ is homogeneously distributed across all zones.

We can use the same formalism of Eq.~\eqref{eq:entropy_space} to compute the \emph{mobility diversity} (\ie, \emph{diversity of accessibility to amenities}) by travellers of type $Y$, $M^{Y}$, over such an area $A^{Y}_{i}$, as:
\begin{equation}
\label{eq:entropy_mobility}
M^{Y} = - \frac{1}{\log_2 N_{Z}} \sum^{N_Z}_{i=1} p^{Y}_{i} \; \log_2 \frac{A^{Y}_{i}}{p^{Y}_{i}} \,.
\end{equation}
Where---following the structure of Eq.~\eqref{eq:prob_spatial_feature}---the probability $p^{Y}_{i}$ corresponds to the ratio between the product of the number of amenities in zone $i$, $n_{i}$, and the number of travellers of type $Y$ whose destination zone is $i$, $T^{Y}_{i}$ (\ie, $n^{Y}_{i} = n_{i} \; T^{Y}_{i}$), and its sum over all zones $N^{Y} = \sum_{i}^{N_z} n^{Y}_{i}$. Eventually, one could also calculate the mobility diversity of accessibility of travellers of type $Y$ to amenities of type $X$, $M^{XY}$. It is worth noting that the values of $M$ and $H$ in the above equations are computed 5{,}000 times using bootstrap resampling of 80\% of the data. The average of these realisations is then used as the empirical estimate.

The spatial diversity $H^X$ (Equation~\ref{eq:entropy_space}) quantifies the \emph{static} distribution of amenities or population across residential zones, representing where features are located without considering movement. In contrast, the mobility diversity $M^Y$ (Equation~\ref{eq:entropy_mobility}) quantifies the \emph{dynamic} accessibility to amenities based on where travellers commute to work, weighted by amenity availability at destination zones. This distinction allows us to separate residential accessibility (amenities near home) from mobility-based accessibility (amenities accessed through commuting). Figure~\ref{fig:spatial_characterisation} presents static distributions (\textcolor{black}{aided also by Figure S3}), whilst Figure~\ref{fig:mobility_characterisation} presents dynamic mobility patterns.

\subsection*{Computing differences of cost and diversity}
\label{ssec:differences_c_and_m}

The difference between the average costs sustained by travellers of type $X$ and $Y$, $\Delta C_{X}$, is computed as:
\begin{equation}
\label{eq:diff_costs}
\Delta C_{X} = C^{X} - C^{Y} \,.
\end{equation}
A positive value of $\Delta C_{X}$ indicates that travellers of type $X$ sustain, on average, higher costs than those of type $Y$, whereas $\Delta C_{X} < 0$ indicates the opposite condition. In analogy with Eq.~\eqref{eq:diff_costs}, we can compute also the difference in the mobility diversity of travellers of type $X$ and $Y$, $\Delta M_{X}$.

% The \emph{relative ratio} of cost, ${R^C}_{XY}$, sustained by travellers of type $X$ and $Y$, is defined as:
% %
% \begin{equation}
% \label{eq:rel_diff_cost}
% %
% {R^C}_{XY} = \dfrac{C^{X} - C^{Y}}{\avg{\widetilde{C}^{X} - \widetilde{C}^{Y}}} \,,
% %
% \end{equation}
% %
% where $\widetilde{C}$ is the value of cost computed in a suitable null case (check the Supplementary Materials for the details) and the $\avg{\, \cdot \,}$ operator denotes the ensemble average. A value of ${R^C}_{XY} > 0$ denotes that both the numerator and the denominator have the same sign, whereas ${R^C}_{XY} < 0$ denotes the opposite case. If $0 < \left\lvert {R^C}_{XY} \right\rvert < 1$, then the empirical difference is smaller than its null case counterpart, whereas $\left\lvert {R^C}_{XY} \right\rvert > 1$ indicates the opposite (\ie, empirical difference bigger than the null one). Lastly, $\left\lvert {R^C}_{XY} \right\rvert = 1$ indicates that the difference is (up to a sign) exactly the same in both the empirical data and the null case. Analogously, we can define the same quantities for mobility diversity. It is worth noting that the values of $C$ and $\widetilde{C}$ appearing in the above equations correspond to those of the average value resulting from the bootstrap sampling.

%\end{linenumbers} % end of line numbering

%
% BIBLIOGRAPHY
%
%\bibliography{biblio}

\section*{Declaration}

For the purpose of open access, the authors have applied a Creative Commons Attribution (CC BY) licence to any accepted manuscript version arising from this submission.

\section*{Acknowledgements}

The authors are grateful to Laura Lotero for helpful discussions during the preliminary stages of the work. AC acknowledges financial support from the Ram\'on y Cajal program through the grant RYC2023-044587-I. AC acknowldeges financial support from the Spanish Ministerio de Ciencia e Innovaci\'on, through project No. PID2024-158120NB-C22.

Numerical analysis has been carried out using the NumPy, Powerlaw and GeoPandas Python packages~\cite{oliphant-book-2006,vanderwalt-compscieng-2011,alstott2014powerlaw,geopandas-2024}. Graphics have been prepared using the Matplotlib and Seaborn Python package~\cite{hunter-matplotlib-2007,waskom2021seaborn}.

\section*{Author contributions statement}

MM and RM developed the original ideas; All authors designed the study; AC supervised the development of the experiments; AC wrote the formalisms; MM collected, curated, and integrated the raw data; MM performed the analysis; All authors analysed the results; All authors wrote the paper; MM and AC prepared the graphics. All authors read, reviewed, and approved the final manuscript.

\section*{Additional information}

\begin{description}
\item[Competing interests]
The authors declare no competing interests.
\item[Availability of data and materials]
The data will be made publicly available upon acceptance of the paper.
\item[Ethical approval]
This article does not contain any studies with human participants performed by any of the authors.
\item[Informed consent]
This article does not contain any studies with human participants performed by any of the authors.
\end{description}

%\end{document}
% end of document

% SUPPLEMENTARY MATERIAL

%\documentclass[11pt,a4paper,oneside,dvipsnames]{article}

%\usepackage[utf8]{inputenc}
%\usepackage{graphicx,epsfig}% Include figure files
% estos producen otro tipo de letra
%\usepackage{times}
%\usepackage{listings}

%\lstset{
%    language=Python,          % Highlight as Python code
%    basicstyle=\scriptsize\ttfamily,  % Use small font + typewriter style
%    breaklines=true,          % Automatically wrap long lines
%    tabsize=4,                % Set default tab size
%    columns=fullflexible,     % Better handling of variable-width fonts
%    keepspaces=true,          % Keep indentation
%    showstringspaces=false    % Don't visually mark spaces in strings
%}

%\usepackage{dcolumn,bm,epic,eepic,float}
%\usepackage{amssymb,amsmath,multirow,rotate}
%\usepackage{xcolor}
%\usepackage{colortbl}
%\usepackage[mathscr]{eucal}
%\usepackage{paralist}
%%\usepackage{cite}
%\usepackage{multirow}
%\usepackage{booktabs,tabularx,ltablex,longtable}
%\usepackage{url}
%%\usepackage{capt-of}
%\usepackage{soul}
%\usepackage{fancybox}
%\usepackage{textcomp}
%\usepackage[normalem]{ulem}
%\usepackage[margin=1.75cm]{geometry}
%\usepackage{enumitem}
%\usepackage{calc} 
%\usepackage{blindtext}
%\usepackage{lineno}
%\usepackage{bbold} % for mathbb numbers
%\usepackage{xr}
%\usepackage{xr-hyper}
%\usepackage[hidelinks]{hyperref}
%\usepackage{graphicx}
%\usepackage{subfig}
%\usepackage{pythonhighlight}

%%%%%%%%%%%%%%%%%%%%%%%%%%%%%%%%%%%%%%%%%%%%

% Definition commands
\newcommand*{\setfont}[1]{\ensuremath{\mathcal{#1}}}
\newcommand*{\setsym}[1]{\ensuremath{\mathscr{#1}}}
\newcommand*{\set}[1]{\ensuremath{\left \lbrace #1 \right \rbrace}}

\definecolor{lightgray}{gray}{0.7}
\definecolor{alehighlightgray}{gray}{0.8}

\newcommand{\all}{\texttt{all}\xspace}
\newcommand{\work}{\texttt{work}\xspace}
\newcommand{\nonwork}{\texttt{nonwork}\xspace}

\newcommand{\kde}{{\rm KDE}}

\newcommand{\nullm}[1]{\textsc{NM#1}\xspace}

%%%%%%%%%% Merge with supplemental materials %%%%%%%%%%
%%%%%%%%%% Prefix a "S" to all equations, figures, tables and reset the counter %%%%%%%%%%
% \setcounter{equation}{0}
% \setcounter{figure}{0}
% \setcounter{table}{0}
% \setcounter{page}{1}
% \makeatletter
% \renewcommand{\theequation}{S\arabic{equation}}
% \renewcommand{\thefigure}{S\arabic{figure}}
% \renewcommand{\bibnumfmt}[1]{[S#1]}
% \renewcommand{\citenumfont}[1]{S#1}
%%%%%%%%%% Prefix a "S" to all equations, figures, tables and reset the counter %%%%%%%%%%

% COMMANDS TO PUT A S IN FRONT OF EVERY TABLE/FIGURE NUMBER
\renewcommand{\thesection}{S\arabic{section}}
\renewcommand{\theequation}{S\arabic{equation} Eq}
% \renewcommand{\thesection}{S\arabic{section}}f
% \renewcommand{\thetable}{S\arabic{table}}
% \renewcommand{\tablename}{Table}
% \renewcommand{\figurename}{Fig}

% graphic paths
\graphicspath{{figures/}}

% TABLE COLUMN ALIGNED WITH THE DECIMAL POINT
\newcolumntype{d}[1]{D{.}{.}{#1}}

\definecolor{mariana}{HTML}{CC79A7}
\DeclareRobustCommand{\magenta}[1]{{\sethlcolor{mariana}\hl{#1}}}
\DeclareRobustCommand{\mmcom}[1]{\noindent{\sethlcolor{mariana}\hl{\textbf{MM COMMENT:}  #1}}}

\definecolor{alessio}{HTML}{53a84a}
\DeclareRobustCommand{\green}[1]{{\sethlcolor{alessio}\hl{#1}}}
\DeclareRobustCommand{\accom}[1]{\noindent{\sethlcolor{alessio}\hl{\textbf{AC COMMENT:}  #1}}}
\newcommand{\actext}[1]{\textcolor{alessio}{#1}}

\definecolor{ronaldo}{HTML}{6faee3}
\DeclareRobustCommand{\blue}[1]{{\sethlcolor{ronaldo}\hl{#1}}}
\DeclareRobustCommand{\rmcom}[1]{\noindent{\sethlcolor{ronaldo}\hl{\textbf{RM COMMENT:}  #1}}}

% % TO ALLOW CROSS-REFERENCING IN OVERLEAF
% \makeatletter
% \newcommand*{\addFileDependency}[1]{% argument=file name and extension
%   \typeout{(#1)}
%   \@addtofilelist{#1}
%   \IfFileExists{#1}{}{\typeout{No file #1.}}
% }
% \makeatother

% \newcommand*{\myexternaldocument}[2]{%
%     \externaldocument[#2]{#1}%
%     \addFileDependency{#1.tex}%
%     \addFileDependency{#1.aux}%
% }

% % include main text to correctly display references
% \myexternaldocument{plos_latex_template}{M-}

% % TO ALLOW CROSS-REFERENCING IN OVERLEAF

%\makeatletter
%\newcommand*{\addFileDependency}[1]{% argument=file name and extension
%	\typeout{(#1)}
%	\@addtofilelist{#1}
%	\IfFileExists{#1}{}{\typeout{No file #1.}}
%}

\renewcommand{\thefigure}{S\arabic{figure}}
\renewcommand{\figurename}{Figure}
\renewcommand{\fnum@figure}{\textbf{\thefigure~\figurename}}

\renewcommand{\thetable}{S\arabic{table}}
\renewcommand{\tablename}{Table}
\def\fnum@table{\textbf{\thetable~\tablename}}

\makeatother

% resetting counters

\setcounter{equation}{0}
\setcounter{figure}{0}
\setcounter{table}{0}

%\newcommand*{\myexternaldocument}[2]{%
%	\externaldocument[#2]{#1}%
%	\addFileDependency{#1.tex}%
%	\addFileDependency{#1.aux}%
%}
% %

% % include main text to correctly display references
% \externaldocument[M-]{main}
%\myexternaldocument{main}{M-}

\section*{Supplementary Materials for the manuscript entitled:\\ \textit{The parenthood effect in urban mobility}}

%\author{Mariana Macedo, Ronaldo Menezes, and Alessio Cardillo}

%\date{}

%\begin{document}
	
	% SPACE BETWEEN EQUATIONS AND TEXT
	
	\setlength{\abovedisplayskip}{10pt}
	\setlength{\belowdisplayskip}{10pt}
	
%	\maketitle

	\section{Data}
	\label{sec:data}

        We consider two sources of data: one accounting for travellers' mobility and features, and another concerning the location of amenities. Mobility data combines information from commuting mobility and on the location of amenities. The former data are collected by the United States Census Bureau (American Community Survey, ACS)~\cite{data_usa_acs}, and can be either extracted and manipulated through their APIs or downloaded directly from the ACS website. The location of amenities, instead, is available via the Open Street Map platform and the corresponding APIs and datasets~\cite{OpenStreetMap2017}.

        Our analysis includes 17 metropolitan areas from the United States. The selection was based on the availability of sufficient spatial resolution after merging home and workplace zone tessellations. As the spatial definitions of residential and workplace zones in the ACS do not align, we mapped zones to ensure consistency between home and work locations. We retained only those metropolitan areas that had at least 6 zones post-mapping, which was necessary to enable meaningful calculation of entropy-based diversity measures. Some large metropolitan areas were excluded because they did not meet this threshold due to how workplace zones are aggregated in the ACS data for privacy protection purposes. The 17 metropolitan areas included in our study, along with their characteristics, are listed in Table~3 of the main manuscript.

        In this study, we Core-Based Statistical Areas (CBSAs) as defined by the U.S. Census Bureau\footnote{Housing Patterns and Core-Based Statistical Areas. \url{https://www.census.gov/topics/housing/housing-patterns/about/core-based-statistical-areas.html} (Accessed: 01/Sep/2025)} to delineate metropolitan regions. CBSAs are standard statistical geographic entities representing metropolitan and micropolitan areas, defined based on commuting patterns and economic ties. In the American Community Survey data, CBSAs are identified by the variable 'MET2013\_label', which corresponds one-to-one with CBSA definitions. Each urban area in our analysis can thus be identified using either its 'MET2013\_label' descriptor or its equivalent CBSA code.

        From the ACS data, we extracted information about the individuals (\ie, travellers) such as their: age, gender, marital status, cohabitation status, household location, and income level. We also extracted the travellers' commuting patterns together with information about the travel time needed to go from home to work, and the locations (\ie, zones) of the travellers' home and workplaces. 

        The spatial tessellation of workplace locations in the ACS data does not align with that of home locations. Therefore, to establish a consistent mapping between home and workplace areas, we merged the home zones to match the workplace definitions. Such a choice has the downside that the workplace tessellation has a smaller spatial granularity (\ie, the number of zones available per metropolitan area is smaller than for the other tessellation) but it allow us to map home and work locations to the same IDs ensuring consistency across our spatial characterisation and mobility analysis. The merge of the two tessellations brought us from having between 30 and 60 zones to having between 5 and 23 zones per metropolitan area. 

        Using Open Street Map (OSM), we can pinpoint how many amenities per type we have within each zone. We collected amenities using OSM APIs from the following categories: \texttt{amenity}, \texttt{highway}, \texttt{building}, and \texttt{healthcare}. Then, we grouped amenities into the following custom categories: \texttt{work}, \texttt{residential}, \texttt{leisure}, \texttt{health}, \texttt{food}, \texttt{transport}, \texttt{religious}, \texttt{education}, and \texttt{services}. In Sec.~\ref{sec:dictionary}, we report the complete list of OSM categories belonging to each of our custom categories. It is worth noting that we considered also a set of amenities to be excluded from our analysis, which we grouped into the category named \texttt{remove\_categories}. The criteria used to discard a type of amenity were based, for instance, on the detection of typos in their name (\eg, curch instead of church), on their little relevance in our context (\eg, roof), or on their classification's ambiguity (\eg{,} \texttt{apartments;hotel;office}).

        The number of amenities kept in each metropolitan area is displayed in Tab.~\ref{tab:amenities_per_area}. New York and Dallas present the highest number of amenities across categories, whereas Indianapolis, Nashville, and Virginia Beach are the areas having the smallest number of amenities across categories. We note that the results reported in the manuscript are not correlated with the number of amenities observed in each metropolitan region. Thus, a higher number of amenities does not necessarily correspond to greater differences in mobility diversity between parents and non-parents, $\Delta M_P$, or between married and non-married individuals, $\Delta M_M$.

        %
        %  TABLE: NR. OF AMENITIES PER CATEGORY AND PER URBAN AREA
        %
        \begin{table}[h]
            \centering
            \caption{Number of amenities per category for each metropolitan area considered in our study and ordered by number of zones. Each urban area in our analysis can be identified using either `MET2013\_label' or `CBSA'; each `CBSA' corresponds to a single, unique `MET2013\_label'.}
            \label{tab:amenities_per_area}
            \resizebox{0.8\textwidth}{!}{%
            \begin{tabular}{r|rrrrrrrrr}
            \toprule
                  & \multicolumn{9}{c}{Amenity categories} \\
            
            Metropolitan area & education &   food & health & leisure & religious & residential & services & transport &   work \\
            \midrule

            Pittsburgh     & 2,605  & 3,052  & 1,393  & 1,033 & 1,917 & 354,302   & 146,717  & 23,602  & 14,921 \\
            Houston        & 3,780  & 5,720  & 2,618  & 1,136 & 2,788 & 210,622   & 333,251  & 35,860  & 7,969  \\
            Baltimore      & 2,641  & 4,192  & 2,176  & 1,328 & 2,121 & 341,244   & 191,810  & 40,333  & 11,839 \\
            Kansas City    & 2,728  & 2,858  & 1,245  & 800   & 1,780 & 158,079   & 164,216  & 32,394  & 4,975  \\
            Indianapolis   & 1,308  & 1,923  & 947    & 611   & 1,289 & 174,281   & 162,118  & 28,270  & 3,513  \\
            Cincinnati     & 1,980  & 2,889  & 986    & 1,473 & 1,780 & 309,783   & 191,697  & 20,849  & 5,385  \\
            Virginia Beach & 1,606  & 2,207  & 1,048  & 1,529 & 1,587 & 538,878   & 94,111   & 16,059  & 21,045 \\
            Nashville      & 1,811  & 2,237  & 582    & 691   & 2,401 & 107,552   & 215,135  & 10,888  & 2,367  \\
            Dallas         & 6,641  & 8,037  & 4,449  & 4,300 & 3,732 & 1,552,158 & 433,361  & 54,439  & 29,115 \\
            St. Louis      & 2,924  & 2,911  & 1,142  & 835   & 2,220 & 165,813   & 198,888  & 36,589  & 3,826  \\
            Charlotte      & 2,520  & 3,046  & 1,005  & 940   & 2,805 & 204,868   & 281,034  & 17,216  & 5,952  \\
            Minneapolis    & 2,472  & 4,025  & 1,646  & 1,337 & 1,188 & 436,152   & 245,928  & 86,069  & 7,248  \\
            Chicago        & 8,807  & 11,235 & 4,299  & 4,551 & 5,121 & 433,504   & 423,858  & 121,270 & 29,331 \\
            Philadelphia   & 5,664  & 6,200  & 2,878  & 1,989 & 3,226 & 253,976   & 341,700  & 42,914  & 8,240  \\
            Washington DC  & 5,583  & 9,126  & 3,244  & 2,311 & 3,434 & 416,985   & 357,768  & 76,469  & 11,497 \\
            Atlanta        & 3,590  & 5,851  & 2,174  & 1,537 & 4,357 & 427,380   & 464,576  & 34,184  & 12,182 \\
            New York       & 14,063 & 23,027 & 11,294 & 5,814 & 7,904 & 1,403,777 & 573,505  & 182,545 & 45,152 \\
            \bottomrule
            \end{tabular}%
            }
        \end{table}

        \section{Null Models}
        \label{sec:null_models}

\begin{figure}[ht]
            \centering
            \includegraphics[width=0.45\linewidth]{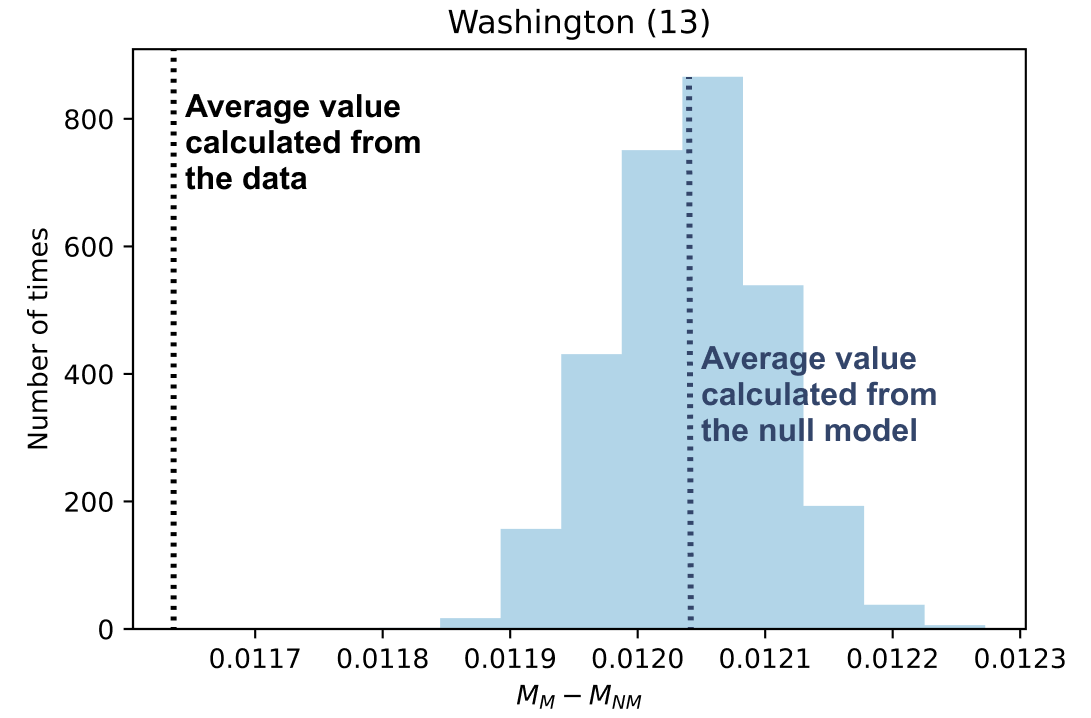}
            \caption{Comparison between null models and empirical values. We compare the distribution of values extracted 5,000 times from each null model (blue histogram) against the average value obtained by bootstrapping 80\% of the data (line in black on the left), also repeated 5,000 times. If the $p$-value is equal to 0, it indicates that the probability of obtaining the observed value (black line) using the null model is equal to zero. If the $p$-value is greater than 0, it indicates that there is a non-zero probability of obtaining the empirical average using the null model, and if its value is higher than 0.1, we consider that we cannot reject the hypothesis that the average empirical value originates from a mechanism similar to that implemented in the null model.}
            \label{fig:example_washington}
        \end{figure}

        To ensure that the phenomenology observed in our work is not the result of chance, we considered five null models and generated the corresponding set of synthetic travels. For a given metropolitan area, for each traveller we collected the following information: \texttt{ID}, \texttt{marital status}, \texttt{parental status}, \texttt{home zone ID}, \texttt{work zone ID}, \texttt{commuting travel time}, and \texttt{commuting travel distance}. The latter is computed as the distance (in metres) between the centroids of the home and work zones. The five null models (henceforth \NM{X} with $X \in \{1, \ldots, 5\}$) act on each travel in the following way:
        \begin{description}
            \item[\NM{1}] We shuffle uniformly at random the travellers' \texttt{marital status}. Here, we can rule out the likelihood that the results arise from random fluctuations or errors related to group classification.
            \item[\NM{2}] We replace the \texttt{commuting travel time} with another extracted randomly from a truncated power law distribution obtained by fitting the commuting times of the whole population. This allows us to rule out the possibility that the observed results are solely driven by the structure of the travel time distribution.
            \item[\NM{3}] We replace the \texttt{commuting travel distance} with another one extracted randomly from a truncated power law distribution obtained by fitting the commuting distances of the whole population. Such an operation implies that we also update the ID of the traveller's work zone to avoid inconsistencies. If multiple zones are located at the same distance from the origin, we select uniformly at random one of them as the destination zone. This allows us to rule out the possibility that the observed results are solely driven by the structure of the travel distance distribution.
            \item[\NM{4}] We shuffle uniformly at random the travellers' \texttt{commuting travel time}. Here, we can rule out the likelihood that the results arise from random fluctuations or errors related to the travel time feature.
            \item[\NM{5}] We shuffle uniformly at random the travellers' \texttt{work zone ID}. Such an operation is equivalent to altering the \texttt{commuting travel distance}.  Here, we can rule out the likelihood that the results arise from random fluctuations or errors related to the \texttt{work zone ID} feature.
        \end{description}
        It is worth mentioning that \NM{2} and \NM{3} are generalisations of \NM{4} and \NM{5}, respectively. For each null model, we have a distribution of 5,000 realisations/values. To ensure the robustness of our findings, we compute the $p$-value by comparing the average values of cost ($C$) and diversity ($H$,$M$) with the distributions obtained from the null models. Taking \figurename~\ref{fig:example_washington} as an example, the observed average value lies outside the distribution of the corresponding null model, resulting in a $p$-value equal to zero, indicating that such a result is highly unlikely to occur under the null hypothesis. If the observed value lies within the distribution with a probability higher than 10\%, we consider that it could have been drawn from a mechanism similar to that represented by the corresponding null model. In our findings, we observe that most of the empirical values lay far away from the null model distribution, providing statistical significance to our findings.

        %
        %  FIG: EXAMPLE OF P-VALUE COMPUTATION FOR WASHINGTON
        %
        %

\subsection{Complete statistics for the null models}
\label{sec:nm_complete_stats}

In this section, we report the complete statistics for the values of $\Delta C_P$ (Tables \ref{tab:fullstats_dcp_nm1} -- \ref{tab:fullstats_dcp_nm5}) and $\Delta C_M$ (Tables \ref{tab:fullstats_dcm_nm1} -- \ref{tab:fullstats_dcm_nm5}) computed using our null models. Specifically, for each case, we computed the average, $\avg{\,\cdot\,}$, and standard deviation, $\sigma (\, \cdot \,)$, of both the empirical and synthetic data, the $p$-value used to compare the null models with the empirical data, the value of Welch's $t$-test~\cite{welchbiometrika1947}, and the 95\% confidence interval, CI. These are standard measures to ensure the robustness of our results. In particular, Welch's $t$-test is a generalization of Student's $t$-test which does not assume equal population variances. High values of $t$ (\ie{} $\abs{t} \gg 100$) indicate that $\Delta C_P$ (or $\Delta C_M$) for the empirical data and the null model have different mean values. By looking at the values of $t$ in Tables \ref{tab:fullstats_dcp_nm1}--\ref{tab:fullstats_dcm_nm5}, we can say that except for \NM{3}, all the other \NM{s} perform relatively well (with few exceptions).

    %
    %  TABLE STATISTICS DELTA C_P NM1
    %
    %
    \begin{table}[th!]
        \centering
        \caption{Description of the statistical measures reported in Table 1 about the comparison between the empirical average mobility cost differences between parents and non-parents travellers, $\Delta C_P$, for the null model \NM{1}. For each urban area we report: the average value of the cost difference, $\langle  \Delta C_P \rangle$, and its standard deviation, $\sigma(\Delta C_P)$, for both empirical values and those obtained using the null model. Concerning the statistics, instead, we report the $p$-value, the boundaries of the 95\% confidence interval (CI), and the Welch's $t$-value.}
        \label{tab:fullstats_dcp_nm1}
        \resizebox{0.99\textwidth}{!}{%
        \begin{tabular}{rrrrrrrrr}
            \toprule
            \multicolumn{1}{c}{\multirow{2}{*}{Urban area}} & \multicolumn{2}{c}{DATA} & \multicolumn{2}{c}{\NM{1}} & \multicolumn{4}{c}{STATISTICS} \\
             & \multicolumn{1}{c}{$\langle  \Delta C_P \rangle$} & \multicolumn{1}{c}{$\sigma(\Delta C_P)$} & \multicolumn{1}{c}{$\langle \Delta C_P \rangle$} & \multicolumn{1}{c}{$\sigma(\Delta C_P )$} & \multicolumn{1}{c}{$p$-value} & \multicolumn{2}{c}{CI (95\%)} & \multicolumn{1}{c}{$t$-value} \\
            \midrule
            Baltimore (6) & -6.10e-03 & 5.67e-05 & -1.27e-06 & 1.03e-04 & 0.0000 & -1.94e-04 & 2.12e-04 & -3668.55 \\
            Houston (6) & 9.70e-03 & 6.69e-05 & 4.32e-06 & 1.21e-04 & 0.0000 & -2.36e-04 & 2.35e-04 & 4952.02 \\
            Pittsburgh (6) & 8.50e-03 & 7.43e-05 & -1.55e-05 & 1.37e-04 & 0.0000 & -2.60e-04 & 2.60e-04 & 3860.80 \\
            Cincinnati (7) & -7.00e-03 & 5.27e-05 &  -1.23e-05 & 9.67e-05 & 0.0000 & -1.88e-04 & 1.82e-04 & -4475.69 \\
            Indianapolis (7) & 4.10e-03 & 6.82e-05 & -1.59e-05 & 1.17e-04 & 0.0000 & -2.34e-04 & 2.29e-04 & 2135.94 \\
            Kansas City (7) & 4.80e-03 & 6.06e-05 & -1.24e-05 & 1.04e-04 & 0.0000 & -2.11e-04 & 1.99e-04 & 2785.35 \\
            Virginia Beach (7) & 6.00e-03 & 5.08e-05 & -1.16e-05 & 9.60e-05 & 0.0000 & -1.79e-04 & 1.90e-04 & 3876.43 \\
            Nashville (8) & -2.80e-03 & 5.62e-05 & -4.48e-06 & 9.98e-05 & 0.0000 & -1.94e-04 & 1.85e-04 & -1710.46 \\
            Charlotte (9) & -2.00e-04 & 5.71e-05 & 1.05e-05 & 9.94e-05 & 0.0124 & -1.98e-04 & 1.90e-04 & -137.25 \\
            Dallas (9) & 2.30e-03 & 5.48e-05 & -3.40e-06 & 9.27e-05 & 0.0000 & -1.85e-04 & 1.75e-04 & 1507.93 \\
            St. Louis (9) & 1.00e-04 & 4.84e-05 & 8.69e-06 & 9.06e-05 & 0.0976 & -1.77e-04 & 1.78e-04 & 80.35 \\
            Minneapolis (10) & -4.40e-03 & 4.17e-05 & 6.73e-06 & 7.47e-05 & 0.0000 & -1.42e-04 & 1.46e-04 & -3636.16 \\
            Chicago (11) & -6.60e-03 & 4.24e-05 & -5.26e-06 & 7.67e-05 & 0.0000 & -1.51e-04 & 1.47e-04 & -5298.74 \\
            Philadelphia (11) & -2.00e-03 & 2.48e-03 & -3.09e-06 & 2.44e-03 & 0.4430 & -6.63e-03 & 1.15e-04 & 21.75 \\
            Washington (13) & 1.32e-02 & 4.80e-05 & -8.51e-06 & 8.34e-05 & 0.0000 & -1.67e-04 & 1.61e-04 & 9734.17 \\
            Atlanta (17) & 6.00e-04 & 4.84e-05 & -1.78e-06 & 8.71e-05 & 0.0000 & -1.72e-04 & 1.63e-04 & 391.82 \\
            New York (23) & 6.00e-04 & 2.80e-05 & -2.46e-05 & 5.42e-05 & 0.0000 & -1.04e-04 & 1.03e-04 & 710.52 \\
            \bottomrule
        \end{tabular}
        }
    \end{table}

    %
    %  TABLE STATISTICS DELTA C_P NM2
    %
    %
    \begin{table}[h]
        \centering
        \caption{Description of the statistical measures reported in Table 1 about the comparison between the empirical average mobility cost differences between parents and non-parents travellers, $\Delta C_P$, for the null model \NM{2}. The reader can look at the caption of Tab.~\ref{tab:fullstats_dcp_nm1} for the details on the meaning of each column.}
        \label{tab:fullstats_dcp_nm2}
        \resizebox{0.99\textwidth}{!}{%
        \begin{tabular}{rrrrrrrrr}
            \toprule
            \multicolumn{1}{c}{\multirow{2}{*}{Urban area}} & \multicolumn{2}{c}{DATA} & \multicolumn{2}{c}{\NM{2}} & \multicolumn{4}{c}{STATISTICS} \\
             & \multicolumn{1}{c}{$\langle  \Delta C_P \rangle$} & \multicolumn{1}{c}{$\sigma(\Delta C_P)$} & \multicolumn{1}{c}{$\langle \Delta C_P \rangle$} & \multicolumn{1}{c}{$\sigma(\Delta C_P )$} & \multicolumn{1}{c}{$p$-value} & \multicolumn{2}{c}{CI (95\%)} & \multicolumn{1}{c}{$t$-value}\\
            \midrule
            Baltimore (6) & -6.10e-03 & 5.74e-05 & -1.39e-06 & 1.07e-04 & 0.0000 & -2.02e-04 & 2.26e-04 & -3538.85 \\
            Houston (6) & 9.70e-03 & 6.68e-05 & -4.34e-05 & 1.21e-04 & 0.0000 & -2.38e-04 & 2.55e-04 & 4953.03 \\
            Pittsburgh (6) & 8.50e-03 & 7.50e-05 & 7.87e-06 & 1.37e-04 & 0.0000 & -2.74e-04 & 2.65e-04 & 3850.51 \\
            Cincinnati (7) & -7.00e-03 & 5.19e-05 & -1.98e-05 & 9.37e-05 & 0.0000 & -1.80e-04 & 1.79e-04 & -4601.71 \\
            Indianapolis (7) & 4.10e-03 & 6.75e-05 & 8.46e-06 & 1.22e-04 & 0.0000 & -2.30e-04 & 2.39e-04 & 2067.16 \\
            Kansas City (7) & 4.80e-03 & 6.22e-05 & 5.32e-06 & 1.04e-04 & 0.0000 & -1.93e-04 & 2.05e-04 & 2763.86 \\
            Virginia Beach (7) & 6.00e-03 & 5.06e-05 & -6.88e-06 & 9.56e-05 & 0.0000 & -1.89e-04 & 1.80e-04 & 3895.56 \\
            Nashville (8) & -2.80e-03 & 5.63e-05 & 4.01e-06 & 9.99e-05 & 0.0000 & -2.04e-04 & 1.94e-04 & -1707.64 \\
            Charlotte (9) & -2.00e-04 & 5.84e-05 & 2.28e-05 & 9.77e-05 & 0.0150 & -1.91e-04 & 1.85e-04 & -138.03 \\
            Dallas (9) & 2.30e-03 & 5.49e-05 & 2.93e-06 & 9.57e-05 & 0.0000 & -1.92e-04 & 1.89e-04 & 1472.63 \\
            St. Louis (9) & 1.00e-04 & 4.83e-05 & 2.34e-06 & 9.10e-05 & 0.1100 & -1.69e-04 & 1.83e-04 & 76.46 \\
            Minneapolis (10) & -4.40e-03 & 4.22e-05 & -1.33e-05 & 7.64e-05 & 0.0000 &-1.44e-04 & 1.53e-04 & -3566.80 \\
            Chicago (11) & -6.60e-03 & 4.16e-05 & 2.66e-06 & 7.06e-05 & 0.0000 & -1.45e-04 & 1.40e-04 & -5663.87 \\
            Philadelphia (11) & -2.00e-03 & 2.48e-03 & -2.17e-05 & 2.29e-03 & 0.4420 & -6.04e-03 & 1.29e-04 & 21.04 \\
            Washington (13) & 1.32e-02 & 4.82e-05 & 1.25e-05 & 8.47e-05 & 0.0000 & -1.56e-04 & 1.60e-04 & 9602.29 \\
            Atlanta (17) & 6.00e-04 & 4.85e-05 & -6.37e-06 & 8.47e-05 & 0.0000 & -1.65e-04 & 1.72e-04 & 397.86 \\
            New York (23) & 6.00e-04 & 2.89e-05 & -9.91e-06 & 5.05e-05 & 0.0000 & -1.03e-04 & 9.63e-05 & 748.04 \\
            \bottomrule
        \end{tabular}
        }
    \end{table}

    %
    %  TABLE STATISTICS DELTA C_P NM3
    %
    %
    \begin{table}[h]
        \centering
        \caption{Description of the statistical measures reported in Table 1 about the comparison between the empirical average mobility cost differences between parents and non-parents travellers, $\Delta C_P$, for the null model \NM{3}. The reader can look at the caption of Tab.~\ref{tab:fullstats_dcp_nm1} for the details on the meaning of each column.}
        \label{tab:fullstats_dcp_nm3}
        \resizebox{0.99\textwidth}{!}{%
        \begin{tabular}{rrrrrrrrr}
            \toprule
            \multicolumn{1}{c}{\multirow{2}{*}{Urban area}} & \multicolumn{2}{c}{DATA} & \multicolumn{2}{c}{\NM{3}} & \multicolumn{4}{c}{STATISTICS} \\
             & \multicolumn{1}{c}{$\langle  \Delta C_P \rangle$} & \multicolumn{1}{c}{$\sigma(\Delta C_P)$} & \multicolumn{1}{c}{$\langle \Delta C_P \rangle$} & \multicolumn{1}{c}{$\sigma(\Delta C_P )$} & \multicolumn{1}{c}{$p$-value} & \multicolumn{2}{c}{CI (95\%)} & \multicolumn{1}{c}{$t$-value} \\
            \midrule
            Baltimore (6) & -6.10e-03 & 5.78e-05 & 1.59e-03 & 5.13e-03 & 0.0702 & -8.23e-03 & 1.14e-02 & -99.77 \\
            Houston (6) & 9.70e-03 & 6.60e-05 & 7.39e-04 & 6.15e-03 & 0.1540 & -8.50e-03 & 1.58e-02 & 70.87 \\
            Pittsburgh (6) & 8.50e-03 & 7.49e-05 & 8.69e-03 & 6.29e-03 & 0.4510 & -4.97e-03 & 2.07e-02 & 9.02 \\
            Cincinnati (7) & -7.00e-03 & 5.23e-05 & 2.20e-02 & 6.35e-03 & 0.0000 & 8.82e-03 & 3.42e-02 & -314.05 \\
            Indianapolis (7) & 4.10e-03 & 6.63e-05 & -4.02e-03 & 6.71e-03 & 0.0652 & -1.79e-02 & 8.27e-03 & 103.68 \\
            Kansas City (7) & 4.80e-03 & 6.18e-05 & 1.21e-02 & 7.22e-03 & 0.1570 & -1.49e-03 & 2.61e-02 & -73.74 \\
            Virginia Beach (7) & 6.00e-03 & 5.02e-05 & 9.61e-03 & 5.35e-03 & 0.1830 & -4.24e-04 & 2.11e-02 & -64.22 \\
            Nashville (8) & -2.80e-03 & 5.58e-05 & 2.15e-02& 7.75e-03 & 0.0002 & 7.46e-03 & 3.84e-02 & -229.81 \\
            Charlotte (9) & -2.00e-04 & 5.90e-05 & -7.34e-03 & 5.63e-03 & 0.2140 & -1.54e-02 & 6.81e-03 & 54.20 \\
            Dallas (9) & 2.30e-03 & 5.43e-05 & 2.73e-03 & 4.37e-03 & 0.4900 & -6.19e-03 & 1.09e-02 & -1.18 \\
            St. Louis (9) & 1.00e-04 & 4.94e-05 & 1.29e-02 & 7.24e-03 & 0.0036 & 3.13e-03 & 3.13e-02 & -143.98 \\
            Minneapolis (10) & -4.40e-03 & 4.19e-05 & 6.49e-03 & 5.54e-03 & 0.0026 & -1.18e-03 & 2.06e-02 & -175.63 \\
            Chicago (11) & -6.60e-03 & 4.12e-05 & 1.60e-02 & 5.16e-03 & 0.0000 & 4.37e-03 & 2.44e-02 & -291.35 \\
            Philadelphia (11) & -2.00e-03 & 2.47e-03 &  9.89e-03 & 4.72e-03 & 0.0012 & 1.72e-03 & 2.07e-02 & -162.82 \\
            Washington (13) & 1.32e-02 & 4.85e-05 & 5.23e-03 & 4.57e-03 & 0.0524 & -2.87e-03 & 1.51e-02 & 115.97 \\
            Atlanta (17) & 6.00e-04 & 4.84e-05 & 8.96e-03 & 5.49e-03 & 0.1300 & -3.77e-03 & 1.77e-02 & -78.20 \\
            New York (23) & 6.00e-04 & 2.86e-05 & 6.14e-03 & 3.19e-03 & 0.1470 & -1.96e-03 & 1.03e-02 & -76.44 \\
            \bottomrule
        \end{tabular}
        }
    \end{table}

    %
    %  TABLE STATISTICS DELTA C_P NM4
    %
    %
    \begin{table}[h]
        \centering
        \caption{Description of the statistical measures reported in Table 1 about the comparison between the empirical average mobility cost differences between parents and non-parents travellers, $\Delta C_P$, for the null model \NM{4}. The reader can look at the caption of Tab.~\ref{tab:fullstats_dcp_nm1} for the details on the meaning of each column.}
        \label{tab:fullstats_dcp_nm4}
        \resizebox{0.99\textwidth}{!}{%
        \begin{tabular}{rrrrrrrrr}
            \toprule
            \multicolumn{1}{c}{\multirow{2}{*}{Urban area}} & \multicolumn{2}{c}{DATA} & \multicolumn{2}{c}{\NM{4}} & \multicolumn{4}{c}{STATISTICS} \\
             & \multicolumn{1}{c}{$\langle  \Delta C_P \rangle$} & \multicolumn{1}{c}{$\sigma(\Delta C_P)$} & \multicolumn{1}{c}{$\langle \Delta C_P \rangle$} & \multicolumn{1}{c}{$\sigma(\Delta C_P )$} & \multicolumn{1}{c}{$p$-value} & \multicolumn{2}{c}{CI (95\%)} & \multicolumn{1}{c}{$t$-value} \\
            \midrule
            Baltimore (6) & -6.10e-03 & 5.72e-05 &  9.02e-06 & 1.06e-04 & 0.0000 & -1.96e-04 & 2.12e-04 & -3584.16 \\
            Houston (6) & 9.70e-03 & 6.68e-05 & 1.55e-05 & 1.21e-04 & 0.0000 & -2.33e-04 & 2.39e-04 & 4966.85 \\
            Pittsburgh (6) & 8.50e-03 & 7.37e-05 & -9.08e-06 & 1.40e-04 & 0.0000 & -2.83e-04 & 2.72e-04 & 3806.55 \\
            Cincinnati (7) & -7.00e-03 & 5.24e-05 & 1.13e-06 & 9.39e-05 & 0.0000 & -1.86e-04 & 1.90e-04 & -4581.46 \\
            Indianapolis (7) & 4.10e-03 & 6.84e-05 & 1.90e-05 & 1.14e-04 & 0.0000 & -2.26e-04 & 2.26e-04 & 2168.41 \\
            Kansas City (7) & 4.80e-03 & 6.19e-05 & 1.57e-05 & 1.06e-04 & 0.0000 & -2.01e-04 & 2.12e-04 & 2746.51 \\
            Virginia Beach (7) & 6.00e-03 & 5.01e-05 & -1.47e-05 & 9.35e-05 & 0.0000 & -1.86e-04 & 1.89e-04 & 3968.31 \\
            Nashville (8) & -2.80e-03 & 5.59e-05 & -3.75e-06 & 9.97e-05 & 0.0000 & -1.97e-04 & 1.96e-04 & -1715.27 \\
            Charlotte (9) & -2.00e-04 & 5.71e-05 & 1.29e-05 & 9.47e-05 & 0.0096 & -1.84e-04 & 1.88e-04 & -141.52 \\
            Dallas (9) & 2.30e-03 & 5.44e-05 & 4.43e-06 & 9.12e-05 & 0.0000 & -1.78e-04 & 1.83e-04 & 1528.33 \\
            St. Louis (9) & 1.00e-04 & 4.91e-05 &  1.29e-05 & 8.98e-05 & 0.1070 & -1.83e-04 & 1.83e-04 & 78.98 \\
            Minneapolis (10) & -4.40e-03 & 4.16e-05 & 2.70e-05 & 7.41e-05 & 0.0000 & -1.49e-04 & 1.46e-04 & -3662.80 \\
            Chicago (11) & -6.60e-03 & 4.17e-05 & -1.33e-05 & 7.26e-05 & 0.0000 & -1.46e-04 & 1.45e-04 & -5547.59 \\
            Philadelphia (11) & -2.00e-03 & 2.48e-03 & -3.74e-03 & 2.13e-03 & 0.4790 & -6.01e-03 & 1.14e-04 & 14.12 \\
            Washington (13) & 1.32e-02 & 4.86e-05 & -5.86e-06 & 8.44e-05 & 0.0000 & -1.63e-04 & 1.63e-04 & 9608.07 \\
            Atlanta (17) & 6.00e-04 & 4.76e-05 & -1.15e-05 & 8.74e-05 & 0.0000 & -1.64e-04 & 1.74e-04 & 391.14 \\
            New York (23) & 6.00e-04 & 2.88e-05 & 6.97e-06 & 4.94e-05 & 0.0000 & -9.68e-05 & 1.01e-04 & 756.39 \\
            \bottomrule
        \end{tabular}
        }
    \end{table}

    %
    %  TABLE STATISTICS DELTA C_P NM5
    %
    %
    \begin{table}[h]
        \centering
        \caption{Description of the statistical measures reported in Table 1 about the comparison between the empirical average mobility cost differences between parents and non-parents travellers, $\Delta C_P$, for the null model \NM{5}. The reader can look at the caption of Tab.~\ref{tab:fullstats_dcp_nm1} for the details on the meaning of each column.}
        \label{tab:fullstats_dcp_nm5}
        \resizebox{0.99\textwidth}{!}{%
        \begin{tabular}{rrrrrrrrr}
            \toprule
            \multicolumn{1}{c}{\multirow{2}{*}{Urban area}} & \multicolumn{2}{c}{DATA} & \multicolumn{2}{c}{\NM{5}} & \multicolumn{4}{c}{STATISTICS} \\
             & \multicolumn{1}{c}{$\langle  \Delta C_P \rangle$} & \multicolumn{1}{c}{$\sigma(\Delta C_P)$} & \multicolumn{1}{c}{$\langle \Delta C_P \rangle$} & \multicolumn{1}{c}{$\sigma(\Delta C_P )$} & \multicolumn{1}{c}{$p$-value} & \multicolumn{2}{c}{CI (95\%)} & \multicolumn{1}{c}{$t$-value} \\
            \midrule
            Baltimore (6) & -6.10e-03 & 5.77e-05 & -4.79e-03 & 6.09e-05 & 0.0000 & -4.93e-03 & -4.69e-03 & -1081.82 \\
            Houston (6) & 9.70e-03 & 6.58e-05 & 9.99e-03 & 7.57e-05 & 0.0002 & 9.81e-03 & 1.01e-02 & -191.79 \\
            Pittsburgh (6) & 8.50e-03 & 7.52e-05 & 1.65e-02 & 3.90e-03 & 0.3160 & 8.06e-03 & 1.67e-02 & -97.79 \\
            Cincinnati (7) & -7.00e-03 & 5.17e-05 & -2.50e-03 & 4.64e-05 & 0.0000 & -2.58e-03 & -2.39e-03 & -4569.17 \\
            Indianapolis (7) & 4.10e-03 & 6.65e-05 & 4.43e-03 & 6.93e-05 & 0.0000 & 4.29e-03 & 4.57e-03 & -251.93 \\
            Kansas City (7) & 4.80e-03 & 6.23e-05 & 4.44e-03 & 5.44e-05 & 0.0000 & 4.38e-03 & 4.60e-03 & 229.47 \\
            Virginia Beach (7) & 6.00e-03 & 5.07e-05 & 6.73e-03 & 4.17e-05 & 0.0000 & 6.64e-03 & 6.80e-03 & -834.60 \\
            Nashville (8) & -2.80e-03 & 5.58e-05 & 2.36e-04 & 6.04e-05 & 0.0000 & 1.23e-04 & 3.67e-04 & -2597.91 \\
            Charlotte (9) & -2.00e-04 & 5.75e-05 & -8.00e-05 & 5.81e-05 & 0.0104 & -2.08e-04 & 1.96e-05 & -115.46 \\
            Dallas (9) & 2.30e-03 & 5.42e-05 & 2.75e-03 & 6.03e-05 & 0.0000 & 2.64e-03 & 2.87e-03 & -402.07 \\
            St. Louis (9) & 1.00e-04 & 4.87e-05 & 3.45e-03 & 6.20e-05 & 0.0000 & 3.34e-03 & 3.59e-03 & -3005.37 \\
            Minneapolis (10) & -4.40e-03 & 4.17e-05 & -2.05e-03 & 4.97e-05 & 0.0000 & -2.16e-03 & -1.96e-03 & -2548.08 \\
            Chicago (11) & -6.60e-03 & 4.15e-05 & -6.33e-03 & 5.87e-05 & 0.0000 & -6.46e-03 & -6.24e-03 & -214.48 \\
            Philadelphia (11) & -2.00e-03 & 2.48e-03 & -3.43e-03 & 4.45e-05 & 0.0000 & -3.52e-03 & -3.35e-03 & 55.11 \\
            Washington (13) & 1.32e-02 & 4.85e-05 & 1.19e-02 & 5.35e-05 & 0.0000 & 1.18e-02 & 1.20e-02 & 1282.69 \\
            Atlanta (17) & 6.00e-04 & 4.95e-05 & 3.61e-03 & 6.14e-05 & 0.0000 & 3.48e-03 & 3.72e-03 & -2739.16 \\
            New York (23) & 6.00e-04 & 2.83e-05 & 6.00e-04 & 2.72e-05 & 0.4530 & 5.62e-04 & 6.66e-04 & 3.57 \\
            \bottomrule
        \end{tabular}
        }
    \end{table}

    %
    %  TABLE STATISTICS DELTA C_M NM1
    %
    %
    \begin{table}[h]
            \centering
            \caption{Description of the statistical measures reported in Table 2 about the comparison between the empirical average mobility cost differences between married and non-married travellers, $\Delta C_M$, for the null model \NM{1}. For each urban area we report: the average value of the cost difference, $\langle  \Delta C_M \rangle$, and its standard deviation, $\sigma(\Delta C_M)$, for both empirical values and those obtained using the null model. Concerning the statistics, instead, we report the $t$-value, $p$-value, and the boundaries of the 95\% confidence interval (CI).}
            \label{tab:fullstats_dcm_nm1}
            \resizebox{0.99\textwidth}{!}{%
            \begin{tabular}{rrrrrrrrr}
                \toprule
                \multicolumn{1}{c}{\multirow{2}{*}{Urban area}} & \multicolumn{2}{c}{DATA} & \multicolumn{2}{c}{\NM{1}} & \multicolumn{4}{c}{STATISTICS} \\
                 & \multicolumn{1}{c}{$\langle  \Delta C_M \rangle$} & \multicolumn{1}{c}{$\sigma(\Delta C_M)$} & \multicolumn{1}{c}{$\langle \Delta C_M \rangle$} & \multicolumn{1}{c}{$\sigma(\Delta C_M )$} & \multicolumn{1}{c}{$p$-value} & \multicolumn{2}{c}{CI (95\%)} & \multicolumn{1}{c}{$t$-value} \\
                \midrule
                Baltimore (6) & -4.40e-03 & 5.58e-05 & 1.69e-05& 9.80e-05 & 0.0000 & -1.87e-04 & 1.93e-04 & -2727.84 \\
                Houston (6) & 1.52e-02 & 6.62e-05 & 2.64e-05 & 1.14e-04 & 0.0000 & -2.16e-04 & 2.34e-04 & 8150.16 \\
                Pittsburgh (6) & -4.20e-03 & 6.85e-05 &  7.24e-06 & 1.16e-04 & 0.0000 & -2.27e-04 & 2.30e-04 & -2190.75 \\
                Cincinnati (7) & 6.10e-03 & 5.03e-05 & -1.71e-05 & 8.97e-05 & 0.0000 & -1.77e-04 & 1.58e-04 & 4217.93 \\
                Indianapolis (7) & 2.20e-03 & 6.19e-05 & 2.45e-05 & 1.12e-04 & 0.0000 & -2.13e-04 & 2.25e-04 & 1196.75 \\
                Kansas City (7) & 1.04e-02 & 5.62e-05 & -6.96e-06 & 1.05e-04 & 0.0000 & -1.93e-04 & 2.07e-04 & 6173.75 \\
                Virginia Beach (7) & 1.05e-02 & 4.95e-05 & -1.21e-05 & 8.53e-05 & 0.0000 & -1.65e-04 & 1.63e-04 & 7551.80 \\
                Nashville (8) & -5.90e-03 & 5.97e-05 & -5.48e-06 & 9.81e-05 & 0.0000 & -1.87e-04 & 1.79e-04 & -3610.96 \\
                Charlotte (9) & -2.00e-03 & 5.24e-05 & -4.23e-06 & 9.39e-05 & 0.0000 & -1.83e-04 & 1.78e-04 & -1283.61 \\
                Dallas (9) & 3.50e-03 & 5.13e-05 & 2.11e-05 & 9.09e-05 & 0.0000 & -1.87e-04 & 1.70e-04 & 2364.09 \\
                St. Louis (9) & -1.10e-03 & 4.96e-05 & -7.57e-06 & 8.81e-05 & 0.0000 & -1.67e-04 & 1.76e-04 & -777.39 \\
                Minneapolis (10) & -1.50e-03 & 4.02e-05 & -1.63e-05 & 7.34e-05 & 0.0000 & -1.40e-04 & 1.41e-04 & -1260.66 \\
                Chicago (11) & 5.10e-03 & 1.23e-02 & -1.43e-05 & 7.07e-05 & 0.0000 & -1.41e-04 & 1.37e-04 & 60.77 \\
                Philadelphia (11) & 8.00e-03 & 3.97e-05 & -5.91e-06 & 6.99e-05 & 0.0000 & -1.33e-04 & 1.42e-04 & 7050.19 \\
                Washington (13) & 1.01e-02 & 4.66e-05 & 7.73e-06 & 7.87e-05 & 0.0000 & -1.55e-04 & 1.51e-04 & 7837.22 \\
                Atlanta (17) & -5.90e-03 & 5.09e-05 & -7.56e-06 & 8.65e-05 & 0.0000 & -1.72e-04 & 1.68e-04 & -4133.03 \\
                New York (23) & 6.40e-03 & 2.64e-05 & -1.81e-06 & 4.58e-05 & 0.0000 & -8.79e-05 & 9.15e-05 & 8551.55 \\
                \bottomrule
            \end{tabular}
            }
        \end{table}

    %
    %  TABLE STATISTICS DELTA C_M NM2
    %
    %
    \begin{table}[h]
        \centering
        \caption{Description of the statistical measures reported in Table 1 about the comparison between the empirical average mobility cost differences between married and non-married travellers, $\Delta C_M$, for the null model \NM{2}. The reader can look at the caption of Tab.~\ref{tab:fullstats_dcm_nm1} for the details on the meaning of each column.}
        \label{tab:fullstats_dcm_nm2}
        \resizebox{0.99\textwidth}{!}{%
        \begin{tabular}{rrrrrrrrr}
            \toprule
            \multicolumn{1}{c}{\multirow{2}{*}{Urban area}} & \multicolumn{2}{c}{DATA} & \multicolumn{2}{c}{\NM{2}} & \multicolumn{4}{c}{STATISTICS} \\
             & \multicolumn{1}{c}{$\langle  \Delta C_M \rangle$} & \multicolumn{1}{c}{$\sigma(\Delta C_M)$} & \multicolumn{1}{c}{$\langle \Delta C_M \rangle$} & \multicolumn{1}{c}{$\sigma(\Delta C_M )$} & \multicolumn{1}{c}{$p$-value} & \multicolumn{2}{c}{CI (95\%)} & \multicolumn{1}{c}{$t$-value} \\
            \midrule
            Baltimore (6) & -4.40e-03 & 5.62e-05 & -1.00e-05 & 9.80e-05 & 0.0000 & -1.92e-04 & 1.96e-04 & -2726.29 \\
            Houston (6) & 1.52e-02 & 6.51e-05 & -1.03e-05 & 1.13e-04 & 0.0000 & -2.20e-04 & 2.29e-04 & 8232.19 \\
            Pittsburgh (6) & -4.20e-03 & 6.93e-05 & 2.34e-05 & 1.20e-04 & 0.0000 & -2.29e-04 & 2.22e-04 & -2121.59 \\
            Cincinnati (7) & 6.10e-03 & 5.03e-05 & -5.54e-06 & 8.94e-05 & 0.0000 & -1.77e-04 & 1.68e-04 & 4228.89 \\
            Indianapolis (7) & 2.20e-03 & 6.15e-05 &  7.25e-06 & 1.13e-04 & 0.0000 & -2.14e-04 & 2.18e-04 & 1191.95 \\
            Kansas City (7) & 1.04e-02 & 5.58e-05 & -2.98e-05 & 9.94e-05 & 0.0000 & -1.99e-04 & 1.94e-04 & 6460.59 \\
            Virginia Beach (7) & 1.05e-02 & 5.02e-05 & 4.88e-06 & 8.74e-05 & 0.0000 & -1.80e-04 & 1.71e-04 & 7393.12 \\
            Nashville (8) & -5.90e-03 & 6.00e-05 & 2.21e-05 & 9.74e-05 & 0.0000 & -1.93e-04 & 1.95e-04 & -3631.11 \\
            Charlotte (9) & -2.00e-03 & 5.24e-05 & 6.01e-06 & 9.81e-05 & 0.0000 & -1.94e-04 & 1.83e-04 & -1242.33 \\
            Dallas (9) & 3.50e-03 & 5.16e-05 & -1.29e-05 & 9.21e-05 & 0.0000 & -1.84e-04 & 1.78e-04 & 2336.77 \\
            St. Louis (9) & -1.10e-03 & 4.98e-05 &  -2.30e-05 & 8.64e-05 & 0.0000 & -1.71e-04 & 1.66e-04 & -777.84 \\
            Minneapolis (10) & -1.50e-03 & 4.14e-05 &  9.37e-06 & 7.29e-05 & 0.0000 & -1.38e-04 & 1.42e-04 & -1257.61 \\
            Chicago (11) & 5.10e-03 & 1.22e-02 &  -5.45e-06 & 6.91e-05 & 0.0000 & -1.38e-04 & 1.32e-04 & 60.58 \\
            Philadelphia (11) & 8.00e-03 & 3.90e-05 & -1.50e-05 & 6.85e-05 & 0.0000 & -1.29e-04 & 1.32e-04 & 7190.86 \\
            Washington (13) & 1.01e-02 & 4.64e-05 & 1.36e-05 & 7.90e-05 & 0.0000 & -1.51e-04 & 1.61e-04 & 7823.03 \\
            Atlanta (17) & -5.90e-03 & 5.11e-05 & -1.53e-06 & 8.65e-05 & 0.0000 & -1.68e-04 & 1.73e-04 & -4131.14 \\
            New York (23) & 6.40e-03 & 2.63e-05 & -5.91e-07 & 4.55e-05 & 0.0000 & -9.10e-05 & 9.49e-05 & 8604.80 \\
            \bottomrule
        \end{tabular}
        }
    \end{table}

    %
    %  TABLE STATISTICS DELTA C_M NM3
    %
    %
    \begin{table}[h]
        \centering
        \caption{Description of the statistical measures reported in Table 1 about the comparison between the empirical average mobility cost differences between married and non-married travellers, $\Delta C_M$, for the null model \NM{3}. The reader can look at the caption of Tab.~\ref{tab:fullstats_dcm_nm1} for the details on the meaning of each column.}
        \label{tab:fullstats_dcm_nm3}
        \resizebox{0.99\textwidth}{!}{%
        \begin{tabular}{rrrrrrrrr}
            \toprule
            \multicolumn{1}{c}{\multirow{2}{*}{Urban area}} & \multicolumn{2}{c}{DATA} & \multicolumn{2}{c}{\NM{3}} & \multicolumn{4}{c}{STATISTICS} \\
             & \multicolumn{1}{c}{$\langle  \Delta C_M \rangle$} & \multicolumn{1}{c}{$\sigma(\Delta C_M)$} & \multicolumn{1}{c}{$\langle \Delta C_M \rangle$} & \multicolumn{1}{c}{$\sigma(\Delta C_M )$} & \multicolumn{1}{c}{$p$-value} & \multicolumn{2}{c}{CI (95\%)} & \multicolumn{1}{c}{$t$-value} \\
            \midrule
            Baltimore (6) & -4.40e-03 & 5.65e-05 & -4.56e-03 & 5.82e-03 & 0.4640 & -1.68e-02 & 5.52e-03 & 9.69 \\
            Houston (6) & 1.52e-02 & 6.55e-05 & 1.60e-03 & 6.58e-03 & 0.0194 & -1.16e-02 & 1.48e-02 & 143.58 \\
            Pittsburgh (6) & -4.20e-03 & 6.82e-05 & 1.94e-03 & 5.77e-03 & 0.1390 & 8.55e-03 & 1.39e-02 & -74.31 \\
            Cincinnati (7) & 6.10e-03 & 5.03e-05 & 6.75e-03 & 5.54e-03 & 0.4540 & -4.36e-03 & 1.75e-02 & -7.53 \\
            Indianapolis (7) & 2.20e-03 & 6.22e-05 & -1.86e-02 & 7.45e-03 & 0.0026 & -3.28e-02 & -3.51e-03 & 183.17 \\
            Kansas City (7) & 1.04e-02 & 5.56e-05 &  9.01e-04 & 5.09e-03 & 0.0322 & -9.10e-03 & 1.10e-02 & 130.43 \\
            Virginia Beach (7) & 1.05e-02 & 5.08e-05 & 7.72e-03 & 4.75e-03 & 0.2220 & -2.39e-03 & 1.67e-02 & 51.04 \\
            Nashville (8) & -5.90e-03 & 5.94e-05 & -2.02e-03 & 5.29e-03 & 0.2230 & -1.31e-02 & 7.54e-03 & -52.56 \\
            Charlotte (9) & -2.00e-03 & 5.21e-05 & -3.55e-03 & 5.66e-03 & 0.3720 & -1.44e-02 & 6.11e-03 & 24.30 \\
            Dallas (9) & 3.50e-03 & 5.14e-05 & 7.58e-03 & 4.29e-03 & 0.1620 & -1.20e-04 & 1.68e-02 & -70.11 \\
            St. Louis (9) & -1.10e-03 & 5.01e-05 & 2.02e-02 & 5.92e-03 & 0.0000 & 1.11e-02 & 3.46e-02 & -278.20 \\
            Minneapolis (10) & -1.50e-03 & 4.08e-05 & -7.75e-03 & 5.50e-03 & 0.1110 & -1.94e-02 & 2.09e-03 & 85.95 \\
            Chicago (11) & 5.10e-03 & 1.22e-02 & 1.21e-02 & 3.97e-03 & 0.4050 & 3.40e-03 & 1.90e-02 & -4.88 \\
            Philadelphia (11) & 8.00e-03 & 3.93e-05 & 2.96e-03 & 4.55e-03 & 0.0626 & -9.52e-03 & 9.94e-03 & 103.18 \\
            Washington (13) & 1.01e-02 & 4.59e-05 &  1.11e-02 & 4.58e-03 & 0.5000 & 1.14e-03 & 1.95e-02 & -0.38 \\
            Atlanta (17) & -5.90e-03 & 5.07e-05 & -5.04e-04 & 5.00e-03 & 0.0708 & -7.92e-03 & 1.13e-02 & -103.29 \\
            New York (23) & 6.40e-03 & 2.69e-05 &  8.09e-03 & 2.95e-03 & 0.2880 & 2.21e-03 & 1.35e-02 & -39.24 \\
            \bottomrule
        \end{tabular}
        }
    \end{table}

    %
    %  TABLE STATISTICS DELTA C_M NM4
    %
    %
    \begin{table}[h]
        \centering
        \caption{Description of the statistical measures reported in Table 1 about the comparison between the empirical average mobility cost differences between married and non-married travellers, $\Delta C_M$, for the null model \NM{4}. The reader can look at the caption of Tab.~\ref{tab:fullstats_dcm_nm1} for the details on the meaning of each column.}
        \label{tab:fullstats_dcm_nm4}
        \resizebox{0.99\textwidth}{!}{%
        \begin{tabular}{rrrrrrrrr}
            \toprule
            \multicolumn{1}{c}{\multirow{2}{*}{Urban area}} & \multicolumn{2}{c}{DATA} & \multicolumn{2}{c}{\NM{4}} & \multicolumn{4}{c}{STATISTICS} \\
             & \multicolumn{1}{c}{$\langle  \Delta C_M \rangle$} & \multicolumn{1}{c}{$\sigma(\Delta C_M)$} & \multicolumn{1}{c}{$\langle \Delta C_M \rangle$} & \multicolumn{1}{c}{$\sigma(\Delta C_M )$} & \multicolumn{1}{c}{$p$-value} & \multicolumn{2}{c}{CI (95\%)} & \multicolumn{1}{c}{$t$-value} \\
            \midrule
            Baltimore (6) & -4.40e-03 & 5.69e-05 & -3.17e-05 & 9.77e-05 & 0.0000 & -1.86e-04&1.86e-04 & -2724.41 \\
            Houston (6) & 1.52e-02 & 6.57e-05 & 2.34e-05 & 1.13e-04 & 0.0000 & -2.25e-04&2.31e-04 & 8208.94 \\
            Pittsburgh (6) & -4.20e-03 & 6.79e-05 & 1.81e-05 & 1.15e-04 & 0.0000 & -2.30e-04 & 2.27e-04 & -2202.22 \\
            Cincinnati (7) & 6.10e-03 & 5.08e-05 & 1.55e-05 & 8.82e-05 & 0.0000 & -1.71e-04 & 1.76e-04 & 4252.05 \\
            Indianapolis (7) & 2.20e-03 & 6.19e-05 & -2.16e-06 & 1.12e-04 & 0.0000 & -2.12e-04 & 2.18e-04 & 1203.56 \\
            Kansas City (7) & 1.04e-02 & 5.58e-05 & -9.98e-06 & 1.03e-04 & 0.0000 & -2.06e-04 & 2.04e-04 & 6288.07 \\
            Virginia Beach (7) & 1.05e-02 & 5.05e-05 & -2.97e-05 & 8.84e-05 & 0.0000 & -1.67e-04 & 1.67e-04 & 7320.96 \\
            Nashville (8) & -5.90e-03 & 5.99e-05 & 8.75e-06 & 9.94e-05 & 0.0000 & -2.07e-04 & 1.83e-04 & -3572.62 \\
            Charlotte (9) & -2.00e-03 & 5.27e-05 & 9.90e-06 & 9.05e-05 & 0.0000 & -1.74e-04 & 1.81e-04 & -1315.89 \\
            Dallas (9) & 3.50e-03 & 5.07e-05 & 1.00e-05 & 8.73e-05 & 0.0000 & -1.74e-04 & 1.67e-04 & 2445.60 \\
            St. Louis (9) & -1.10e-03 & 5.06e-05 & -5.37e-06 & 8.34e-05 & 0.0000 & -1.74e-04 & 1.62e-04 & -794.70 \\
            Minneapolis (10) & -1.50e-03 & 4.07e-05 &  2.39e-05 & 7.15e-05 & 0.0000 & -1.37e-04 & 1.45e-04 & -1280.95 \\
            Chicago (11) & 5.10e-03 & 1.23e-02 & 1.05e-05 & 7.36e-05 & 0.0000 & -1.40e-04 & 1.51e-04 & 60.49 \\
            Philadelphia (11) & 8.00e-03 & 3.83e-05 & 1.62e-05 & 6.90e-05 & 0.0000 & -1.26e-04 & 1.42e-04 & 7181.12 \\
            Washington (13) & 1.01e-02 & 4.69e-05 & -5.87e-06 & 7.79e-05 & 0.0000 & -1.59e-04 & 1.48e-04 & 7883.74 \\
            Atlanta (17) & -5.90e-03 & 5.04e-05 &  7.76e-06 & 8.50e-05 & 0.0000 & -1.70e-04 & 1.60e-04 & -4194.22 \\
            New York (23) & 6.40e-03 & 2.68e-05 & -5.42e-06 & 4.73e-05 & 0.0000 & -9.26e-05 & 9.31e-05 & 8318.47 \\
            \bottomrule
        \end{tabular}
        }
    \end{table}

    %
    %  TABLE STATISTICS DELTA C_M NM5
    %
    %
    \begin{table}[h]
        \centering
        \caption{Description of the statistical measures reported in Table 1 about the comparison between the empirical average mobility cost differences between married and non-married travellers, $\Delta C_M$, for the null model \NM{5}. The reader can look at the caption of Tab.~\ref{tab:fullstats_dcm_nm1} for the details on the meaning of each column.}
        \label{tab:fullstats_dcm_nm5}
        \resizebox{0.99\textwidth}{!}{%
        \begin{tabular}{rrrrrrrrr}
            \toprule
            \multicolumn{1}{c}{\multirow{2}{*}{Urban area}} & \multicolumn{2}{c}{DATA} & \multicolumn{2}{c}{\NM{5}} & \multicolumn{4}{c}{STATISTICS} \\
             & \multicolumn{1}{c}{$\langle  \Delta C_M \rangle$} & \multicolumn{1}{c}{$\sigma(\Delta C_M)$} & \multicolumn{1}{c}{$\langle \Delta C_M \rangle$} & \multicolumn{1}{c}{$\sigma(\Delta C_M )$} & \multicolumn{1}{c}{$p$-value} & \multicolumn{2}{c}{CI (95\%)} & \multicolumn{1}{c}{$t$-value} \\
            \midrule
            Baltimore (6) & -4.40e-03 & 5.66e-05 & -4.16e-03 & 5.22e-05 & 0.0000 & -4.27e-03 & -4.06e-03 & -171.17 \\
            Houston (6) & 1.52e-02 & 6.56e-05 & 1.55e-02 & 6.76e-05 & 0.0000 & 1.54e-02 & 1.56e-02 & -230.25 \\
            Pittsburgh (6) & -4.20e-03 & 6.83e-05 & -3.39e-03 & 7.86e-05 & 0.0000 & -3.53e-03 & -3.21e-03 & -548.01 \\
            Cincinnati (7) & 6.10e-03 & 5.04e-05 & 6.33e-03 & 5.33e-05 & 0.0000 & 6.25e-03 & 6.46e-03 & -214.33 \\
            Indianapolis (7) & 2.20e-03 & 6.17e-05 & 6.79e-03 & 6.56e-05 & 0.0000 & 6.66e-03 & 6.91e-03 & -3618.40 \\
            Kansas City (7) & 1.04e-02 & 5.60e-05 & 9.26e-03 & 5.71e-05 & 0.0000 & 9.11e-03 & 9.33e-03 & 1046.61 \\
            Virginia Beach (7) & 1.05e-02 & 5.07e-05 & 1.25e-02 & 4.38e-05 & 0.0000 & 1.25e-02 & 1.26e-02 & -2142.32 \\
            Nashville (8) & -5.90e-03 & 5.89e-05 & -7.51e-04 & 6.59e-05 & 0.0000 & -8.81e-04 & -6.24e-04 & -4093.98 \\
            Charlotte (9) & -2.00e-03 & 5.19e-05 & 8.38e-04 & 6.04e-05 & 0.0000 & 7.22e-04 & 9.53e-04 & -2475.83 \\
            Dallas (9) & 3.50e-03 & 5.16e-05 & 7.25e-03 & 6.46e-05 & 0.0000 & 7.10e-03 & 7.35e-03 & -3202.80 \\
            St. Louis (9) & -1.10e-03 & 5.02e-05 & 6.00e-04 & 5.32e-05 & 0.0000 & 4.91e-04 & 7.00e-04 & -1638.77 \\
            Minneapolis (10) & -1.50e-03 & 4.07e-05 & 5.38e-04 & 4.53e-05 & 0.0000 & 4.54e-04 & 6.32e-04 & -2355.80 \\
            Chicago (11) & 5.10e-03 & 1.24e-02 & 4.01e-03 & 5.54e-05 & 0.0000 & 3.88e-03 & 4.10e-03 & 37.64 \\
            Philadelphia (11) & 8.00e-03 & 3.92e-05 & 5.43e-03 & 4.81e-05 & 0.0000 & 5.32e-03 & 5.52e-03 & 2948.36 \\
            Washington (13) & 1.01e-02 & 4.66e-05 & 1.02e-02 & 5.19e-05 & 0.1750 & 1.01e-02 & 1.03e-02 & -49.34 \\
            Atlanta (17) & -5.90e-03 & 5.09e-05 & -6.78e-04 & 5.33e-05 & 0.0000 & -7.75e-04 & -5.66e-04 & -4980.27 \\
            New York (23) & 6.40e-03 & 2.71e-05 & 6.39e-03 & 2.77e-05 & 0.4690 & 6.34e-03 & 6.45e-03 & 3.54 \\
            \bottomrule
        \end{tabular}
        }
        
    \end{table}

\clearpage

\subsection{About confounding from socioeconomic covariates}
\label{sec:confounding}

An important consideration in our analysis is that household arrangements (parenthood and marital status) may correlate with other sociodemographic factors that independently influence mobility patterns. For instance, socioeconomic status is known to correlate with both family size \cite{fingerman2015ll} and residential location choices. Individuals with lower socioeconomic status tend to have more children and may face different mobility constraints due to factors such as vehicle ownership, access to public transportation, and residential options. Similarly, our analysis does not account for employment type, work schedule flexibility, or other factors that may interact with household arrangements.

Our null model framework provides a systematic approach to mitigate confounding from such correlated covariates, though it cannot eliminate these effects entirely. Each null model disrupts specific relationships whilst preserving others, allowing us to test whether observed differences persist under different assumptions about the underlying mechanisms:

\begin{description}
\item [\NM{1}] (shuffling household arrangement labels). If affluent zones have fewer parents and better amenity access, and if this spatial sorting alone explains the observed patterns, then randomly reassigning household labels would eliminate the differences. The fact that differences remain statistically significant under \NM{1} suggests that the relationship between household arrangements and mobility patterns is not solely driven by spatial sorting correlated with unmeasured socioeconomic factors.

\item [\NM{2} / \NM{4}] (randomising travel times from fitted distributions) and (shuffling travel times), respectively. These null models help assess whether household-specific time constraints (which may correlate with employment type or income) drive the observed patterns.

\item [\NM{3}] (randomising travel distances from fitted distributions). This is particularly relevant for addressing socioeconomic confounding, as lower-income individuals may live farther from employment centres due to housing affordability constraints. If the observed parenthood effect were entirely driven by income-related residential sorting, \NM{3} would likely eliminate the differences.

\item [\NM{5}] (shuffling destination zones) This disrupts potential correlations between household arrangements, socioeconomic status, and workplace locations.
\end{description}

The robustness of our findings across these multiple null models, where each testing different potential confounding mechanisms, strengthens our confidence that the observed patterns reflect genuine differences in how household arrangements experience urban mobility. However, we acknowledge important limitations. Our null models cannot completely disentangle all correlated factors, particularly those that are spatially clustered. For example, if affluent neighbourhoods have both fewer parents and systematically better infrastructure, some portion of the observed `parenthood effect' may reflect underlying socioeconomic gradients. A complete disentanglement would require individual-level data linking household arrangements, income, employment characteristics, and mobility patterns and to our knowledge this data is currently unavailable at the spatial resolution needed for our analysis.

Despite these limitations, our approach represents a substantial improvement over analyses that do not account for potential confounding at all. The consistency of our findings across multiple null models, combined with the statistical significance observed in most metropolitan areas, provides reasonable confidence that household arrangements genuinely influence urban mobility experiences, even if the magnitude of effects may be partially mediated by correlated socioeconomic factors.

        \section{Quantifying spatial diversity}
        \label{sec:spatial_diversity}

        It is important to start by clarifying that our use of `diversity' refers to the spatial evenness of distributions across zones, measured via Shannon entropy, rather than individual-level mobility diversity (\ie{}, the variety of trips made by individuals). In spatial diversity, $H$, (high spatial entropy~\cite{wang2018spatial}), high values indicate homogeneous distributions (\eg{}, amenities), whereas in individual mobility diversity, $M$, high values indicate varied trip patterns. Thus, we note that this term and concept may be referred to differently across fields and research papers~\cite{Pappalardo2015, chen2024achieving}.

        Given a metropolitan area divided into $N_Z$ zones, we compute the \emph{diversity} $H^{X}$ of the spatial coverage of a given amenity $X$ (e.g., hospitals, restaurants) across the area~\cite{cocchi2014spatial,batty2010space}. This diversity measure is based on the Shannon entropy of the spatial distribution, normalized by the maximum possible entropy, yielding:
        \begin{equation}
        \label{eq:entropy_space}
        H^{X} = - \frac{1}{\log_2 N_{Z}} \sum^{N_Z}_{i=1} p^{X}_{i} \; \log_2 \frac{A^{X}_{i}}{p^{X}_{i}} \,.
        \end{equation}
        where $p^{X}_{i}$ represents the probability of observing type $X$ in zone $i$, defined as:
        \begin{equation}
        \label{eq:prob_spatial_feature}
        p^{X}_{i} = \frac{n^{X}_{i}}{N^{X}} \,,
        \end{equation}

        Here, $n^{X}_{i}$ denotes the number of entities (\ie, amenities or travellers) of type $X$ in zone $i$, and $N^{X} = \sum_{i}^{N_z} n^{X}_{i}$ represents the total number of such entities across the entire metropolitan area.

        The diversity measure $H^{X}$ ranges from 0 to 1. A value of $H^{X} = 0$ indicates complete spatial concentration, where feature $X$ is present in only a single zone. Conversely, $H^{X} = 1$ indicates perfect spatial homogeneity, where feature $X$ is evenly distributed across all zones. Intermediate values ($0 < H^{X} < 1$) reflect non-homogeneous spatial distributions, with lower values signifying greater concentration and higher values indicating greater dispersion. The normalization by $\log_2 N_{Z}$ ensures that diversity values remain comparable across metropolitan areas with different numbers of zones, as discussed in the main manuscript.

\begin{figure}[h!]
            \centering
            \includegraphics[width=0.55\linewidth]{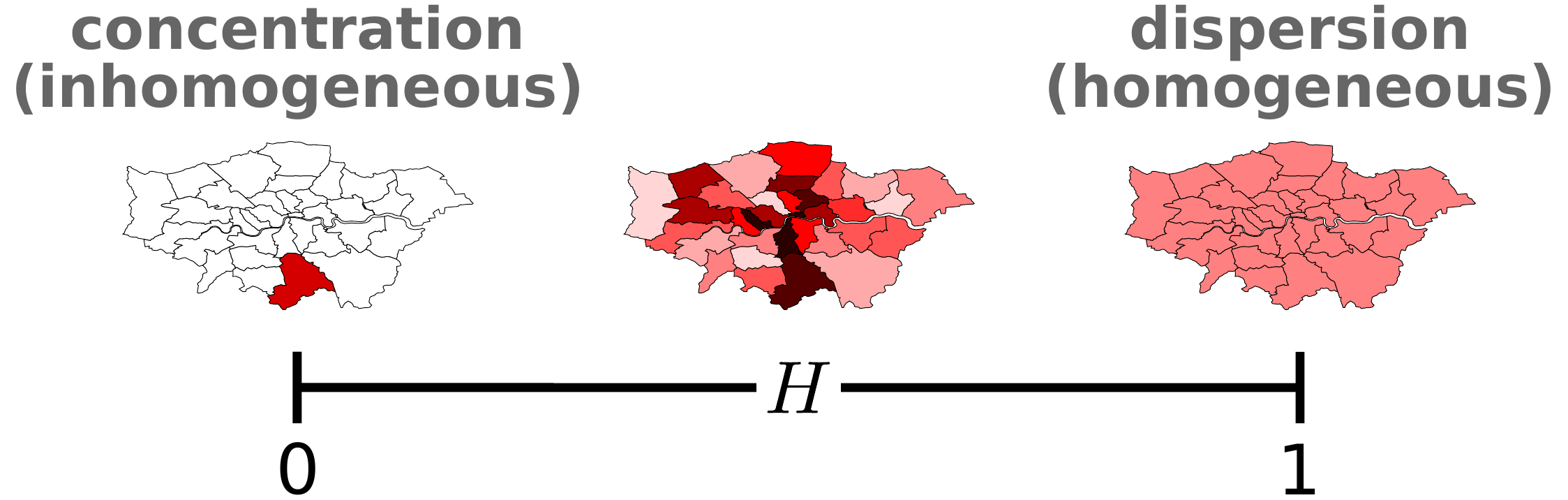}
            \caption{Schematic representation of the relationship between diversity, $H$, and spatial dispersion. Given a metropolitan area divided into zones, if the quantity under scrutiny is concentrated into a single zone, then $H = 0$. Conversely, if the quantity under scrutiny is dispersed uniformly, then $H = 1$. Non-homogeneous spatial distributions correspond to intermediate values of $H$ (\ie, $0 < H < 1$).}
            \label{fig:entropy_meaning}
        \end{figure}

        In \figurename~\ref{fig:entropy_meaning}, we display a simplified view of the mapping existing between the values of \emph{diversity} (\ie, entropy), $H$, and the spatial concentration of some quantity. From such a diagram, one can see that extreme values of $H$ correspond to concentrated ($H = 0$) and homogeneously spread ($H = 1$) configurations, whereas intermediate values of $H$ correspond to less homogeneous configurations. It is also important to note that we account for the area size of each location in this analysis, capturing how evenly the quantity is distributed relative to the area~\cite{battyjourgeosys2014}.

        %
        %  FIGURE ENTROPY MEANING
        %
        %
        %
        % 

        % where $n^{X}_{i}$ is the number of entities (\ie, amenities or travellers) of type $X$ in zone $i$, and $N^{X} = \sum_{i}^{N_z} n^{X}_{i}$ is the total number of such entities in the whole metropolitan area. Diversity $H^{X} \in [0,1]$, where $H^{X} = 0$ corresponds to the case in which feature $X$ is concentrated in a single zone, whereas $H^{X} = 1$ denotes the case in which feature $X$ is homogeneously distributed across all zones.

        Thus, in our analysis, we apply this diversity measure to multiple types of features. For amenities, we calculate $H^{X}$ for different amenity categories (education, food, health, leisure, religious, residential, services, transport, and work). For travellers, we calculate $H^{X}$ for different household arrangements (parent, non-parent, married, and non-married). The results for amenity distributions are presented in \figurename~\ref{fig:spatial_characterisation}A of the main manuscript, whilst traveller distributions are shown in \figurename~\ref{fig:spatial_characterisation}C. We observe that amenity diversity values span approximately between 0.1 and 0.6 across metropolitan areas, whereas traveller diversity values range from 0.5 to 1.0, indicating that travellers are more homogeneously distributed than amenities.

        We apply the same formalism of Eq.~\eqref{eq:entropy_space} to compute the \emph{mobility diversity} (\ie, \emph{diversity of accessibility to amenities}) by travellers of type $Y$, $M^{Y}$, within the urban area $A^{Y}_{i}$ as:
        \begin{equation}
        \label{eq:entropy_mobility}
        M^{Y} = - \frac{1}{\log_2 N_{Z}} \sum^{N_Z}_{i=1} p^{Y}_{i} \; \log_2 \frac{A^{Y}_{i}}{p^{Y}_{i}} \,.
        %
        %
        % H^{X} = - \frac{1}{\log_2 N_{Z}} \sum^{N_Z}_{i=1} p^{X}_{i} \; \log_2 \frac{A^{X}_{i}}{p^{X}_{i}} \,.
        %
        \end{equation}

        where the probability $p^{Y}_{i}$ represents the ratio between the product of the number of amenities in zone $i$, $n_{i}$, and the number of travellers of type $Y$ whose destination zone is $i$, $T^{Y}_{i}$ (\ie, $n^{Y}_{i} = n_{i} \; T^{Y}_{i}$), and its sum over all zones $N^{Y} = \sum_{i}^{N_z} n^{Y}_{i}$. This formulation captures how travellers of a given household arrangement access amenities across the metropolitan area, with higher values of $M^{Y}$ indicating that travellers visit destinations more evenly distributed across zones. 

        Eventually, one could also calculate the mobility diversity of accessibility of travellers of type $Y$ to amenities of type $X$, $M^{XY}$, by restricting the amenities considered to only those of category $X$. This allows for analysis of how specific household types access specific amenity categories, though in the main manuscript we focus primarily on aggregate accessibility to all amenities.

        It is worth noting that the values of $M$ and $H$ appearing in the above equations and throughout our analysis correspond to those of the average value calculated from the bootstrap sampling procedure. We bootstrap 80\% of the data over 5,000 realisations to obtain robust estimates of these diversity measures and their distributions. This approach accounts for sampling variability and provides confidence in the observed patterns across metropolitan areas and household types.

        %
        %  SECTION ON MOBILITY 
        %
        %
        \section{Mobility}
	    \label{sec:mobility}

        \figurename~\ref{fig:mobility_characterisation} of the main manuscript presents the relationship between the average mobility cost, $C$, and mobility diversity, $M$, for all travellers and across household arrangements. Tables~\ref{tab:values_c} and \ref{tab:values_m} contain the values displayed in \figurename~\ref{fig:mobility_characterisation}. In Table~\ref{tab:values_h}, we also provide the values of the spatial characterisation $H$ for comparison. We can notice that $C$ varies between 0.12 and 0.21, $M$ varies between 0.47 and 0.95, and $H$ varies between 0.54 and 0.95.

\begin{figure}[ht]
            \centering
            \includegraphics[width=\linewidth]{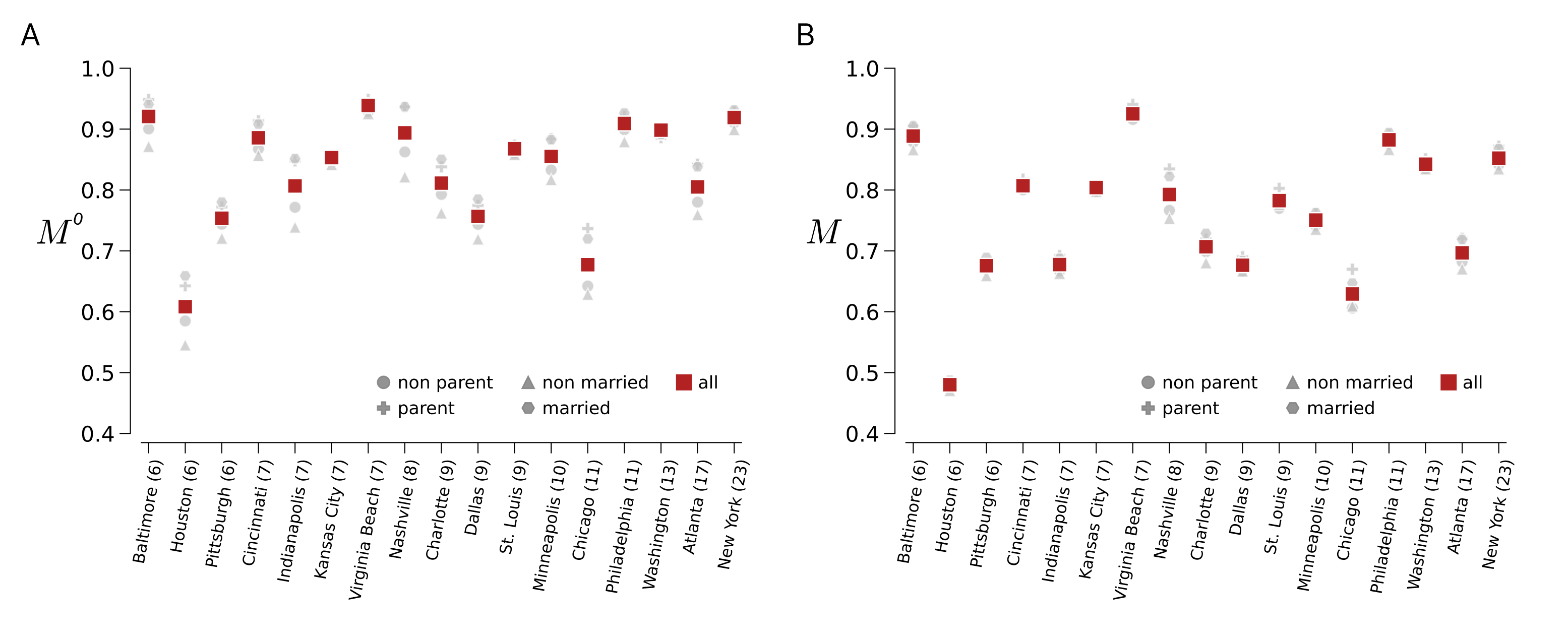}
            \caption{Characterisation of the mobility diversity for all travellers. Mobility diversity, $M$, measures how evenly travellers access amenities across destination zones (typically workplaces) based on their commuting patterns. For each urban area, we display: the mobility diversity in relation to the home location $M^0$ (panel A) and to the workplace $M$ (panel B) for travellers of different household arrangements.}
            \label{fig:M0results}
        \end{figure}

        We further examine how the values of $M$ relate to travellers’ places of residence, highlighting whether mobility diversity increases or decreases with movement. \figurename~\ref{fig:M0results} presents the mobility diversity values associated with home locations ($M^0$, panel~A) and work destinations ($M$, panel~B), while Table~\ref{tab:m0_minus_m} details their differences. Across all areas, mobility diversity decreases from home to work locations, indicating that travellers tend to converge spatially at their destinations compared to their origins—consistent with previous findings in the literature~\cite{macedo2022differences}. When we focus our attention to the differences between parents and non-parents and married and non-married travellers, we see that only in Washington the mobility diversity of non-parents and non-married are higher than the ones for parents and married individuals. This is in line with the spatial charaterisation of amenities displayed in Table~\ref{tab:values_h}.

        %
        %  FIG:  MOBILITY DIVERSITY RESIDENTIAL VS WORK ZONES
        %
        %

        %
        %  TABLE VALUES OF MOBILITY COST FOR ALL AREAS
        %
        %
        \begin{table}[ht]
            \centering
            \caption{Values of the average mobility cost, $C$, for each metropolitan area considered in our study (see \figurename~\ref{fig:mobility_characterisation} of the main manuscript). Columns labelled $C_{\text{All}}$, $C_{\text{P}}$, $C_{\text{NP}}$, $C_{\text{M}}$, and $C_{\text{NM}}$ represent the average mobility cost of all travellers (regardless of their type), parents, non-parents, married, and non-married travellers, respectively. The column named $\Delta{C_{\text{P}}}$ represents the difference between $C_{\text{P}}$ and $C_{\text{NP}}$, whereas the column named $\Delta{C_{\text{M}}}$ represents the difference between $C_{\text{M}}$ and $C_{\text{NM}}$.}
            % \accom{I propose to replace $\Delta_{C_{\text{P}}}$ with ${\Delta C}^{\text{P}}$ (a similar reasoning applies also to $\Delta_{C_{\text{M}}}$}
            \label{tab:values_c}
            \resizebox{0.6\textwidth}{!}{%
            \begin{tabular}{r|rrrrrrr}
                \toprule
                \multicolumn{1}{c|}{Metropolitan area} & \multicolumn{1}{c}{$C_{\text{All}}$} & \multicolumn{1}{c}{$C_{\text{P}}$} & \multicolumn{1}{c}{$C_{\text{NP}}$} & \multicolumn{1}{c}{$\Delta{C_{\text{P}}}$} &  \multicolumn{1}{c}{$C_{\text{M}}$} & \multicolumn{1}{c}{$C_{\text{NM}}$} & \multicolumn{1}{c}{$\Delta{C_{\text{M}}}$} \\
                \midrule
                Baltimore (6) & 0.1714 & 0.1677 & 0.1738 & -0.0061 & 0.1697 & 0.1740 & -0.0044 \\
                Houston (6) & 0.1607 & 0.1667 & 0.1570 & 0.0097 & 0.1684 & 0.1532 & 0.0152 \\
                Pittsburgh (6) & 0.1506 & 0.1566 & 0.1481 & 0.0085 & 0.1486 & 0.1528 & -0.0042 \\
                Cincinnati (7) & 0.1360 & 0.1317 & 0.1386 & -0.0070 & 0.1392 & 0.1331 & 0.0061 \\
                Indianapolis (7) & 0.1318 & 0.1344 & 0.1303 & 0.0041 & 0.1330 & 0.1308 & 0.0022 \\
                Kansas City (7) & 0.1307 & 0.1339 & 0.1291 & 0.0048 & 0.1352 & 0.1248 & 0.0104 \\
                Virginia Beach (7) & 0.1464 & 0.1503 & 0.1443 & 0.0060 & 0.1516 & 0.1411 & 0.0105 \\
                Nashville (8) & 0.1428 & 0.1412 & 0.1440 & -0.0028 & 0.1410 & 0.1469 & -0.0059 \\
                Charlotte (9) & 0.1379 & 0.1378 & 0.1380 & -0.0002 & 0.1372 & 0.1392 & -0.0020 \\
                Dallas (9) & 0.1449 & 0.1464 & 0.1441 & 0.0023 & 0.1462 & 0.1427 & 0.0035 \\
                St. Louis (9) & 0.1317 & 0.1318 & 0.1317 & 0.0001 & 0.1313 & 0.1324 & -0.0011 \\
                Minneapolis (10) & 0.1303 & 0.1276 & 0.1320 & -0.0044 & 0.1297 & 0.1312 & -0.0015 \\
                Chicago (11) & 0.1567 & 0.1526 & 0.1592 & -0.0066 & 0.1653 & 0.1548 & 0.0105 \\
                Philadelphia (11) & 0.1693 & 0.1689 & 0.1709 & -0.0020 & 0.1733 & 0.1653 & 0.0080 \\
                Washington (13) & 0.1947 & 0.2036 & 0.1904 & 0.0132 & 0.1997 & 0.1895 & 0.0101 \\
                Atlanta (17) & 0.1612 & 0.1621 & 0.1615 & 0.0006 & 0.1596 & 0.1655 & -0.0059 \\
                New York (23) & 0.2064 & 0.2070 & 0.2064 & 0.0006 & 0.2097 & 0.2033 & 0.0064 \\
                \bottomrule
            \end{tabular}
            } % end of resizebox
        \end{table}

        %
        %  TABLE VALUES OF MOBILITY diversity FOR ALL AREAS
        %
        %
        \begin{table}[ht]
            \centering
            \caption{Values of the mobility diversity, $M$, for each metropolitan area considered in our study (see \figurename~\ref{fig:mobility_characterisation} of the main manuscript). Columns labelled $M_{\text{All}}$, $M_{\text{P}}$, $M_{\text{NP}}$, $M_{\text{M}}$, and $M_{\text{NM}}$ represents the mobility diversity of amenities accessible to all travellers (regardless of their type), parents, non-parents, married, and non-married travellers, respectively. The column named $\Delta{M_{\text{P}}}$ represents the difference between $M_{\text{P}}$ and $M_{\text{NP}}$, whereas column named $\Delta{M_{\text{M}}}$ represents the difference between $M_{\text{M}}$ and $M_{\text{NM}}$, respectively.}
            \label{tab:values_m}
            \resizebox{0.6\textwidth}{!}{%
            \begin{tabular}{r|rrrrrrr}
                \toprule
                \multicolumn{1}{c|}{Metropolitan area} & \multicolumn{1}{c}{$M_{\text{All}}$} & \multicolumn{1}{c}{$M_{\text{P}}$} & \multicolumn{1}{c}{$M_{\text{NP}}$} & \multicolumn{1}{c}{$\Delta{M_{\text{P}}}$} &  \multicolumn{1}{c}{$M_{\text{M}}$} & \multicolumn{1}{c}{$M_{\text{NM}}$} & \multicolumn{1}{c}{$\Delta{M_{\text{M}}}$} \\
                \midrule
                Baltimore (6) & 0.8886 & 0.9053 & 0.8786 & 0.0267 & 0.9051 & 0.8658 & 0.0392 \\
                Houston (6) & 0.4801 & 0.4844 & 0.4770 & 0.0073 & 0.4876 & 0.4703 & 0.0172 \\
                Pittsburgh (6) & 0.6755 & 0.6752 & 0.6745 & 0.0007 & 0.6892 & 0.6591 & 0.0301 \\
                Cincinnati (7) & 0.8070 & 0.8180 & 0.8002 & 0.0178 & 0.8077 & 0.8055 & 0.0022 \\
                Indianapolis (7) & 0.6774 & 0.6921 & 0.6668 & 0.0253 & 0.6878 & 0.6627 & 0.0250 \\
                Kansas City (7) & 0.8040 & 0.7971 & 0.8068 & -0.0097 & 0.8080 & 0.7975 & 0.0105 \\
                Virginia Beach (7) & 0.9251 & 0.9405 & 0.9159 & 0.0246 & 0.9254 & 0.9242 & 0.0012 \\
                Nashville (8) & 0.7926 & 0.8347 & 0.7665 & 0.0682 & 0.8221 & 0.7534 & 0.0687 \\
                Charlotte (9) & 0.7068 & 0.7195 & 0.6985 & 0.0209 & 0.7288 & 0.6803 & 0.0486 \\
                Dallas (9) & 0.6764 & 0.6904 & 0.6668 & 0.0236 & 0.6840 & 0.6668 & 0.0172 \\
                St. Louis (9) & 0.7825 & 0.8027 & 0.7697 & 0.0329 & 0.7869 & 0.7761 & 0.0108 \\
                Minneapolis (10) & 0.7506 & 0.7615 & 0.7414 & 0.0201 & 0.7626 & 0.7354 & 0.0272 \\
                Chicago (11) & 0.6291 & 0.6698 & 0.6060 & 0.0638 & 0.6471 & 0.6093 & 0.0378 \\
                Philadelphia (11) & 0.8824 & 0.8904 & 0.8759 & 0.0145 & 0.8942 & 0.8665 & 0.0278 \\
                Washington (13) & 0.8424 & 0.8513 & 0.8350 & 0.0162 & 0.8467 & 0.8347 & 0.0120 \\
                Atlanta (17) & 0.6967 & 0.7205 & 0.6820 & 0.0385 & 0.7187 & 0.6701 & 0.0486 \\
                New York (23) & 0.8523 & 0.8718 & 0.8416 & 0.0302 & 0.8677 & 0.8344 & 0.0333 \\
            \bottomrule
        \end{tabular}
        } % end of resizebox
    \end{table}

        %
        %  TABLE VALUES OF SPATIAL DIVERSITY FOR ALL AREAS
        %
        %
        \begin{table}[ht]
            \centering
            \caption{Values of the average spatial characterisation, $H$, for each metropolitan area considered in our study (see \figurename~\ref{fig:spatial_characterisation} of the main manuscript). Columns labelled $H_{\text{All}}$, $H_{\text{P}}$, $H_{\text{NP}}$, $H_{\text{M}}$, and $H_{\text{NM}}$ represent the average mobility cost of all travellers (regardless of their type), parents, non-parents, married, and non-married travellers, respectively. The column named $\Delta{H_{\text{P}}}$ represents the difference between $H_{\text{P}}$ and $H_{\text{NP}}$, whereas the column named $\Delta{H_{\text{M}}}$ represents the difference between $H_{\text{M}}$ and $H_{\text{NM}}$.}
            \label{tab:values_h}
            \resizebox{0.6\textwidth}{!}{%
            \begin{tabular}{r|rrrrrrr}
                \toprule
                \multicolumn{1}{c|}{Metropolitan area} & \multicolumn{1}{c}{$H_{\text{All}}$} & \multicolumn{1}{c}{$H_{\text{P}}$} & \multicolumn{1}{c}{$H_{\text{NP}}$} & \multicolumn{1}{c}{$\Delta{H_{\text{P}}}$} &  \multicolumn{1}{c}{$H_{\text{M}}$} & \multicolumn{1}{c}{$H_{\text{NM}}$} & \multicolumn{1}{c}{$\Delta{H_{\text{M}}}$} \\
                \midrule
                Baltimore (6) & 0.9208 & 0.9486 & 0.9004 & 0.0482 & 0.9412 & 0.8715 & 0.0697 \\
                Houston (6) & 0.6082 & 0.6415 & 0.5849 & 0.0566 & 0.6581 & 0.5453 & 0.1129 \\
                Pittsburgh (6) & 0.7541 & 0.7748 & 0.7443 & 0.0305 & 0.7818 & 0.7210 & 0.0608 \\
                Cincinnati (7) & 0.8855 & 0.9132 & 0.8674 & 0.0458 & 0.9075 & 0.8565 & 0.0510 \\
                Indianapolis (7) & 0.8069 & 0.8489 & 0.7718 & 0.0771 & 0.8519 & 0.7391 & 0.1129 \\
                Kansas City (7) & 0.8531 & 0.8577 & 0.8489 & 0.0088 & 0.8589 & 0.8419 & 0.0169 \\
                Virginia Beach (7) & 0.9389 & 0.9483 & 0.9287 & 0.0196 & 0.9344 & 0.9251 & 0.0093 \\
                Nashville (8) & 0.8941 & 0.9369 & 0.8628 & 0.0741 & 0.9367 & 0.8216 & 0.1151 \\
                Charlotte (9) & 0.8112 & 0.8383 & 0.7928 & 0.0455 & 0.8508 & 0.7621 & 0.0888 \\
                Dallas (9) & 0.7567 & 0.7752 & 0.7437 & 0.0315 & 0.7854 & 0.7194 & 0.0660 \\
                St. Louis (9) & 0.8681 & 0.8740 & 0.8612 & 0.0127 & 0.8683 & 0.8590 & 0.0093 \\
                Minneapolis (10) & 0.8552 & 0.8871 & 0.8329 & 0.0542 & 0.8829 & 0.8171 & 0.0658 \\
                Chicago (11) & 0.6774 & 0.7365 & 0.6424 & 0.0941 & 0.7197 & 0.6287 & 0.0910 \\
                Philadelphia (11) & 0.9092 & 0.9255 & 0.8991 & 0.0263 & 0.9267 & 0.8791 & 0.0477 \\
                Washington (13) & 0.8983 & 0.8855 & 0.8962 & -0.0106 & 0.8909 & 0.8915 & -0.0007 \\
                Atlanta (17) & 0.8052 & 0.8423 & 0.7803 & 0.0620 & 0.8381 & 0.7598 & 0.0783 \\
                New York (23) & 0.9191 & 0.9300 & 0.9104 & 0.0196 & 0.9313 & 0.8991 & 0.0322 \\
                \bottomrule
            \end{tabular}
            } % end of resizebox
        \end{table}

    %
    %  TABLE: MOBILITY DIVERSITY FOR HOME LOCATIONS
    %
    %
    \begin{table}[ht]
            \centering
            \caption{Values of the mobility diversity in relation to the home location, $M^0$, for each metropolitan area considered in our study (see \figurename~\ref{fig:M0results}). Columns labelled $M^0_{\text{All}}$, $M^0_{\text{P}}$, $M^0_{\text{NP}}$, $M^0_{\text{M}}$, and $M^0_{\text{NM}}$ represents the mobility diversity of amenities accessible to all travellers (regardless of their type), parents, non-parents, married, and non-married travellers, respectively. The column named $\Delta{M^0_{\text{P}}}$ represents the difference between $M^0_{\text{P}}$ and $M^0_{\text{NP}}$, whereas column named $\Delta{M^0_{\text{M}}}$ represents the difference between $M^0_{\text{M}}$ and $M^0_{\text{NM}}$, respectively.}
            \label{tab:m0}
            \resizebox{0.6\textwidth}{!}{
            \begin{tabular}{r|rrrrrrr}
            \toprule
             \multicolumn{1}{c|}{Metropolitan area} & \multicolumn{1}{c}{$M^0_{\text{All}}$} & \multicolumn{1}{c}{$M^0_{\text{P}}$} & \multicolumn{1}{c}{$M^0_{\text{NP}}$} & \multicolumn{1}{c}{$\Delta{M^0_{\text{P}}}$} &  \multicolumn{1}{c}{$M^0_{\text{M}}$} & \multicolumn{1}{c}{$M^0_{\text{NM}}$} & \multicolumn{1}{c}{$\Delta{M^0_{\text{M}}}$} \\
            \midrule
            Baltimore (6) & 0.9208 & 0.9485 & 0.9004 & 0.0480 & 0.9407 & 0.8715 & 0.0692 \\
            Houston (6) & 0.6083 & 0.6425 & 0.5849 & 0.0576 & 0.6590 & 0.5452 & 0.1138 \\
            Pittsburgh (6) & 0.7535 & 0.7720 & 0.7437 & 0.0283 & 0.7801 & 0.7205 & 0.0596 \\
            Cincinnati (7) & 0.8858 & 0.9137 & 0.8677 & 0.0460 & 0.9082 & 0.8569 & 0.0513 \\
            Indianapolis (7) & 0.8067 & 0.8476 & 0.7716 & 0.0761 & 0.8515 & 0.7389 & 0.1125 \\
            Kansas City (7) & 0.8531 & 0.8568 & 0.8489 & 0.0079 & 0.8579 & 0.8419 & 0.0160 \\
            Virginia Beach (7) & 0.9389 & 0.9486 & 0.9287 & 0.0199 & 0.9343 & 0.9252 & 0.0091 \\
            Nashville (8) & 0.8939 & 0.9365 & 0.8625 & 0.0740 & 0.9365 & 0.8214 & 0.1151 \\
            Charlotte (9) & 0.8112 & 0.8381 & 0.7928 & 0.0453 & 0.8504 & 0.7621 & 0.0883 \\
            Dallas (9) & 0.7565 & 0.7746 & 0.7436 & 0.0310 & 0.7849 & 0.7193 & 0.0656 \\
            St. Louis (9) & 0.8677 & 0.8736 & 0.8606 & 0.0129 & 0.8678 & 0.8585 & 0.0093 \\
            Minneapolis (10) & 0.8552 & 0.8842 & 0.8329 & 0.0513 & 0.8825 & 0.8171 & 0.0654 \\
            Chicago (11) & 0.6772 & 0.7366 & 0.6422 & 0.0944 & 0.7195 & 0.6286 & 0.0910 \\
            Philadelphia (11) & 0.9092 & 0.9212 & 0.8991 & 0.0221 & 0.9264 & 0.8791 & 0.0474 \\
            Washington (13) & 0.8983 & 0.8858 & 0.8962 & -0.0104 & 0.8911 & 0.8915 & -0.0004 \\
            Atlanta (17) & 0.8051 & 0.8423 & 0.7800 & 0.0622 & 0.8380 & 0.7595 & 0.0785 \\
            New York (23) & 0.9191 & 0.9299 & 0.9104 & 0.0195 & 0.9313 & 0.8991 & 0.0322 \\
            \bottomrule
            \end{tabular}
            }
            
        \end{table}

        %
        %  TABLE DIFFERENCES BETWEEN THE MOBILITY DIVERSITY OF HOME AND WORK ZONE FOR 
        %  DIFFERENT HOUSEHOLD ARRANGEMENTS
        %
        %
        \begin{table}[ht]
            \centering
            \caption{Differences in mobility diversity between home locations ($M^0$) and workplaces ($M$) for all travellers ($All$), as well as for parents ($P$), non-parents ($NP$), married ($M$), and non-married ($NM$) individuals. We also illustrate the differences in the differences due to parenthood and marriage effects, comparing those derived from home locations ($\Delta M^0_{P|M}$) with those derived from workplaces ($\Delta M_{P|M}$).}
            \resizebox{0.9\textwidth}{!}{%
            \begin{tabular}{rrrrrrrr}
            \toprule
             Metropolitan Area & $M^0_{All} - M_{All}$ & $M^0_P - M_P$ & $M^0_{NP} - M_{NP}$ & $\Delta M^0_P - \Delta M_{P}$ & $M^0_M - M_M$ & $M^0_{NM} - M_{NM}$ & $\Delta M^0_M - \Delta M_{M}$ \\
            \midrule
            Atlanta (17) & 0.1083 & 0.1218 & 0.0980 & 0.0238 & 0.1194 & 0.0894 & 0.0299 \\
            Baltimore (6) & 0.0322 & 0.0432 & 0.0218 & 0.0214 & 0.0356 & 0.0056 & 0.0300 \\
            Charlotte (9) & 0.1044 & 0.1186 & 0.0943 & 0.0244 & 0.1216 & 0.0818 & 0.0398 \\
            Chicago (11) & 0.0481 & 0.0668 & 0.0362 & 0.0307 & 0.0724 & 0.0193 & 0.0531 \\
            Cincinnati (7) & 0.0787 & 0.0956 & 0.0674 & 0.0282 & 0.1004 & 0.0514 & 0.0491 \\
            Dallas (9) & 0.0802 & 0.0842 & 0.0768 & 0.0074 & 0.1009 & 0.0525 & 0.0484 \\
            Houston (6) & 0.1281 & 0.1581 & 0.1078 & 0.0503 & 0.1714 & 0.0749 & 0.0965 \\
            Indianapolis (7) & 0.1293 & 0.1555 & 0.1048 & 0.0507 & 0.1637 & 0.0762 & 0.0875 \\
            Kansas City (7) & 0.0492 & 0.0598 & 0.0421 & 0.0176 & 0.0499 & 0.0444 & 0.0055 \\
            Minneapolis (10) & 0.1046 & 0.1228 & 0.0915 & 0.0313 & 0.1199 & 0.0817 & 0.0382 \\
            Nashville (8) & 0.1013 & 0.1019 & 0.0960 & 0.0059 & 0.1144 & 0.0680 & 0.0464 \\
            New York (23) & 0.0667 & 0.0581 & 0.0687 & -0.0107 & 0.0636 & 0.0646 & -0.0011 \\
            Philadelphia (11) & 0.0269 & 0.0308 & 0.0232 & 0.0076 & 0.0322 & 0.0126 & 0.0196 \\
            Pittsburgh (6) & 0.0780 & 0.0969 & 0.0692 & 0.0276 & 0.0909 & 0.0614 & 0.0295 \\
            St. Louis (9) & 0.0852 & 0.0709 & 0.0909 & -0.0200 & 0.0809 & 0.0824 & -0.0014 \\
            Virginia Beach (7) & 0.0138 & 0.0081 & 0.0128 & -0.0047 & 0.0089 & 0.0010 & 0.0079 \\
            Washington (13) & 0.0559 & 0.0345 & 0.0611 & -0.0266 & 0.0444 & 0.0569 & -0.0124 \\
            \bottomrule
            \end{tabular}
            }
            \label{tab:m0_minus_m}
        \end{table}

    To confirm that the distributions of $M$ are statistically different, we also calculate the differences and apply the two-sample Kolmogorov--Smirnov test in Table~\ref{tab:mobility_stats}. For all the metropolitan areas, the distributions are statistically different with $p$-value less than 0.001 when considering the parenthood and cohabitation statuses. We also replicate the same analysis for mobility cost in Table~\ref{tab:cost_stats}, where all $p$-values are below 0.001, confirming that our results are statistically significant. We can also replicate the same analysis for the spatial characterisation, $H$, in Tables~\ref{tab:spatial_comparison_people}-~\ref{tab:spatial_comparison_amenities}, and we draw the same conclusions.

    Next, we compare the average values of $M$ with the distributions obtained from the null models \NM{1}, \NM{3}, and \NM{5}, as shown in Tables~\ref{tab:parents_diversity_nullmodels}–\ref{tab:marriage_diversity_nullmodels}. The null models \NM{2} and \NM{4} are not included here, as they only affect the mobility cost $C$ (results reported in the main manuscript). Once again, we find that most $p$-values are below 0.001, with only three exceptions. These results confirm that the observed patterns of mobility diversity are robust across the three null models considered.

    %
    %  TABLE KS TEST MOBILITY diversity
    %
    %
    \begin{table}[h!]
        \centering
        \caption{Comparison of the distribution of $M$ between travellers' type for all the metropolitan areas considered in our study, using two-sample Kolmogorov-Smirnov test. For each area, we report the comparisons between \texttt{non-parent} and \texttt{parent} travellers, as well as between \texttt{married} and \texttt{non-married}. Then, we report the values of the differences between the average values, displayed as $\Delta M_P$ and $\Delta M_M$ in \figurename~\ref{fig:mobility_characterisation}. Finally, we report that the $p$-values are below 0.001 (***).}
        % \accom{Perhaps we can swap the case of non-parent vs parent (and change the sing of the difference) to make the table more "readable". What do you think?}}
        \label{tab:mobility_stats}
        \resizebox{0.7\textwidth}{!}{%
        \begin{tabular}{r|rlrc}
        \toprule
        \multicolumn{1}{c|}{Metropolitan Area} & \multicolumn{1}{c}{Group 1} & \multicolumn{1}{c}{Group 2} & \multicolumn{1}{c}{Difference} & \multicolumn{1}{c}{$p$-value} \\
        \midrule
            \multirow[c]{2}{*}{Atlanta} & \texttt{married} & \texttt{non-married} & 0.0486 & \multirow[c]{34}{*}{***} \\
                & \texttt{parent} & \texttt{non-parent} & 0.0385  & \\ \cline{1-4}
            \multirow[c]{2}{*}{Baltimore} & \texttt{married} & \texttt{non-married} & 0.0392 &  \\
             & \texttt{parent} & \texttt{non-parent} & 0.0267  & \\ \cline{1-4}
            \multirow[c]{2}{*}{Charlotte} & \texttt{married} & \texttt{non-married} & 0.0486 & \\
             & \texttt{parent} & \texttt{non-parent} & 0.0209  & \\ \cline{1-4}
            \multirow[c]{2}{*}{Chicago} & \texttt{married} & \texttt{non-married} & 0.0378 &  \\
             & \texttt{parent} & \texttt{non-parent} & 0.0638  & \\ \cline{1-4}
            \multirow[c]{2}{*}{Cincinnati} & \texttt{married} & \texttt{non-married} & 0.0022 & \\
             & \texttt{parent} & \texttt{non-parent} & 0.0178  & \\ \cline{1-4}
            \multirow[c]{2}{*}{Dallas} & \texttt{married} & \texttt{non-married} & 0.0172 & \\
             & \texttt{parent} & \texttt{non-parent} & 0.0236  & \\ \cline{1-4}
            \multirow[c]{2}{*}{Houston} & \texttt{married} & \texttt{non-married} & 0.0172 & \\
             & \texttt{parent} & \texttt{non-parent} & 0.0073  & \\ \cline{1-4}
            \multirow[c]{2}{*}{Indianapolis} & \texttt{married} & \texttt{non-married} & 0.0250 & \\
             & \texttt{parent} & \texttt{non-parent} & 0.0253  & \\ \cline{1-4}
            \multirow[c]{2}{*}{Kansas City} & \texttt{married} & \texttt{non-married} & 0.0105 &  \\
             & \texttt{parent} & \texttt{non-parent} & -0.0097  & \\ \cline{1-4}
            \multirow[c]{2}{*}{Minneapolis} & \texttt{married} & \texttt{non-married} & 0.0272 & \\
             & \texttt{parent} & \texttt{non-parent} & 0.0201  & \\ \cline{1-4}
            \multirow[c]{2}{*}{Nashville} & \texttt{married} & \texttt{non-married} & 0.0687 &  \\
             & \texttt{parent} & \texttt{non-parent} & 0.0682  & \\ \cline{1-4}
            \multirow[c]{2}{*}{New York} & \texttt{married} & \texttt{non-married} & 0.0333 &  \\
             & \texttt{parent} & \texttt{non-parent} & 0.0302  & \\ \cline{1-4}
            \multirow[c]{2}{*}{Philadelphia} & \texttt{married} & \texttt{non-married} & 0.0278 &  \\
             & \texttt{parent} & \texttt{non-parent} & 0.0145  & \\ \cline{1-4}
            \multirow[c]{2}{*}{Pittsburgh} & \texttt{married} & \texttt{non-married} & 0.0301 &  \\
             & \texttt{parent} & \texttt{non-parent} & 0.0007 &  \\ \cline{1-4}
            \multirow[c]{2}{*}{St. Louis} & \texttt{married} & \texttt{non-married} & 0.0108 & \\
             & \texttt{parent} & \texttt{non-parent} & 0.0329  & \\ \cline{1-4}
            \multirow[c]{2}{*}{Virginia Beach} & \texttt{married} & \texttt{non-married} & 0.0012 &  \\
             & \texttt{parent} & \texttt{non-parent} & 0.0246  & \\ \cline{1-4}
            \multirow[c]{2}{*}{Washington DC} & \texttt{married} & \texttt{non-married} & 0.0120 &  \\
             & \texttt{parent} & \texttt{non-parent} & 0.0162 & \\
        \bottomrule
        \end{tabular}
        } % end resizebox
    \end{table}

    %
    %  TABLE KS TEST MOBILITY COST
    %
    %
    \begin{table}[h!]
        \centering
        \caption{Comparison of the distribution of $C$ between travellers' type for all the metropolitan areas considered in our study, using two-sample Kolmogorov-Smirnov test. For each area, we report the comparisons between \texttt{non-parent} and \texttt{parent} travellers, as well as between \texttt{married} and \texttt{non-married}. Then, we report the values of the differences between the average values, displayed as $\Delta C_P$ and $\Delta C_M$ in \figurename~\ref{fig:mobility_characterisation}. Finally, we report that the $p$-values are below 0.001 (***).
        % Comparison of the distribution of $C$ between travellers' type for all the metropolitan areas considered in our study, using two-sample Kolmogorov-Smirnov test. For each area, we report the comparisons between \texttt{non-parent} and \texttt{parent} travellers, as well as between \texttt{married} and \texttt{non-married}. Then, we report the values of the differences between the values of $C$ corresponding to the average values of $C$. Finally, we report the $p$-values that are below 0.001 using ***. 
        }
        \label{tab:cost_stats}
        \resizebox{0.7\textwidth}{!}{%
        \begin{tabular}{r|rlrc}
        \toprule
        \multicolumn{1}{c|}{Metropolitan Area} & \multicolumn{1}{c}{Group 1} & \multicolumn{1}{c}{Group 2} & \multicolumn{1}{c}{Difference} & \multicolumn{1}{c}{$p$-value} \\
        \midrule
            \multirow[c]{2}{*}{Atlanta} & \texttt{married} & \texttt{non-married} & -0.0059 & \multirow[c]{34}{*}{***} \\
                & \texttt{parent} & \texttt{non-parent} & 0.0006  & \\ \cline{1-4}
                
            \multirow[c]{2}{*}{Baltimore} & \texttt{married} & \texttt{non-married} & -0.0044 &  \\
             & \texttt{parent} & \texttt{non-parent} & -0.0061  & \\ \cline{1-4}
             
            \multirow[c]{2}{*}{Charlotte} & \texttt{married} & \texttt{non-married} & -0.0020 &  \\
             & \texttt{parent} & \texttt{non-parent} & -0.0002  & \\ \cline{1-4}
             
            \multirow[c]{2}{*}{Chicago} & \texttt{married} & \texttt{non-married} & 0.0051 &  \\
             & \texttt{parent} & \texttt{non-parent} & -0.0066  & \\ \cline{1-4}
             
            \multirow[c]{2}{*}{Cincinnati} & \texttt{married} & \texttt{non-married} & 0.0061 & \\
             & \texttt{parent} & \texttt{non-parent} & -0.0070  & \\ \cline{1-4}
             
            \multirow[c]{2}{*}{Dallas} & \texttt{married} & \texttt{non-married} & 0.0035 &  \\
             & \texttt{parent} & \texttt{non-parent} & 0.0023  & \\ \cline{1-4}
             
            \multirow[c]{2}{*}{Houston} & \texttt{married} & \texttt{non-married} & 0.0152 &  \\
             & \texttt{parent} & \texttt{non-parent} & 0.0097  & \\ \cline{1-4}
             
            \multirow[c]{2}{*}{Indianapolis} & \texttt{married} & \texttt{non-married} & 0.0022 &  \\
             & \texttt{parent} & \texttt{non-parent} & 0.0041  & \\ \cline{1-4}
             
            \multirow[c]{2}{*}{Kansas City} & \texttt{married} & \texttt{non-married} & 0.0104 &  \\
             & \texttt{parent} & \texttt{non-parent} & 0.0048  & \\ \cline{1-4}
             
            \multirow[c]{2}{*}{Minneapolis} & \texttt{married} & \texttt{non-married} & -0.0015 &  \\
             & \texttt{parent} & \texttt{non-parent} & -0.0044  & \\ \cline{1-4}
             
            \multirow[c]{2}{*}{Nashville} & \texttt{married} & \texttt{non-married} & -0.0059 &  \\
             & \texttt{parent} & \texttt{non-parent} & -0.0028  & \\ \cline{1-4}
             
            \multirow[c]{2}{*}{New York} & \texttt{married} & \texttt{non-married} & 0.0064 &  \\
             & \texttt{parent} & \texttt{non-parent} & 0.0006  & \\ \cline{1-4}
             
            \multirow[c]{2}{*}{Philadelphia} & \texttt{married} & \texttt{non-married} & 0.0080 &  \\
             & \texttt{parent} & \texttt{non-parent} & -0.0020  & \\ \cline{1-4}
             
            \multirow[c]{2}{*}{Pittsburgh} & \texttt{married} & \texttt{non-married} & -0.0042 &  \\
             & \texttt{parent} & \texttt{non-parent} & 0.0085 &  \\ \cline{1-4}
             
            \multirow[c]{2}{*}{St. Louis} & \texttt{married} & \texttt{non-married} & -0.0011 & \\
             & \texttt{parent} & \texttt{non-parent} & 0.0001  & \\ \cline{1-4}
             
            \multirow[c]{2}{*}{Virginia Beach} & \texttt{married} & \texttt{non-married} & 0.0105 &  \\
             & \texttt{parent} & \texttt{non-parent} & 0.0060  & \\ \cline{1-4}
             
            \multirow[c]{2}{*}{Washington DC} & \texttt{married} & \texttt{non-married} & 0.0101 &  \\
             & \texttt{parent} & \texttt{non-parent} & 0.0132 & \\
        \bottomrule
        \end{tabular}
        } % end resizebox
    \end{table}

    %
        %  TABLE KS-TEST HOUSEHOLD STATUS
        %
        %
        \begin{table}[ht]
            \centering
            \caption{Comparison of the distribution of $H$ between travellers' type for all the metropolitan areas considered in our study, using two-sample Kolmogorov-Smirnov test. For each area, we report the comparisons between \texttt{non-parent} and \texttt{parent} travellers, as well as between \texttt{married} and \texttt{non-married}. Then, we report the values of the differences between the average values, $\Delta H_P$ and $\Delta H_M$. Finally, we report that the $p$-values are below 0.001 (***).
            }
            \label{tab:spatial_comparison_people}
            \resizebox{0.7\textwidth}{!}{%
            \begin{tabular}{r|rlrc}
            \toprule
            \multicolumn{1}{c|}{Metropolitan Area} & \multicolumn{1}{c}{Group 1} & \multicolumn{1}{c}{Group 2} & \multicolumn{1}{c}{Difference} & \multicolumn{1}{c}{$p$-value}  \\
            \midrule
            
            \multirow[c]{2}{*}{Atlanta} & \texttt{married} & \texttt{non-married} & 0.0783 & \multirow[c]{34}{*}{***} \\
                                        & \texttt{parent} & \texttt{non-parent} & 0.0620   & \\ \cline{1-4}
                                        
            \multirow[c]{2}{*}{Baltimore} & \texttt{married} & \texttt{non-married} & 0.0697  & \\
                                          & \texttt{parent} & \texttt{non-parent} & 0.0482   & \\ \cline{1-4}
                                          
            \multirow[c]{2}{*}{Charlotte} & \texttt{married} & \texttt{non-married} & 0.0888   & \\
                                          & \texttt{parent} & \texttt{non-parent} & 0.0455 &   \\ \cline{1-4}
                                          
            \multirow[c]{2}{*}{Chicago} & \texttt{married} & \texttt{non-married} & 0.0910 &   \\
                                        & \texttt{parent} & \texttt{non-parent} & 0.0941 &   \\ \cline{1-4}
                                        
            \multirow[c]{2}{*}{Cincinnati} & \texttt{married} & \texttt{non-married} & 0.0510 &   \\
                                           & \texttt{parent} & \texttt{non-parent} & 0.0458 &   \\ \cline{1-4}
                                           
            \multirow[c]{2}{*}{Dallas} & \texttt{married} & \texttt{non-married} & 0.0660 &   \\
                                       & \texttt{parent} & \texttt{non-parent} & 0.0315 & \\ \cline{1-4}
                                       
            \multirow[c]{2}{*}{Houston} & \texttt{married} & \texttt{non-married} & 0.1129 &   \\
                                        & \texttt{parent} & \texttt{non-parent} & 0.0566 &   \\ \cline{1-4}
                                        
            \multirow[c]{2}{*}{Indianapolis} & \texttt{married} & \texttt{non-married} & 0.1129 &   \\
                                             & \texttt{parent} & \texttt{non-parent} & 0.0771 &   \\ \cline{1-4}
                                             
            \multirow[c]{2}{*}{Kansas City} & \texttt{married} & \texttt{non-married} & 0.0169 &   \\
                                            & \texttt{parent} & \texttt{non-parent} & 0.0088 &   \\ \cline{1-4}
                                            
            \multirow[c]{2}{*}{Minneapolis} & \texttt{married} & \texttt{non-married} & 0.0659 &   \\
                                            & \texttt{parent} & \texttt{non-parent} & 0.0542 &   \\ \cline{1-4}
                                            
            \multirow[c]{2}{*}{Nashville} & \texttt{married} & \texttt{non-married} & 0.1151 &   \\
                                          & \texttt{parent} & \texttt{non-parent} & 0.0741 &   \\ \cline{1-4}
                                          
            \multirow[c]{2}{*}{New York} & \texttt{married} & \texttt{non-married} & 0.0322 &   \\
                                         & \texttt{parent} & \texttt{non-parent} & 0.0196 &   \\ \cline{1-4}
                                         
            \multirow[c]{2}{*}{Philadelphia} & \texttt{married} & \texttt{non-married} & 0.0477 &   \\
                                             & \texttt{parent} & \texttt{non-parent} & 0.0263 &   \\ \cline{1-4}
                                             
            \multirow[c]{2}{*}{Pittsburgh} & \texttt{married} & \texttt{non-married} & 0.0608 &   \\
                                           & \texttt{parent} & \texttt{non-parent} & 0.0305 &   \\ \cline{1-4}
                                           
            \multirow[c]{2}{*}{St. Louis} & \texttt{married} & \texttt{non-married} & 0.0093 &   \\
                                          & \texttt{parent} & \texttt{non-parent} & 0.0127 &   \\ \cline{1-4}
                                          
            \multirow[c]{2}{*}{Virginia Beach} & \texttt{married} & \texttt{non-married} & 0.0093 &   \\
                                               & \texttt{parent} & \texttt{non-parent} & 0.0196 &   \\ \cline{1-4}
                                               
            \multirow[c]{2}{*}{Washington DC} & \texttt{married} & \texttt{non-married} & -0.0007 &   \\
                                              & \texttt{parent} & \texttt{non-parent} & -0.0106 &   \\
            \bottomrule
            \end{tabular}
            } % end of resizebox
        \end{table}

        %
        %  TABLE KS-TEST AMENITY TYPES FOR ATLANTA, BALTIMORE, & CHARLOTTE
        %
        %        
        \begin{table}[ht]
            \centering
            \caption{Comparison of the average value of $H$ between amenities type exemplified by Atlanta, Baltimore, and Charlotte metropolitan areas via a two-sample Kolmogorov-Smirnov test. We report the comparisons between every pair of amenities. Then, we report the values of the differences between the values of $H$ corresponding to the average values of the respective $H$. Finally, we report the $p$-value associated with the test. The column labelled as `Conclusion' denotes whether we can accept strictly (***) or not (**) the test's outcome.}
            \label{tab:spatial_comparison_amenities}
            \resizebox{0.85\textwidth}{!}{%            
            \begin{tabular}{rl|rc|rc|rc}
                \toprule
                & &  \multicolumn{2}{c|}{Atlanta} & \multicolumn{2}{c|}{Baltimore} & \multicolumn{2}{c}{Charlotte} \\
                
                \multicolumn{1}{c}{Group 1} & \multicolumn{1}{c|}{Group 2} & \multicolumn{1}{c}{Difference} & \multicolumn{1}{c|}{$p$-value} & \multicolumn{1}{c}{Difference} & \multicolumn{1}{c|}{$p$-value} & \multicolumn{1}{c}{Difference} & \multicolumn{1}{c}{$p$-value} \\
                \midrule
                \multirow[c]{9}{*}{all} &   education & -0.0102 &  
                \multirow[c]{9}{*}{***} & -0.0243 &  
                \multirow[c]{9}{*}{***} & 0.0173 &  \multirow[c]{9}{*}{***} \\
                
                    &        food & -0.0039 &  & -0.0291 &  & 0.0324 &  \\
                    &      health & -0.0005 &  & -0.0196 &  & 0.0023 &  \\
                    &     leisure & -0.0039 &  & -0.0054 &  & 0.0285 &  \\
                    &   religious & -0.0226 &  & -0.0180 &  & -0.0469 &  \\
                    & residential & 0.1041 &  & 0.0596 &  & 0.0821 &  \\
                    &    services & -0.0666 &  & -0.0447 &  & -0.0337	 &  \\
                    &   transport & -0.0038 &  & -0.0205 &  & 0.0551	 &  \\
                    &        work & 0.0216 &  & 0.0713 &  & 0.0895	&   \\ \hline

                \multirow[c]{8}{*}{education} &        food & 0.0069 &  \multirow[c]{8}{*}{***}
                & -0.0049 &  \multirow[c]{8}{*}{***}
                & 0.0160 &  \multirow[c]{8}{*}{***} \\
                                              &      health & 0.0100 &  & 0.0049 &  & -0.0154 &  \\
                                              &     leisure & 0.0068 &  & 0.0194 &  & 0.0121 &  \\
                                              &   religious & -0.0125 &  & 0.0067 &  & -0.0667 &  \\
                                              & residential & 0.1193 &  & 0.0873 &  & 0.0675 &  \\
                                              &    services & -0.0583 &  & -0.0212 &  & -0.0530 &  \\
                                              &   transport & 0.0070 &  & 0.0040 &  & 0.0393 &  \\
                                              &        work & 0.0334 &  & 0.0997 &  & 0.0754 &  \\ \hline

                \multirow[c]{7}{*}{food} &      health & 0.0034 &  \multirow[c]{7}{*}{***}
                & 0.0095 &  \multirow[c]{7}{*}{***}
                & -0.0301 &  \multirow[c]{7}{*}{***} \\
                                         &     leisure & -0.0001 &  & 0.0237 &  & -0.0039 &  \\
                                         &   religious & -0.0186 &  & 0.0111 &  & -0.0793 &  \\
                                         & residential & 0.1080 &  & 0.0887 &  & 0.0497 &  \\
                                         &    services & -0.0627 &  & -0.0156 &  & -0.0661 &  \\
                                         &   transport & 0.0002 	 & & 0.0087 &  & 0.0227 &  \\
                                         &        work & 0.0255 &  & 0.1004 &  & 0.0571 &  \\ \hline

                \multirow[c]{6}{*}{health} &     leisure & -0.0034 &  \multirow[c]{6}{*}{***}
                & 0.0142 &  \multirow[c]{6}{*}{***}
                & 0.0262 &  \multirow[c]{6}{*}{***} \\
                                           &   religious & -0.0220 &  & 0.0016 &  & -0.0492 &  \\
                                           & residential & 0.1046 &  & 0.0792 &  & 0.0798 &  \\
                                           &    services & -0.0661 &  & -0.0251 &  & -0.0360 &  \\
                                           &   transport & -0.0032 &  & -0.0008 &  & 0.0548 &  \\
                                           &        work & 0.0221 &  & 0.0909 &  & 0.0872 &   \\ \hline

                \multirow[c]{5}{*}{leisure} &   religious & -0.0194 &  \multirow[c]{5}{*}{***} & -0.0127 &  \multirow[c]{5}{*}{***} & -0.0788 &  \multirow[c]{5}{*}{***} \\
                                            & residential & 0.1124 &  & 0.0679 &  & 0.0554 &  \\
                                            &    services & -0.0652 &  & -0.0406 &  & -0.0650 &  \\
                                            &   transport & 0.0002 &  & -0.0154 &  & 0.0273 &  \\
                                            &        work & 0.0266 &  & 0.0803 &  & 0.0633 &  \\ \hline

                \multirow[c]{4}{*}{religious} & residential & 0.1267	 &  \multirow[c]{4}{*}{***} 
                & 0.0776		 &  \multirow[c]{4}{*}{***} 
                & 0.1290		 &  \multirow[c]{4}{*}{***} \\
                
                                              &    services & -0.0441	 &  & -0.0267	 &  & 0.0132 &  \\
                                              &   transport & 0.0188	 &  & -0.0024	 &  & 0.1020	 &  \\
                                              &        work & 0.0441 &  & 0.0893 &  & 0.1364 &  \\ \hline

                \multirow[c]{3}{*}{residential} &    services & -0.1707	 &  \multirow[c]{3}{*}{***} & -0.1043	 &  \multirow[c]{3}{*}{***} & -0.1158	 &  \multirow[c]{3}{*}{***} \\
                                                &   transport & -0.1079	 &  & -0.0801	 &  & -0.0270	 &  \\
                                                &        work & -0.0825	 &  & 0.0117 &  & 0.0074 &  \\ \hline

                \multirow[c]{2}{*}{services} &   transport & 0.0629 &  \multirow[c]{2}{*}{***} & 0.0242	 &  \multirow[c]{2}{*}{***} & 0.0888	 &  \multirow[c]{2}{*}{***} \\
                                             &        work & 0.0882 &  & 0.1160 &  & 0.1232 &  \\ \hline

                transport &        work & 0.0253 &  *** & 0.0918 &  *** & 0.0344 &  *** \\
                \bottomrule
                \end{tabular}
                } % end of resizebox
        \end{table}

            %
            %  TAB: DIFFERENCES AND NULL MODELS FOR PARENTHOOD
            %
            %
            \begin{table}[ht]
            \centering
            \caption{Comparison between the empirical average mobility diversity differences, $\Delta M$, between parents and non-parents and the values from each null model \NM{x} (with $x \in \{1,3,5\}$). We compare the average empirical value against the distribution of null model values.}
            \label{tab:parents_diversity_nullmodels}
            \resizebox{0.85\linewidth}{!}{
             \begin{tabular}{r|rrrrrrr}
            \toprule
            \multicolumn{1}{c|}{\multirow{2}{*}{Urban area}} & \multicolumn{1}{c}{DATA} & \multicolumn{2}{c}{\NM{1}} & \multicolumn{2}{c}{\NM{3}} & \multicolumn{2}{c}{\NM{5}} \\
            & \multicolumn{1}{c}{$\Delta M_P$} & \multicolumn{1}{c}{$\Delta M_P$} & $p$-val & \multicolumn{1}{c}{$\Delta M_P$} & $p$-val & \multicolumn{1}{c}{$\Delta M_P$} & $p$-val \\
            \midrule
            Baltimore (6) & 2.67e-02 & -4.65e-05 & 0.0000 & 3.10e-02 & 0.1140 & 1.46e-02 & 0.0000 \\
            Houston (6) & 7.31e-03 & -4.72e-06 & 0.0000 & 5.24e-02 & 0.0000 & -5.44e-02 & 0.0000 \\
            Pittsburgh (6) & 7.05e-04 & -1.13e-05 & 0.0332 & 5.81e-02 & 0.0000 & -6.34e-02 & 0.0000 \\
            Cincinnati (7) & 1.78e-02 & 3.65e-05 & 0.0000 & 8.11e-02 & 0.0000 & -1.75e-02 & 0.0000 \\
            Indianapolis (7) & 2.53e-02 & 2.75e-05 & 0.0000 & 1.23e-01 & 0.0000 & -2.47e-02 & 0.0000 \\
            Kansas City (7) & -9.72e-03 & -1.12e-05 & 0.0000 & -4.62e-03 & 0.0526 & -1.94e-02 & 0.0000 \\
            Virginia Beach (7) & 2.46e-02 & 1.57e-05 & 0.0000 & 6.01e-03 & 0.0000 & 1.75e-02 & 0.0000 \\
            Nashville (8) & 6.81e-02 & -1.77e-05 & 0.0000 & 1.22e-01 & 0.0000 & 1.83e-02 & 0.0000 \\
            Charlotte (9) & 2.09e-02 & 2.85e-05 & 0.0000 & 6.20e-02 & 0.0000 & -7.94e-03 & 0.0000 \\
            Dallas (9) & 2.37e-02 & -1.29e-05 & 0.0000  & 4.14e-02 & 0.0000 & -8.24e-03 & 0.0000 \\
            St. Louis (9) & 3.29e-02 & 1.48e-05 & 0.0000 & -8.55e-03 & 0.0000 & 2.11e-02 & 0.0000 \\
            Minneapolis (10) & 2.01e-02 & -1.42e-05 & 0.0000 & 7.45e-02 & 0.0000 & -9.62e-03 & 0.0000 \\
            Chicago (11) & 6.38e-02 & 1.39e-05 & 0.0000 & 7.88e-02 & 0.0000 & -5.36e-03 & 0.0000 \\
            Philadelphia (11) & 1.45e-02 & 3.26e-05 & 0.0000 & 6.84e-03 & 0.0000 & 3.24e-03 & 0.0000 \\
            Washington (13) & 1.62e-02 & 5.05e-06 & 0.0000 & -1.33e-02 & 0.0000 & 9.21e-04 & 0.0000 \\
            Atlanta (17) & 3.85e-02 & 1.45e-05 & 0.0000 & 7.71e-02 & 0.0000 & 3.09e-03 & 0.0000 \\
            New York (23) & 3.02e-02 & 9.14e-06 & 0.0000 & 2.18e-03 & 0.0000 & 9.46e-03 & 0.0000 \\
            \bottomrule
            \end{tabular}
            }
        \end{table}

        %
        %  TAB: DIFFERENCES AND NULL MODELS FOR MARITAL STATUS
        %
        %
        \begin{table}[hb]
            \centering
            \caption{Comparison between the empirical average mobility diversity differences, $\Delta M$, between married and non-married individuals and the values from each null model \NM{x} (with $x \in \{1,3,5\}$). We compare the average empirical value against the distribution of null model values.}
            \label{tab:marriage_diversity_nullmodels}
            \resizebox{0.85\linewidth}{!}{
            \begin{tabular}{r|rrrrrrr}
            \toprule
            \multicolumn{1}{c|}{\multirow{2}{*}{Urban area}} & \multicolumn{1}{c}{DATA} & \multicolumn{2}{c}{\NM{1}} & \multicolumn{2}{c}{\NM{3}} & \multicolumn{2}{c}{\NM{5}} \\
            & \multicolumn{1}{c}{$\Delta M_M$} & \multicolumn{1}{c}{$\Delta M_M$} & $p$-val & \multicolumn{1}{c}{$\Delta M_M$} & $p$-val & \multicolumn{1}{c}{$\Delta M_M$} & $p$-val \\
            %
            %Zone & S & Difference & S1 & NM1 & PNM1 & S3 & NM3 & PNM3 & S5 & NM5 & PNM5 \\
            \midrule
            Baltimore (6) & 3.92e-02 & -3.20e-05 & 0.0000 & 6.57e-02 & 0.0000 & 2.84e-02 &  0.0000 \\
            Houston (6) & 1.73e-02 & -1.54e-05 & 0.0000 & 1.27e-01 & 0.0000 & -1.51e-03 &  0.0000 \\
            Pittsburgh (6) & 3.01e-02 & -4.09e-05 & 0.0000 & 1.12e-01 & 0.0000 & 1.74e-02 &  0.0000 \\
            Cincinnati (7) & 2.23e-03 & 9.68e-06 & 0.0000 & 9.89e-02 & 0.0000 & 9.91e-04 & 0.1760 \\
            Indianapolis (7) & 2.51e-02 & 1.15e-04 & 0.0000 & 1.62e-01 & 0.0000 & -2.14e-03 &  0.0000 \\
            Kansas City (7) & 1.05e-02 & -4.86e-06 & 0.0000 & 1.62e-02 & 0.0182 & 1.88e-02 &  0.0000 \\
            Virginia Beach (7) & 1.18e-03 & 4.77e-05 & 0.0000 & 2.05e-02 & 0.0000 & -2.53e-03 &  0.0000 \\
            Nashville (8) & 6.87e-02 & 9.19e-05 & 0.0000 & 1.72e-01 & 0.0000 & 3.56e-02 &  0.0000 \\
            Charlotte (9) & 4.86e-02 & -7.55e-05 & 0.0000 & 1.34e-01 & 0.0000 & 3.26e-02 &  0.0000 \\
            Dallas (9) & 1.72e-02 & 1.29e-05 & 0.0000 & 8.96e-02 & 0.0000 & 7.47e-03 &  0.0000 \\
            St. Louis (9) & 1.08e-02 & 7.42e-06 & 0.0000 & -5.89e-03 & 0.0000 & 3.27e-02 &  0.0000 \\
            Minneapolis (10) & 2.73e-02 & -2.64e-05 & 0.0000 & 9.62e-02 & 0.0000 & 1.11e-02 &  0.0000 \\
            Chicago (11) & 3.78e-02 & 3.58e-06 & 0.0000 & 1.34e-01 & 0.0000 & 7.68e-03 &  0.0000 \\
            Philadelphia (11) & 2.78e-02 & 1.07e-06 & 0.0000 & 9.06e-03 & 0.0000 & 1.71e-02 &  0.0000 \\
            Washington (13) & 1.20e-02 & -9.12e-06 & 0.0000 & -7.55e-03 & 0.0000 & 1.19e-02 & 0.2480 \\
            Atlanta (17) & 4.86e-02 & 2.08e-05 & 0.0000 & 1.17e-01 & 0.0000 & 3.49e-02 &  0.0000 \\
            New York (23) & 3.33e-02 & 1.12e-05 & 0.0000 & -1.68e-02 & 0.0000 & 1.08e-02 &  0.0000 \\
            \bottomrule
            \end{tabular}
            }
        \end{table}

\clearpage

\section{Dictionary of amenities' categories}
\label{sec:dictionary}

In the following, we present the complete list (encoded as a Python dictionary) of amenities that we used in our work, grouped into categories.

%
%  JSON OF AMENITIES BY CATEGORIES
%
%
\begin{lstlisting}
mapping_categories = {
    "remove_categories" : [
        'hospital:historical','funeral_home','funeral home','funeral_directors',
        'hospital (historic)','tomb','recreation','gf','gh','hg','mri','part','m',
        'o','iona_College_Dorm','smoking_area','chr','yb','city','rostrum','stage',
        'bridleway','vending_machine','saint louis city morgue','out','bunker',
        'don chief denmyer facilities building','social_club','em','security_booth',
        'container','residential,_business,_restaurants','no','yes','barn','cabin',
        'canopy','carport','compressed_air','corridor','detached','footway',
        'grandstand','hangar','proposed','caravan','ab','ye','doityourself','raceway',
        'rest_area','roof','ruins','secondary','gaq','ab','gr','secondary_link','shed',
        'static_caravan','tertiary','ind','clubhouse','strip_club','yacht_club',
        'tertiary_link','toll_gantry','unclassified','device_charging_station',
        'gatehouse','gazebo','karaoke_box','payment_centre','polling_station',
        'outbuilding','poolhouse','sty','trolley_bay','fortune_teller','lounger',
        'ventilation_shaft','works','parcel_locker','table','stripclub','car_sharing',
        'give_box','watering_place','training','gaze','rectory','tent','townhouse',
        'guardhouse','relay_box','swingerclub','drinks','nail_salon','stock_exchange',
        'collapsed','research_institute','tower','meditation_centre','graphic_design',
        'subway_entrance','waste_transfer_station','dressing_room','hand_sanitizing',
        'stroller_rental','cafe;bar','wifi;telephone;device_charging_station',
        'gambling','disused','bungalow','community_group_office','manor',
        'derelict,_uninhabited','library_dropoff','disused:restaurant','demolished',
        'binoculars','bus_depot','beauty_school','military','ranger_station',
        'mist_spraying_cooler','ship','mortuary','slaughterhouse','hookah_lounge',
        'aviary','driver_training','lost_property_office','acting_school','tutoring',
        'postal_relay_box','printer','disused:pub','mixed_use','surface','radio station',
        'washing_machine','waste_basket;vending_machine','vending_machine;toilets',
        'vending_machine;waste_basket','toilets;shower;laundry',
        'toilets;drinking_water;bbq','toilets;kitchen;fridge;drinking_water;stove;bbq',
        'toilets; laundry','car_wash; toilets','vending_machine;toilets',
        'dressing_room;toilets','dog_toilet','building_concrete','county_building',
        'public_building;shelter','scrapyard','building_yard','court_yard',
        'building_supply','verizon building','roof;apartments','apartments;hotel;office',
        'roof;kindergarten','artwo','part:roof','biergarten;bar','apartments;retail',
        'airport','airport_terminal','common_area','picnic area','dog_relief_area',
        'swimming_area','banquet_hall','exhibition_hall','kingdom_hall','parish_hall',
        'riding_hall','concert_hall','reception_hall','borough hall','riding_hall;stable',
        'hall','chimney','counselling','dialysis','dome','fixme','fortune_telling',
        'grit_bin','gurdwara','pagoda','photo_booth','rv_storage','silo','storage tank',
        'storage_facility','vacuum_cleaner','water_tank','boat_storage','shade',
        'concrete paving','condominium','nonprofit organizations=','abandoned','ap',
        'bandstand','barracks','batting_cage','bird_bath','book_return','books','closed',
        'concession_stand','concessions_stand','condomnium','contemplation','demolition',
        'field_house','foundation','fraternity','garbage_shed','hospice','laboratory',
        'lifeguard','lighthouse','mausoleum','midwife','motel','motorcycle_rental','off',
        'pain_management','presbytery','sch','tank','tele','tool_library','trailer',
        'training_police_fire','vacant','occupational_therapist','physical_therapy',
        'refugee_site','tourism','church;commercial','festival_grounds','nutrition',
        'outdoors','pyramid','therapist','hunting_stand','lockers','parish','salt_dome',
        'showground','teepee','traffic_sign','sleep disorder','duplex',
        'timothy_e._simmons','cosmetic_surgery','heavy_equipment_rental','main','mill',
        'peace_pole','salt_pyramid','self_storage','surgery','concrete contractor',
        'nail salon','natatorium','communications','craniosacral_therapy',"i'",
        'lifejacket','mailrom','mist spraying cooler','mobile','nurse','power',
        'professional services','show_house','soccer field','society','traffic_island',
        'weight_loss','ice_cream;shaved_ice','planned','utility','allotment_house',
        'auxiliary','conservatory','deli','hostel','houseboat','siil','windmill','mailbox',
        'castle_hill_electrical_supply','shooting_stand','argricultural_center',
        'commercial,_storage','country_store','datacenter','boat_stoarage','counseling',
        'cowshed','makerspace','splash_pad','armory','razed','remnant','garage doors',
        'appartment','archive','bleachers','check_cashing','ger','undefined','stilt_house',
        'scoring_box','donation center','plastic surgeon','bear_cache','ga','pulmonology',
        'stables','grand_old_hatchery','visitor_centre','hair replacement','paint supplies',
        'psychologist','recycle glass vases','alleyway','bench;waste_basket','caboose',
        'concussion','condominiums','control_tower','financial_advice','footpath','kitchen',
        'manhole','outdoor_kiosk','plastic_surgery','reception_desk','ticket_validator',
        'tutor','waste_container','wellness_program','sona_dermatology_&_medspa,_inc.',
        'bell_tower','dive_centre','es','garageq','gymnasium','jail','retail;roof','sign',
        'bathroom','auto detailing','delicatessen','dry_cleaner','gallery','nail',
        'organization','photography','recycling;waste_basket','via_ferrata','ballroom',
        'chair','dugout','hitching_post','marquee','roundhouse','sample_collection',
        'senior_center','vaccination_centre','walkway','wall','cigar_bar','events_centre',
        'its being distroyed','medium','podiatry','small','iona_college_dorm','boat_sharing',
        'flat','gabled','municipal','quonset_hut','tennis court','g','hair removal',
        'football','package_room','poultry_house','paving driveways','horse_facility',
        'aircraft_control','dmv','donations_box','realty','permanent makeup training',
        'for-sale','judo','laundromat','mixed','shrine','unknown','unkown','beauty',
        'interior designer','diner','hairdresser','tutoring_centre','concrete_plant',
        'event_venue','occupied','parsonage','gate_house','gas','porch','rain_garden',
        'retail;commercial','paediatrics','playground_structure','event_center','arena',
        'child_amusement_center','detached house','donation_box','flowerpot',
        'psychotherapist;speech_therapist','vaping_lounge','amphitheatre','bicycle_library',
        'crossing;give_way','insurance','support','check_in','construction_equipment_supplier',
        'seating','triplex','beach_hut','boat','castle','drinking_water;watering_place',
        'prefabricated','tree_house','water_tap','alcohol','border_control','burial_vault',
        'catering','cloakroom','crypt','flower_planter','footway;crossing',
        'money_transfer;notary_public','other','personal_trainer','place_of_meditation',
        'refuge','theatre (historic)','tourist','depot','flower_containers','laundry',
        'marker','therapy','moving company cypress','polyurea coatings','arch',
        'bicycle_wash','cheque_cashing','cigar_lounge','construction;store','feeding_place',
        'gasometer','houaw','plant_nursery','plasma_center','restrooms','romney',
        'truck_rental','wellness','barn;house','garage;house','roof;church','deck',
        'renovation','chiropractor','barrel','convention_centre','house;garage','microwave',
        'newsbox','roof;retail','shed;garage;house','shed;house',
        'clinic;laboratory;physiotherapist;occupational_therapist','inflatable','road_depot',
        'shed_and_seasonal_restrooms','eating_disorders','sewage deodorization','shaved_ice',
        'ski_rental','multi-tennant_commercial','church;roof','generic','grill','highrise',
        'house poly','roof;commercial','roof;industrial','roof;university',
        'mainstream automotive','totally_dog','fire_training_facility','amphitheater',
        'fish_cleaning','generator','addiction_treatment_center','chiropratic','cryo',
        'recreational','spa,_sauna','user_defined','escape','picnic shelter',
        'university;chapel','restroom','ruins/foundation','pool entrance and tickets',
        'wedding location','chicken_coop','commercial;detached','disused:church','garden',
        'glasshouse','outer','post_storage_box','res','skybridge','snack_bar','street_vendor',
        'student_accommodation','fish_hatchery','checkpoint','complex','disused:garage','em',
        'housing','retailer','scoreboard','vintage store','christ','exercise','municipial',
        'pharma','viewing_platform','compressed_air;vacuum_cleaner','conrainer','market',
        'amenity','sroof','toliets','ice','outdoor_seating','road_maintenance',
        'science_incubator','conveneince_store','palyground','semi','clinic;physiotherapist',
        'dropbox','attached','charity','hookah','lounge','pest_control','q','realtor',
        'strip_mall','urgent','use=stable','j','sheepfold','cemetery','firstaid','gar',
        'trailer_storage','tunnel','granary','detached;shed','first aid','rescue_squad',
        'book_drop','corn_crib','pump_house',
        'shelter;baseball;basketball;disc golf;volleyball;gardens','sledding','soccer',
        'tractor supply','masons','utilities','archives','customer_service',
        'disused:drinking_water','skywalk','storage_rental','memorial','podium','dining',
        'information_sign','pharmaceutical_company','springhouse','yurt','scj','chabad_house',
        'sewer','tree','fair_booth','harbourmaster','x_and_y_chromosome_variation_center',
        "' crossing",'emissions_testing','farmhouse','guard_tower','bathroom remodeling',
        'airplane','chu','emissions','gunsmith','police_academy','semidetached_houseyes','train',
        'vaccination_centre;sample_collection','excavating contractor','skating_rink',
        'fire_lookout','storefront','fuel oil distributor','airplane_fuselage','detached;house',
        'oncologist','pumps','taxidermy','psychiatrist','snowmobile','auto insurance',
        'shelter;fuel','commer','overlook','roof_overhang','web_development',
        'testosterone_replacement','wellness_center','taxi_point','rehabilitation;alternative',
        'furniture repair','com','senior','lodging','terminal','motor vehicle administration',
        'layer','indoor_range','dj services','family therapy','wellness center',
        'running_specialty','nuclear','furniture maker','concession','fuel;convenience store',
        'convenience','books;cafe','Shelter','restaurant;karaoke','bandshell','street_lamp',
        'motorway','motorway_link','road','convenience_store','flour_mill','fourplex',
        'printing_facility','animal_boarding','animal_training','animal_boarding; animal_training',
        'animal_control','dog_parking','animal_training;training','pet','pet supplies',
        'pet_relief_area','dog_run','boat_rental','power_substation','sanitary_dump_station',
        'police_station','weight_station','weigh_station','utility_suubstation','fire_station;police',
        'fire_station;yes','crossing;stop','car_repair','career_center','church;carport','carillon',
        'in_home_care','mailroom','post_depot','block','bollard','civic','crossing','sto',
        'transportation','crossing;traffic_signals','cycle_barrier','elevator','fence','give_way',
        'kiosk','lift_gate','passing_place','priority','services','speed_camera','speed_display',
        'steps','stile','stop','track','traffic_mirror','traffic_signals','traffic_signals;crossing',
        'trunk','trunk_link','pavilion','primary','primary_link','bureau_de_change','fountain',
        'centre','grave_yard','hostel:homeless','hut','money_transfer','place_of_chanting_daimoku',
        'shoe','storage_units_and_shopping_center','greenhouse','farm','farm_auxiliary','water_tower',
        'water_point','bell','bench','clock','drinking_water','letter_box','locker','luggage_locker',
        'taxi_service','post_box','public','shower','storage_tank','historic','transformer_tower',
        'waste_basket','former hospital','health_insurance','alternative','animal_breeding',
        'school;hospital','school;industrial','place_of_worship;school','disused:school',
        'public_bookcase',"police; townhall; clerk's office",'parking;post_office','grass',
        'water_tower','storage','terrace','living_street','commercial;residential',
        'residential;retail','residential;commercial','sport_hall','sports_centre','martial_arts',     
        'sport','gymnasium','athletic','sports_hall','sport_centre','sports',
        'soccer','pool,_tennis','judo','pool,_Tennis','greenhouse','farm','farm_auxiliary','stable',
        'botanical_gardens'
    ],

    "food" : [
        'bbq','cafe','fast_food','food_court','ice_cream','food_and_drink','restaurant',
        'restaurant;cafe','juice_bar','bakery','food_sharing','cafeteria','pub;cafe',
        'ice_cream;chocolate;cake','fast_food;ice_cream','food_court;atm;toilets','food','market',
        'restaurant;coffee','donuts','ice_cream;shaved_ice','restaurant;grocery;butcher;',
        'bar;pub;restaurant','coffee_shop','coffee shop','restaurant;banquet','restaurant;catering',
        'bakehouse','supermarket','diner'
    ],

    "religious": [
        'place_of_worship','mosque','convent','cathedral','religious','church;roof','church_hall',
        'place_of_meditation','synagogue','temple'
    ],

    "transport" : [
        'bus_stop','car_rental','car_park','parking_garage','parking_lot','parking lot',
        'bicycle_parking','cycleway','bus','ferry_terminal','fuel','fuel;pharmacy','garage',
        'garages','gate','milestone','mini_roundabout','motorcycle_barrier','path',
        'motorcycle_parking','motorway_junction','parking','bridge','parking_entrance','platform',
        'taxi','toll_booth','bus_station','trailhead','train_station','parking_space',
        'payment_terminal','pedestrian','turnstile','vehicle_inspection','weighbridge'
    ],

    "leisure" : [
        'shopping_center','arts_centre','bar','biergarten','casino','chapel','church','cinema',
        'brewery','stadium','dojo','exhibition_centre','reail','retial','karaoke','gym',
        'wellness_center','hotel','museum','music_venue','triumphal_arch','internet_cafe',
        'love_hotel','marketplace','fairgrounds','store','nightclub','photo booth','shop',
        'department_store','planetarium','art_gallery','pub','spa','studio','mall','social_center',
        'art_studio','arts_center','freeshop','shopping','hobby_shop','shops','surf shop',
        'antique auto repair shop','sign shops','copyshop','nail salon','book_return','books',
        'book_drop','photo_booth','photography','bar; music_venue','music lessons',
        'live music venue','bed_and_breakfast','swimming_school','ski_school','sailing_school',
        'dance_school','dancing_school','theatre'
    ],

    "services" : [
        'auditorium','community_centre','courthouse','crematorium','embassy','events','events_venue',
        'fire_station','funeral_hall','salon','atm','bank','fire department','jobcentre','playground',
        'monastery','police','post_office','massage_chair','massage','masseuse','prison','public_bath',
        'public_council','recycling','service','shelter','social_centre','social_facility','city_hall',
        'town_hall','telephone','toilets','day care','daycare','childcare','warehouse',
        'warehouse;house','animal_shelter','shoe_shine','driving_school','waste_disposal'
    ],

    "health" : [
        'hospital;roof','roof;hospital','yes;hospital','hospital;clinic','home_health_care_service',
        'healthcare','nursing','birthing_centre','clinic','foot_clinic','health_evaluation','health',
        'clinicacentre','dentist','doctor','doctors','audiologist','blood_donation','health_center',
        'first_aid','health_club','medical','first_aid_school','urgent_care','nutrition_counselling',
        'Sleep Disorder','chriopractic','yes;clinic','mental_health_service','doctor;clinic',
        'doctors;pharmacy','rhinoplasty','urologist','radiology','blood_bank','emergency_bay',
        'emergency_access_point','emergency room','emergency_service',
        'clinic;laboratory;physiotherapist;occupational_therapist','fertility_clinic','oral_surgery',
        'plastic_surgeon','orthopaedics','physiotherapy','pyschotherapist','testosterone_replacement',
        'speech_therapist','otolaryngologist','fertility_clinic','physiotherapy','pyschotherapist',
        'physical_therapy','craniosacral_therapy','therapy','physiotherapy','family therapy',
        'general_medicine','podiatry medical billing','medical_practise','medical_equipment',
        'medical_imaging','hospital','nursing_home','optometrist','massage_therapy','pharmacy',
        'mental_health_service','doctor;pharmacy','physiotherapist','podiatrist','psychotherapist',
        'veterinary','veterinary_clinic','ultrasound','ambulance','rehabilitation'
    ],

    "education" : [
        'education_centre','college','kindergarten','health_school','language_school','library',
        'music_school','prep_school','school','university','yes;school','trade_school','flight_school',
        'cooking_school','preschool','school;yes','high school','coding_school','bartending_school',
        'roof;school','school;roof','grade_school','schoolyard','social_facility;school','art_school',
        'business school'
    ],

    "work" : [
        'business','conference_centre','data_center','townhall','government','public_building',
        'loading_dock','factory','coworking_space','manufacture','commercial','construction',
        'engineering','industrial','workshop','working_space','conference_center','office','retail',
        'central_office','retail;commercial;office','government_office','apartments;hotel;office',
        'offices','sales_office','leasing_office','register_office','warehouse;office'
    ],

    "residential" : [
        'mansion','residential','house','boathouse','retirement_home','home','homes for sale',
        'trailer_home','townhome','residential;house','residential_condominium','garage;residential',
        'semidetached_house','residential;roof','roof;residential','residence','apartment','garage',
        'dormitory','apartments'
    ],
}
\end{lstlisting}

% 	\newpage
	
%
% BIBLIOGRAPHY
%
%\bibliographystyle{naturemag-doi} 
%\bibliography{biblio}
	
% 	\begin{filecontents}{mybib.bib}
%     \end{filecontents} 

%\end{document}
% end of document

%\bibliographystyle{naturemag-doi} 
\bibliography{biblio}

\end{document}